\documentclass{aa}
\usepackage{graphicx}
\usepackage{latexsym,lscape,longtable}

\def\hmpc{h^{-1} {\rm Mpc}}
\def\kms{\ {\rm km~s^{-1} }}
\def\kmsmpc{\ {\rm km~s^{-1} Mpc^{-1}}}
\def\d3k{{\displaystyle {\rm d}{\bf k} \over \displaystyle (2\pi)^3}}
\def\greatapprox{\setbox0=\hbox{$>$}\setbox1=\hbox{$\sim$}
     \lower0.5\ht0
     \hbox{\vbox{\baselineskip=0pt\lineskip=0.5pt\box0\box1} }}
\begin{document}
\title{Local Supercluster Dynamics:\\
External Tidal Impact of the PSC$z$ sample\\
traced by \\
Optimized Numerical Least Action Method\\}

\author{E. Romano-D\'{\i}az \inst{1}, E. Branchini \inst{2}
  \and    R. van de Weygaert \inst{1}
}

\offprints{E. Romano-D\'{\i}az, \email{emilio@phys.huji.ac.il}}

\institute{Kapteyn Astronomical Institute, University of Groningen,
  P.O. Box 800, 9700 AV Groningen, The Netherlands.
  \and Dipartimento di Fisica, Universita' Degli Studi di Roma Tre, Via
  della Vasca Navale 84, 00146 Roma, Italy.\\}

\date{Received 15.09.2004; accepted 20.04.2005}

\abstract{We assess the extent to which the flux-limited IRAS PSC$z$
  redshift survey encapsulates the complete or major share of matter
  inhomogeneities responsible for the external tidal forces affecting
  the peculiar velocity flow within the Local Supercluster and its
  immediate surroundings. We here investigate this issue on the basis
  of artificially constructed galaxy catalogs. Two large unconstrained
  $N$-body simulations of cosmic structure formation in two different
  cosmological scenarios form the basis of this study.  From these
  $N$-body simulations a set of galaxy mock catalogs is selected. From
  these a variety of datasets is selected imitating the observational
  conditions of either the local volume-limited Local Supercluster
  mimicking NBG catalog or the deeper magnitude-limited PSCz catalog.
  The mildly nonlinear dynamics in the ``mock'' Local Supercluster and
  PSC$z$ velocities are analyzed by means of the Least Action
  Principle technique in its highly optimized implementation of the
  Fast Action Method. By comparing the velocities in these
  reconstructions with the ``true'' velocities of the corresponding
  galaxy mock catalogs we assess the extent and nature of the external
  tidal influence on the Local Supercluster volume. We find that the
  dynamics in the inner $30\hmpc$ volume is strongly affected by the
  external forces. Most of the external forces can be traced back to a
  depth of no more than $100\hmpc$. This is concluded from the fact
  that the FAM reconstructions of the $100\hmpc$ PSCz volume appear to
  have included most gravitational influences.  In addition, we
  demonstrate that for all considered cosmological models the bulk
  flow and shear components of the tidal velocity field generated by
  the external distribution of PSC$z$ galaxies provides sufficient
  information for representing the full external tidal force field.

  \keywords{Cosmology: theory - large-scale structure of Universe - 
    Methods: numerical - Surveys }
}

\titlerunning{External Tidal Impact on Local Supercluster Dynamics}
\authorrunning{E. Romano-D\'{\i}az et al.}
\maketitle
%


\section{Introduction}

Migration flows of cosmic matter are one of the major physical
manifestations accompanying the emergence and growth of structure out
of the virtually homogeneous primordial Universe. The cosmic flows
displace matter towards the regions where ever more matter
accumulates, ultimately condensing into the objects and structures we
observe in the Universe.

Within the gravitational instability scenario of structure formation,
the displacements are the result of the cumulative gravitational force
exerted by the inhomogeneous spatial matter distribution of
continuously growing density surpluses and deficits throughout the
Universe. This establishes a direct causal link between gravitational
force and the corresponding peculiar velocities. Given a suitably
accurate measurement of peculiar velocities within a well-defined
``internal'' region of space, $V_{int}$, we may invert these
velocities and relate them to the inducing gravitational force. Hence,
the source of the measured motions may be traced and possibly even
reconstructed.  In principle, it may even allow us to infer the total
amount of mass involved and thus estimate the value of the
cosmological density parameter $\Omega_m$ and other fundamental
cosmological parameters.

The practical execution of such studies of cosmic velocity flows is
ridden by various complicating factors. One major complication is that
the cosmic regions in which peculiar velocities have been determined
to sufficient accuracy may have a substantially smaller size than what
may be deemed appropriate for a dynamically representative volume.
Ideally, in order to account for almost the complete flow in our local
cosmic neighbourhood we should have probed the density field in a
sufficiently large cosmic volume. This should involve a region of
space substantially superseding that of the characteristic scale of
the largest coherent structures in the Universe. Only then the
magnitude of the gravitational influence of inhomogeneities at larger
distances will represent a negligible contribution and, as well, start
to even out against each other.

The size of this dynamically effective volume depends sensitively on
the structure formation scenario which is prevailing in our Universe.
Hence, it will be closely affiliated to the spatial distribution,
characteristic size and coherence scale of cosmic structures, and its
size will therefore be in the order of the scale of the largest
pronounced structures in the Universe. Within the conventional
structure formation models, based on Gaussian initial density and
velocity fields, this is fully specified through the scenario's
fluctuations power spectrum $P(k)$. When the power spectrum involves a
substantial large-scale component and the survey volume is rather
limited we have to be aware of significant external influences.
Although not yet exactly determined, observational evidence suggests
its size to be in the range of $\approx 100-200 \hmpc$ (where $h$
denotes the Hubble constant in units of $100 \kmsmpc$).

An equally important consideration concerns the spatial resolution at
which the velocity field is studied. Available samples of galaxy
peculiar velocities extend out to reasonable depth of $\approx
60 \hmpc$. Yet, they involve a rather coarsely and
inaccurately sampled cosmic velocity field.  By absence of precise
distance estimators, more accurately and densely sampled velocity
information is therefore mostly confined to a rather limited region in
and around the Local Supercluster [LS].  As a consequence, most
analyses of large-scale cosmic flows are necessarily confined to
spatial scales at which the evolving cosmic structures are still
residing in a linear phase of development. The dynamics in more
advanced stages of cosmic structure formation are as yet poorly
constrained by measurements.


\subsection{Cosmic Force Fields and Supercluster Dynamics}

In this work we wish to extend the analysis of cosmic flows to the
more advanced evolutionary stages pertaining within supercluster
regions. Only within the local cosmic neighbourhood of our Local
Supercluster, the quality, quantity and spatial coverage of the
peculiar velocity data are sufficiently good to warrant an assessment
of the cosmic velocity field and the corresponding dynamics at a
sufficiently high spatial resolution.  On these quasi-linear or mildly
nonlinear scales we hope to find traces of the onset towards the more
advanced stages of cosmic structure formation.  In order for this to
yield a meaningful and successful analysis, two major questions have to
be addressed. Both form the main focus of this contribution.

The first issue, that of the rather restricted sample volume,
constitutes the major incentive behind this work. The volume of the
galaxy catalog that best samples our local cosmic neighbourhood, the
Nearby Galaxy Catalog (Tully \cite{tullycat}, hereafter [NBG]), is
certainly substantially smaller than what may be considered
dynamically representative.  Any analysis of the (internal) velocity
field in our Local cosmic neighbourhood should therefore take into
account the impact of external gravitational influences.

We focus on two related problems. In the first place, there is the
need to quantify the effect of neglecting the external gravitational
influence ${\bf g}_{ext}$ when modeling the cosmic velocity field on
scales comparable to that of the Local Supercluster. Various studies
have attempted to determine cosmological parameters on the basis of a
comparison between modelled versus measured velocity field in the
Local Universe (Tonry \& Davis \cite{tonry}, Tully \& Shaya
\cite{tully}, Shaya, Peebles \& Tully \cite{shaya}, Tonry {\it et al.}
\cite{tonry00}). For this it is crucial to understand in how far local
density perturbations may account for the local peculiar gravity field
within the Local Supercluster.  Directly related to this is the need
to have a sufficiently accurate description of the external force
field ${\bf g}_{ext}$, in terms of its nature and spatial extent, in
order to properly model the total measured gravity field ${\bf
  g}_{tot}$. For studies intent on a comparison of modelled versus
observed peculiar velocities on scales larger than the Local
Supercluster this is an essential requirement (Faber \& Burstein
\cite{faber}, Han \& Mould \cite{han}, Yahil {\it et al.}\cite{yahil},
Webster {\it et al.} \cite{webster}, Branchini {\it et al.}
\cite{branchini99}). Similar considerations are equally relevant for
the inverse problem, in which one attempts to infer the external
gravitational influence ${\bf g}_{ext}$ from peculiar velocities
measured within the Local Supercluster (e.g. Lilje, Yahil \& Jones
\cite{lilje}, Lynden-Bell \& Lahav \cite{lynden}, Kaiser
\cite{kaiser}, Hoffman {\it et al.}  \cite{hoffman01}). Indeed, both
problems concerning the restricted galaxy sample volume have figured
prominently in previous cosmic velocity field studies and were
addressed in a variety of publications. However, usually these tend to
discard the fact that the local cosmic region in which we have access
to high quality velocity data has already reached an advanced
quasi-linear dynamical state.

Referring to the latter, the second major issue concerns the
innovative way in which we evaluate the dynamical state of
superclusters. These structures reside in a mildly nonlinear
evolutionary stadium, having evolved significantly beyond their
initial linear phase. Unlike the vast majority of previous studies, we
seek to probe into the more detailed and informative kinematic aspects
of these structures.  A conventional linear analysis will not be able
to provide an adequate description, and for the most evolved
circumstances not even the Zel'dovich approximation (Zel'dovich
\cite{zeld70}) may be expected to do so. In order to be able to
optimally exploit the available velocity information -- without
suffering the loss of valuable high-resolution information through a
filtering process -- we apply the Least Action Principle [LAP]
formalism (Peebles \cite{peebles89}) for dealing with the individual
galaxy velocities. To that end, an optimal implementation developed by
Nusser \& Branchini (\cite{nusbran}, hereafter [NB]), the Fast Action
Minimization [FAM] proved an essential tool.

Elaborating on the first issue of the external gravitational
influence, one of the as yet undecided issues is the extent to which a
LAP analysis of a cosmological self-gravitating system is dependent on
a proper representation of the external gravitational influence.
Various strategies have been followed, ranging from a complete neglect
of external forces (Peebles \cite{peebles89}), or taking account of
the influence of merely a few nearby objects (Peebles
\cite{peebles89}, \cite{peebles90}, Dunn \& Laflamme \cite{dunnlaf93},
\cite{peebles2001}), towards methods involving approximate
descriptions of external influences. The latter mostly incorporated
the wider external influence through a frozen, linearly evolving,
external tidal field estimated on the basis of the present-day
locations of an extended sample of objects deemed representative for
the external matter distribution (e.g. Shaya, Peebles \& Tully
\cite{shaya}, Schmoldt \& Saha \cite{schmoldtsaha}, Sharpe {\it et
  al.} \cite{sharpe}). This study does include the influence of force
fields, but does so in a fully systematic and self-consistent fashion,
enabled by the FAM method to take into account the evolution of the
full sample of external matter concentrations.

The principal conclusion of our study is that the gravitational forces
exacted by the matter inhomogeneities encapsulated by the IRAS-PSC$z$
redshift survey sample (Saunders et al. \cite{saunders}) are indeed
able to account for all motions within our local Universe. In
addition, we demonstrate that its external influence may almost
exclusively be ascribed to the bulk and shear flow components.


\subsection{Strategy}

This study is based on a number of artificial galaxy samples mimicking
the properties of genuine catalogs. They consist of several
well-defined and well-selected model catalogs of galaxies and galaxy
peculiar velocities.  These mock samples are extracted from a set of
extensive $N$-body simulations: for the nearby Universe models they
adhere to the characteristics of the Nearby Galaxy Catalog, for the
deep galaxy redshift samples they are modelled after the IRAS PSC$z$
catalog. These model samples allow us to thoroughly investigate the
various strategies forwarded for a successful and conclusive analysis.

Two sets of realistic mock catalogs of galaxies are extracted.  The
first set, the ``local'' one, is meant to mimic the mass distribution
within the LS as traced by galaxies in the Nearby Galaxy Catalog of
Tully (\cite{tullycat}).  It consists of a volume-limited galaxy
sample within a (spherical) interior region with radius $30 \hmpc$.
Each of these interior samples is embedded within a larger mock
sample, the ``extended'' sample which covers a larger cosmic region.
In addition to the interior volume-limited sample in the inner
$30\hmpc$ they contain an enclosing outer flux-limited sample covering
the surrounding spherical region located between $30 \hmpc < x < 100
\hmpc$. This exterior sample mimics a flux-limited galaxy catalog
whose characteristics are modelled after the IRAS PSC$z$ sample.

For both the ``local'' and ``extended'' mock samples we model the
peculiar velocity field at the positions of the particles in the
``local'' cosmic region, i.e. for the objects out to a radius $x <
30\hmpc$. These model predictions result from the application of Fast
Action Minimization method (Nusser \& Branchini \cite{nusbran}). As
our FAM reconstruction procedure only takes into account the
gravitational forces between the particles in the mock samples -- i.e.
does not include contributions from outside objects -- the differences
in predicted velocities between the ``local'' and ``extended'' samples
will reflect the influence of the mass concentrations in the
surrounding region $30 \hmpc < r < 100 \hmpc$. The comparison with the
corresponding $N$-body velocities, representing the ``real''
velocities, will inform us in how far ``sky-covering'' samples of
galaxies with a depth of $\approx 100\hmpc$ may be expected to
represent a proper cosmic region as far as its dynamics are concerned.
The strategy of analyzing and comparing the velocity models obtained
from the small ``local'' mock catalogs, the large ``extended'' mock
catalogs, and the ``real'' velocities in the original $N$-body
simulations, will yield a solid understanding of the effect of
neglecting the externally induced peculiar gravitational acceleration
${\bf g}_{ext}$ (see Eqn.~(1)).

When analyzing a dataset of galaxy peculiar velocities in the local
Universe. The analysis of the larger PSC$z$ mimicking catalogs should
elucidate if and to what extent such samples will be able to account
for ${\bf g}_{ext}$. If the galaxies in these samples indeed appear to
be responsible for the major share of the external forces, we may feel
reassured to use the PSC$z$ sample of galaxies for a proper
representation of ${\bf g}_{ext}$.

One aspect of this question concerns the investigation of the question
whether the external tidal influence may be explicitly framed in an
analytical approximation consisting of a dipolar and quadrupolar term.
Our fully self-consistent FAM reconstructions, in which the
``extended'' mock catalogs are processed with the inclusion of all
external matter concentrations, enable us to estimate the bulk and
shear components in the induced ``local'' galaxy motions. By comparing
the resulting velocity fields in the ``local'' and ``extended''
samples we will be able to judge the quality of the approximate
methods, and quantify and investigate the possible presence of
systematic trends throughout the ``local'' cosmos.

To account for possible systematic effects due to global cosmology,
the mock galaxy catalogs are extracted from $N$-body simulations in
two different cosmic structure formation scenarios. One involves a
$\Lambda CDM$ Universe with a characteristic large-scale dominated
power spectrum, while the other concerns a $\tau CDM$ cosmology.  The
more small-scale dominated character of the latter leads to a
different character of its gravitational field fluctuations, the
smaller coherence scale of the density field fluctuations yielding a
comparatively smaller influence of the external (quadrupolar) tidal
field (the induced bulk flows are similar, as the smaller $\tau CDM$
fluctuations are exactly compensated by the larger mass involved). The
resulting comparisons of FAM velocity field reconstructions are
expected to reflect these velocity field differences.
 
In the end, this study of artificial galaxy samples should allow us to
appreciate the manifestations of the real physical effects we wish to
grasp. In this, we also should learn how to deal with the
complications due to the host of measurement uncertainties which beset
the observational data. The scope is to quantify the systematic errors
which might have affected similar, local, comparisons based on real
data and to judge whether the information on the external mass
distribution available to these analyses is indeed sufficient to
account for ${\bf g}_{ext}$.


\subsection{Outline}

In the next section, we will elaborate on the astrophysical background
of this study, the study of velocity flows on cosmological scales, and
in particular the issue of internal and external gravitational
influences. Ensuingly, we address the specific problem of treating the
dynamics and related cosmic motions within mildly nonlinear structures
such as the Local Supercluster. This brings us to a brief exposition
on the LAP analysis for dealing with the complications of mildly
nonlinear orbits and the technical issue of the FAM technique which
allows us to apply this to a system composed of many objects. Special
emphasis is put on the inclusion of external gravitational influences
within the LAP/FAM formalism. In section 4 we describe the
cosmological setting of the simulations on which this study has been
based. As a guidance towards interpreting our calculations, we address
a variety of theoretical aspects and predictions concerning cosmic
velocity fields in these cosmological scenarios. The basis of this
work is the set of two ``parent'' N-body simulations and the mock
catalogs extracted from these simulations, forty in total. They are
presented in section 5. In the subsequent sections we present the
results obtained from the various FAM computations. In section 6 we
analyze the velocity vector maps for the FAM reconstructions. These
maps allow a direct and visually illuminating appreciation of the
effects we wish to address. This is followed by a first quantitative
assessment in section 7. This consists of a comparison between the FAM
velocity field reconstructions of the Local Supercluster volume($r <
30\hmpc$), the FAM reconstructions for the corresponding PSC$z$ sample
and the complete ``real world'' N-body velocity field. The comparison
is mainly based on a point-by-point evaluation through scatter
diagrams of velocity-related quantities.  To encapsulate these results
into a spatially coherent description of the large scale external
velocity and gravity field, in section 8 we turn to a decomposition of
the peculiar velocity field into multipolar components. In particular,
we demonstrate that a restriction to its dipolar and quadrupolar
components, i.e. the bulk flow and velocity shear, does represent a
good description. Thus having looked at the issue of cosmic velocity
fields from different angles, the summary of section 9 will focus on
the repercussions of our analysis and its relation to the study of the
(relatively nearby) surrounding matter distribution. On the basis of
these conclusions we provide a description of the various projects
which follow up on this work, together with some suggestions for
additional future work.


\section{Cosmic Flows:\\ \ \ \ 
  probes of cosmic matter distribution}


\subsection{the Large-Scale Universe: linear flows}

Over the past two decades a major effort has been directed towards
compiling large samples of galaxy peculiar velocities. These samples
made it possible to obtain a rather impressive view of cosmic dynamics
on scales $\greatapprox 10 \hmpc$. In particular the Mark
III catalog, with an effective depth $\approx 60 \hmpc$,
stands as a landmark achievement (Willick {\it et al.}
\cite{willicka}, also see Dekel \cite{dekel} and Strauss \& Willick
\cite{strauss}).  Further progress has been enabled by the assembly of
additional and partially complementary samples of galaxy peculiar
velocities, of which the SFI late-type galaxy and ENEAR early-type
galaxy samples are noteworthy examples. The SFI Catalog of Peculiar
Velocities of Galaxies (Giovanelli et al. \cite{sfi1}, Giovanelli et
al. \cite{sfi2}, Haynes et al. \cite{sfi3} and Haynes et al.
\cite{sfi4}) consists of around 1300 spiral galaxies with I-band
Tully-Fisher (TF) distances, out to $cz < 7500 \kms$. The ENEAR sample
(da Costa et al. \cite{costaenear}) is an equivalent sample of around
1600 early-type galaxies, out to a distance $cz < 7000 \kms$, with
$D_n-\sigma$ distance estimates available for nearly all of them.
Tracing cosmic motions over larger volumes of space is a rather more
cumbersome affair and attempts to do so are mainly based upon the
peculiar motions of galaxy clusters. The claim of a puzzlingly large
flow over scales of $150\hmpc$ by Lauer \& Postman~\cite{laupost}
could not be corroborated. Nonetheless, flows on such large scales may
indeed be a reality, as has been inferred from the far better defined
``Streaming Motions of Abell Clusters'' (SMAC) sample of Hudson et
al.~\cite{hudsiii}. They did recover a bulk flow in the order of
$687\pm 203 \kms$, of which $225 \kms$ may arise from sources at a
distance larger than $100 \hmpc$ (Hudson et al.~\cite{hudsv}). One
prime objective of most analyses of these large samples of peculiar
velocities has been the determination of the cosmological mass density
parameter $\Omega_m$ (Davis, Nusser \& Willick 1996, Willick {\it et
  al.} \cite{willickb}, Willick \& Strauss \cite{willick}, Nusser {\it
  et al.} \cite{nusseretal}, Branchini {\it et al.}
\cite{branchini01b}).  Such assessments are based on a comparison of
observed velocities to a model velocity field. A basic requirement for
obtaining self-consistent estimates of $\Omega_m$ is that the velocity
samples concern a ``representative'' volume of space.

However, even while such studies appear to succeed in attuning the
large-scale matter distributions and velocity fields in a reasonably
self-consistent fashion, doubts remain with respect to a variety of
practical and systematic problems. Firstly, in these comparisons the
random errors on the observed velocities are substantial, much larger
than those in the structure formation models. Considerable effort has
been directed towards quantifying and minimizing errors on the
observed peculiar velocities (e.g. Dekel \cite{dekel}, Strauss \&
Willick \cite{strauss}, and references therein). These involve random
measurement errors as well as more subtle systematic, yet reasonably
well understood, errors. Secondly, there remain various systematic
effects which have not been addressed and corrected for in an equally
convincing fashion. Even though they also tend to play a role with
respect to the model predictions they are often overlooked.

A major systematic factor concerns the incomplete information on the
spatial mass distribution within the region of the sample itself. This
prevents an adequate treatment of artifacts due to the incomplete sky
coverage and limited depth of the available samples, and effects
systematic errors stemming from luminosity and density effects. These
systematic errors are usually accounted for by using large, all-sky
redshift surveys, such as the Optical Redshift Survey of Santiago {\it
  et al.} (\cite{santiago}) or the 1.2 Jy and PSC$z$ surveys of IRAS
galaxies (Fisher {\it et al.} \cite{fisher}, Saunders {\it et al.}
\cite{saunders}). In particular when using IRAS based surveys the
effects of incomplete sky coverage are greatly reduced.

Even more problematic for a successful handling of luminosity and
density related effects is our incomplete knowledge with respect to
the relationship between the observable galaxy distribution and the
underlying mass distribution. By absence of a compelling theory of
galaxy formation this ``galaxy bias'' is usually encapsulated in
heuristic formulations.  The rather ad-hoc and possibly unrealistic or
inadequate nature of the latter may seriously affect the significance
of the inferred conclusions. Most studies make the simplifying
assumption of a galaxy population fairly tracing the underlying
density field. This is usually embodied in a global and linear
``galaxy bias'' factor. A large variety of results suggest that this
may be a reasonable approximation on scales in excess of a few
Megaparsec. Moreover, while this bias may be problematic in the case
of early-type galaxies, it has proved to be quite successful with
respect to the later type galaxies which figure prominently in IRAS
based samples (Verde et al. \cite{verde2002}).


\subsection{Internal and External influences}

Unlike most studies of cosmic flows which seek to assess and analyze
the nature and source of dynamical influences within a confined region
of space, we try to get an impression of the cosmic dynamics on mildly
nonlinear scales of only a few Megaparsec. We focus on the Local
Supercluster region and its immediate neighbourhood. The galaxy sample
of the NBG catalog is taken to be representative for this region.
Because the catalog entails a volume which is substantially smaller
than what may be considered dynamically representative, the peculiar
velocities of the galaxies are partially due to the gravitational
action by outside matter concentrations. That is, the peculiar
velocities are not only due to the gravitational force induced by the
matter concentrations within the ``internal'' survey volume $V_{int}$,
but also reflect the gravitational influence by the ``external''
matter density distribution, ${\bf g}_{ext}$. Because it does not
constitute a truly representative volume of the Universe, a
meaningful dynamic analysis of the Local Universe on the basis of the
NBG sample is substantially complicated by its limited depth, which is
one of the major systematic problems besetting the analysis of
virtually all available surveys of galaxy peculiar velocities.
Theoretical models of peculiar velocities nearly always involve the
implicit assumption of the mass being homogeneously distributed
outside $V_{int}$, so that its gravitational effect may be neglected.
Even in the case of having a sufficiently large volume at one's
disposal, this approximation is only valid in the central part of
$V_{int}$, certainly not near its edges.

The distinction between external versus internal gravitational force
may be best appreciated by noting that the total (peculiar) gravity
field ${\bf g}_{tot}({\bf x})$ is the netto sum of the individual
contributions by all patches of matter throughout the visible
Universe. At any position within the internal volume $V_{int}$, we may
then decompose the full gravitational field into an ``internally''
induced component ${\bf g}_{int}$ and an ``externally'' generated
contribution ${\bf g}_{ext}$,
\begin{equation}
  {\bf g}_{tot}({\bf x}) = {\bf g}_{int}({\bf x}) + {\bf g}_{ext}({\bf x})\ .
  \label{eq:grav}
\end{equation}
In this way we have defined the internal gravitational force 
${\bf g}_{int}$ as the integrated contribution from the density
fluctuations $\delta({\bf x})$ within the volume $V_{int}$, while the
external gravitational force ${\bf g}_{ext}$ concerns the combined
gravitational force generated by the density fluctuations outside the
realm of $V_{int}$, so that
\begin{eqnarray}
  {\bf g}_{tot}({\bf x},t) & = & {\displaystyle 3 \Omega H^2 \over 
    \displaystyle 8\pi}\,
  \int_{V_{int}} {\rm d}{\bf x}'\,\delta({\bf x}',t)\,{\displaystyle 
    ({\bf x}'-{\bf x})
    \over \displaystyle |{\bf x}'-{\bf x}|^3}\ + \ \nonumber\\
  && {\displaystyle 3 \Omega H^2 \over \displaystyle 8\pi}\,
  \int_{V_{ext}} {\rm d}{\bf x}'\,\delta({\bf x}',t)\,{\displaystyle 
    ({\bf x}'-{\bf x}) \over \displaystyle |{\bf x}'-{\bf x}|^3} \ . 
\end{eqnarray}
The peculiar velocities of galaxies within $V_{int}$ bear the mark of
both the acceleration due to the matter concentrations within the
volume itself, ${\bf g}_{int}$, as well as that of the combined
gravitational influence of the external mass distribution, ${\bf
  g}_{ext}$. A comparison of predicted internally induced velocities
with the observed local velocity field should therefore enable us to
infer the magnitude and nature of the external field ${\bf g}_{ext}$.
This analysis is usually facilitated by the fact that the fine details
of the external force contribution are largely negligible. The
contributions by the various external matter concentrations to the
combined gravitational force mostly average out such that what remains
noticeable is mainly confined to the low order components of the
multipole decomposition of ${\bf g}_{ext}$. This can be most readily
appreciated from a description of the external gravitational force
field in terms of its successive multipole components. When we expand
${\bf g}_{ext}$ around some central location in $V_{int}$ -- here
defined to be the origin of the coordinates ${\bf x}$ -- we find that
to second order
\begin{eqnarray}
  g_{ext,i}({\bf x})&\ =\ &\ g_{bulk,i}\ - \,\sum_{j=1}^3\,{\cal T}_{ij} \, 
  x_j \ .
\end{eqnarray}
In this, we assume that the additional divergence term 
${\displaystyle 1 \over \displaystyle 3} (\nabla \cdot {\bf
  g}_{ext})\, x_i$ has been embedded into the (zeroth) order {\it
  monopole} term. In essence, it corresponds to a ``breathing mode''
affecting the ``local'' Hubble expansion within the volume, and
therefore can not possibly be inferred from the local measurement of
the internal gravity field ${\bf g}_{int}$.

The leading term in the overall external gravitational acceleration is
the bulk gravity term $g_{bulk,i}$. This {\it dipole} term constitutes
the uniform acceleration of the matter within $V_{int}$,
\begin{eqnarray}
  {\bf g}_{bulk}\ = \ {\displaystyle 3 \Omega H^2 \over \displaystyle 8\pi}\,
  \int_{V_{ext}} {\rm d}{\bf x}'\,\delta({\bf x}',t)\,{\displaystyle {\bf x}'
    \over \displaystyle |{\bf x}'|^3}\
  \label{eq:gbulk}
\end{eqnarray}
Evidently, when considering peculiar velocities relative to 
the centre of mass inside the volume $V_{int}$ instead of absolute
velocities this constant vector disappears. The first term whose
magnitude and configuration is independent of the reference frame is
the {\it quadrupolar} term ${\cal T}_{ij}$.

If the contribution to the (peculiar) gravitational potential by the
external mass inhomogeneities is $\phi_{ext}$, the quadrupolar tidal
tensor ${\cal T}_{ij}$ is the trace-free part of $\partial^2
\phi_{ext}/\partial x_i \partial x_j$, evaluated at the centre of
$V_i$. It is determined by the external matter distribution through

\begin{equation}
  {\cal T}_{ij}(t) = {\displaystyle 3 \Omega H^2 \over \displaystyle 8\pi}
  \int_{V_{ext}} {\rm d}{\bf x}'\,\delta({\bf x}',t)\ \left\{
    {\displaystyle 3 x_i' x_j'- |{\bf x}'|^2\,\delta_{ij} \over 
      \displaystyle |{\bf x}'|^5}\right\} 
  \label{eq:tideij}
\end{equation}
\noindent

The integral expressions for the dipole and quadrupole components of
the external gravity field (eqn.~4 and 5), illustrate that it is
unfeasible to exploit the observed local cosmic velocity field to
recover the detailed and complete spatial distribution of the external
matter inhomogeneities. On the other hand, it does indicate how it is
that we can infer some overall characteristics of the external matter
distribution from an analysis of the local velocity field. From this
we may extract interesting and significant information on the nature
and even distribution of the large scale cosmic matter distribution
and set constraints on the values of some of the fundamental
cosmological parameters. The pioneering work by Lilje, Yahil \& Jones
(\cite{lilje}) in which the velocity field of the Local Supercluster was
exploited to infer the presence of a major external source of
gravitational attraction has shown the potential of this approach.
Ultimately, it inspired the analysis of Lynden-Bell et al.
(\cite{samur88}) that lead to the discovery of the Great Attractor.


\section{Cosmic Flows:\\ \ \ \ 
  the mildly nonlinear dynamics of superclusters} 

Even though a structure's evolution may have progressed to a dynamical
stage at which a first-order description of cosmic velocity fields
will no longer be adequate, it may still be possible to find a direct
link to the structure's initial configuration. This is in particularly
true for the early and mildly nonlinear phases of evolution. The
exemplary archetype of a structure in which such mildly nonlinear
circumstances are prevalent is that of superclusters, the filamentary
or wall-shaped elements of the cosmic foamlike matter distribution.

Over the past two decades intriguing foamlike patterns have gained
prominence as a prime characteristic of the cosmic matter
distribution. The first indications for the actual existence of a
foamlike galaxy distribution were provided by CfA2 redshift slices (de
Lapparent, Geller \& Huchra \cite{lappcfa}) and established as a
universal cosmic pattern with the Las Campanas redshift survey
(Shectman et al. \cite{lcrs}). With the arrival of the large recent
and ongoing systematic galaxy redshift surveys, the 2dF galaxy
redshift survey ($\approx 250,000$ redshifts, Colless et
al.~\cite{coll2df}, also see e.g.  Colless~\cite{coll2dfcosm} and
Tegmark et al.~\cite{teg2df} for a discussion on clustering in the
2dFGRS) and the Sloan Digital Sky Survey (SDSS, will determine
$\approx 1,000,000$ redshifts, see e.g. Zehavi et al.~\cite{zehasdss}
and Tegmark et. al.~\cite{tegsdss} for an overview of present-day
status wrt. galaxy clustering), we may hope to have entered the stage
in which we will be enabled to explore the formation and the dynamics
of these characteristic spatial structures in the cosmic matter
distribution. The typical elements of the cosmic foam -- filamentary
and wall-shaped superclusters -- are precisely at the youthful yet
mildly nonlinear phase of development mentioned earlier. They were
identified as such within the context of Zel'dovich' ``pancake''
theory of cosmic structure formation (see e.g. Shandarin \& Zel'dovich
\cite{shandzeld}). The significance of the cosmic foamlike network for
the understanding of the process of cosmic structure formation has
since been generally recognized. This may be appreciated from the
widespread use of the concept of the `cosmic web', coined by Bond,
Kofman \& Pogosyan (\cite{cosmweb}) in their study of the dynamics
underlying its formation (see Van de Weygaert \cite{weyfoam} for a
recent general review).

Mildly nonlinear cosmic features such as superclusters have recently
turned their initial co-expansion into a genuine physical contraction
(or are on the brink of doing so), marking the emerging structure as a
truely identifiable entity.  Once it has surpassed this
``turn-around'' stadium the complexity of its internal kinematics
quickly increases. At first moderately, and ultimately dramatically as
the virialization process advances, the matter orbits inside the
structure become more and more complex. Even in the more moderate
early phases of this process, an appropriately sophisticated treatment
of the mildly nonlinear dynamics appears to be a necessary requisite
for any study based upon kinematic information.  In and around
emerging nonlinear structures a simple linear analysis for
reconstructing initial conditions will therefore no longer suffice. In
other words, a sufficiently detailed and profound understanding of the
emergence of these key elements in the cosmic matter distribution
cannot be obtained without the development of a more elaborate
technique for the analysis of cosmic velocity fields.


\subsection{Structure formation: mildly nonlinear dynamics}

A linear analysis simplifies the dynamical evolution of a system into
an initial conditions problem. It implies the reconstruction of the
primordial density and velocity field by means of a simple linear
inversion of the observed matter distribution and galaxy peculiar
velocity field. Such an approach may even be followed towards a
slightly more advanced stage. The Zel'dovich formalism, a Lagrangian
first-order approximation for gravitationally evolving systems, has
been remarkably successful in describing the early nonlinear evolution
of a supercluster (for a review, see Shandarin \& Zel'dovich
\cite{shandzeld}). Substantially surpassing its formal linear
limitations, it proved to be a highly versatile medium for analyzing
and explaining the overall spatial morphology and characteristics of
emerging structures. The Zel'dovich approximation elucidated and
explained qualitatively the fundamental tendencies of gravitational
contraction in an evolving cosmos. Perhaps most noteworthy this
concerned the tendency of gravitational collapse to proceed
anisotropically, together with its predictive power with respect to
location and timescales of the first phase of collapse into planar
mass concentrations, ``pancakes''. This offered the basic explanation
for the foamlike morphology of the cosmic matter distribution,
stressing its existence many years in advance of its discovery through
observational programs to map the galaxy distribution (for an
extensive review of various nonlinear approximation schemes seeking to
expand upon the Zel'dovich approximation see Sahni \& Coles
\cite{sahnicoles}).

\begin{figure*}[t]
  \sidecaption
  \includegraphics[width=5.5in]{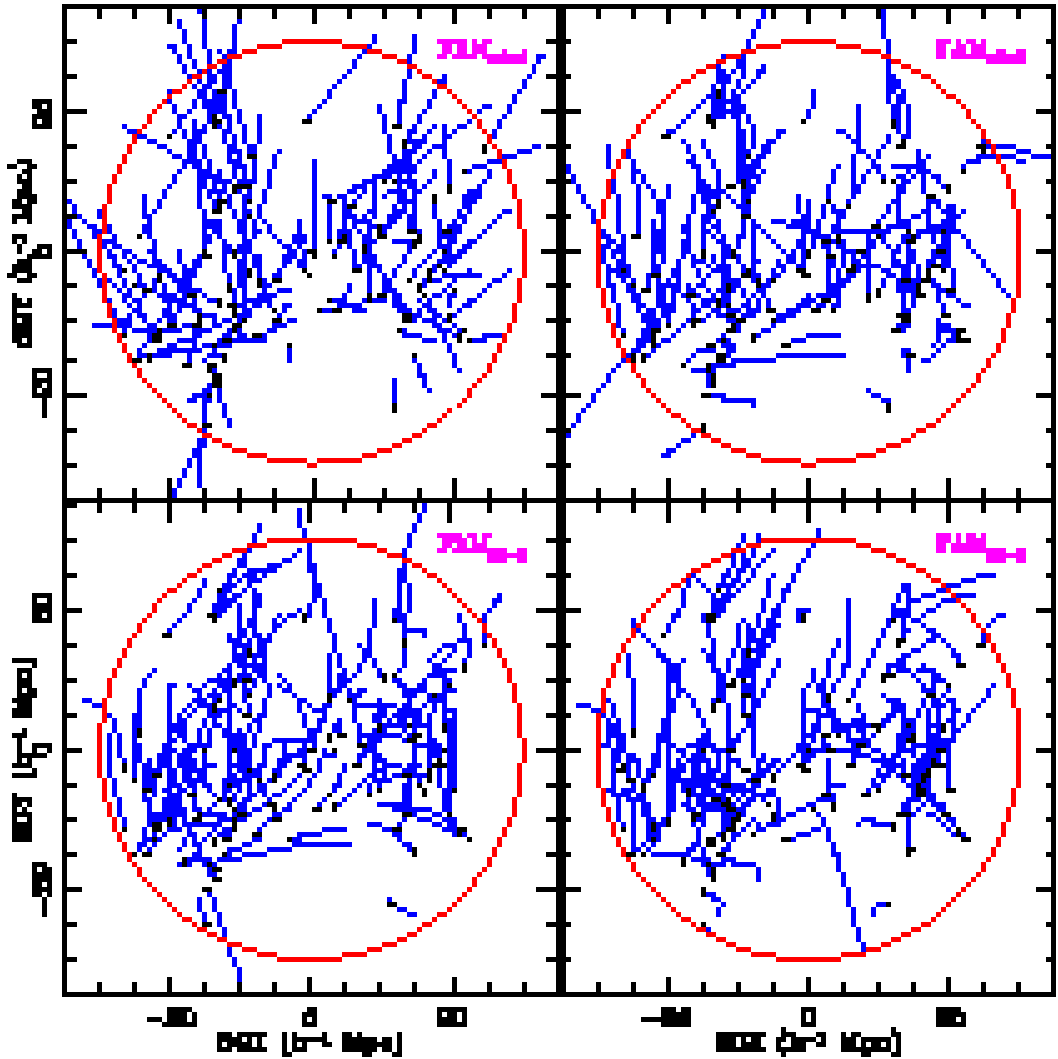}
  \caption{\ \ 2D projected reconstructed FAM orbits for different values 
    of $N_f$.  The black dots represent the final (present) positions
    for each object.  The solid lines indicate the trajectories
    followed by the objects as a function of time. The top-left panel
    shows FAM reconstructed orbits with $N_f=1$ (Zel'dovich
    approximation). The top-right with $N_f=2$, the lower-left $N_f=3$
    and the lower-right $N_f=6$.}
  \label{fig:orbits}
\end{figure*} 

In line with the above, the Zel'dovich approximation proved a highly
versatile tool for the analysis of the cosmic matter flows. It was
successfully applied to the nonlinear situation of mixed boundary
conditions -- tested and calibrated using $N$-body simulations -- by
Nusser et al. (1991) and Nusser \& Dekel (1992).  However, its
validity remains restricted to the early stages of nonlinearity at
which there is still a linear and direct relation between velocity and
gravity field. Once matter inside the emerging structures starts to
reach densities so high that local interactions become dominant, the
Zel'dovich scheme quickly ceases to lose its applicability. Once
matter elements start to cross each each others path the interaction
between the nonlinear matter concentrations within the realm of the
contracting structure will more and more deflect the orbits away from
their initial linear trajectory. The linear kinematics of the
Zel'dovich approximation will therefore no longer be able to follow
the orbits of the matter elements. Higher order approximations based
on perturbation theory have been advocated in order to follow such
more advanced nonlinear circumstances. However, the improvement over
simple first order Zel'dovich approximation turns out to be quite
limited and not warranting the effort invested at each successive
perturbation step. This is particularly so as with the onset of
nonlinearity the rate at which successive perturbative orders terms
become significant rapidly accelerates.


\subsection{Least Action Principle in Cosmology}

In more advanced nonlinear circumstances we encounter a more generic
dynamical situation than a simple {\it initial value problem}.
Typically, one seeks to compute the velocity field consistent with an
observed density structure at the present epoch or, reversely, one
deduces the density from the measured peculiar galaxy velocities. In
the case of generic systems, the dynamical evolution represents a {\it
  mixed boundary condition} problem. This implies the system to be
sufficiently constrained by complementing the incomplete dynamical
information regarding the initial conditions with that pertaining to
the dynamical state of the system at a different epoch. While $N$-body
codes are particularly concerned with the ideal circumstances usually
embodied in terms of {\it initial value problems}, a different kind of
technique needs to be invoked to exploit the typical {\it mixed
  boundary} information yielded by observations.

A more profound and direct exploitation of the available information
to follow the physics of such a cosmological nonlinear system was
forwarded by Peebles (\cite{peebles89}, \cite{peebles90}). He pointed
out that finding the orbits that satisfy initial homogeneity -- and by
implication vanishing initial peculiar velocities -- and match the
(present-day) observed distribution of mass tracers constitutes a
mixed-boundary value problem. Such problems lend themselves naturally
to an application of Hamilton's principle. This naturally leads to the
formulation of the Least Action Principle (also known as ``Numerical 
Action Method''), based on a variational
analysis of the action $S$ of an isolated system of $M$ particles,
which at a cosmic expansion factor $a(t)$ is given by
\begin{equation}
  S\ = \ \int_0^{t_0} L dt \ = \int_0^{t_0} dt\ \sum_i\ [{1 \over 2} m_i a^2 
  {\dot {\bf x}}_i^2 - m_i\,\phi({\bf r}_i)\ ]\ ,  
  \label{lapS}
\end{equation} 
in which $L$ is the Lagrangian for the orbits of particles 
with masses $m_i$ and comoving coordinates $x_i$ and corresponding
peculiar gravitational potential $\phi({\bf x})$. For a system of
particles interacting by gravity alone, embedded within a uniform
cosmological background of density $\rho_b(t)$, this yields the
following explicit expression for the action $S$,
\begin{eqnarray}
  S\ =\ \int_0^{t_0} dt&\ \Bigl\lbrack\,&\sum_i\,{m_i a^2 \over 2} \left(
    {d {\bf x}_i \over d t}\right)^2  +\ {G \over a} \sum_{i \not= j} 
  {m_i m_j \over |{\bf x}_i - {\bf x}_j|}  +\nonumber\\ 
  &&\ +\ {2 \over 3}\pi G \rho_b a^2 \sum m_i {\bf x}_i^2\,\Bigr\rbrack
\end{eqnarray}
The {\it exact} equations of motion for the particles are then
obtained from finding the stationary trajectories
amongst the variations of the action $S$ subject to fixed
boundary conditions at both the initial and final time.

Confining oneself to a feasible approximate evaluation in this 
{\it Least Action Principle} approach, one describes the orbits of particles, 
${\bf x}_i(t)$, as a linear combination of suitably chosen universal
functions of time with unknown coefficients specific to each particle
presently located at a position ${\bf x}_{i,0}$. For instance, by
using the linear growth mode $D(t)$ as time variable (Giavalisco et
al.  \cite{giav93}, Nusser \& Branchini \cite{nusbran}), one can
parametrize the orbit ${\bf x}_i(D)$ of a particle as
\begin{equation}
  {\bf x}_i(D)\ =\ {\bf x}_{i,0}\ +\ \sum_{n=1}^{N_f}\ q_n(D)\ {\bf C}_{i,n}\ ,
\end{equation}
where the functions $q_n(D)$ form a set of $N_f$ time-dependent basis
functions. The factors ${\bf C}_{i,n}$ are then a set of free
parameters, whose value is determined from evaluating the stationary
variations of the action.

The functions $q_{n}(D)$ satisfy both two orbital constraints,
necessary and sufficient to define solutions in agreement with
evolution in the context of the Gravitational Instability theory for
the formation of structure in the Universe: $q_{n}(1)=0$ ensures that
at the present time the galaxies are located at their observed
positions ${\bf x}_{i}(1)={\bf x}_{i,0}$ and $\lim_{D \to
  0}D^{3/2}q_{n}(D) {\bf \theta}(D)=0$ guarantees vanishing peculiar
velocities at early epochs which, in turns, ensures initial
homogeneity.


\subsection{Fast Action Minimization}
\label{sec:fam}

The successful application of the {\it Least Action Principle} towards
probing the kinematics and dynamics of an evolving cosmological system
depends to a large extent on the specific implementation. This will be
dictated by the characteristics of the physical system. In order to
enable a meaningful LAP analysis of large samples of galaxies, like
the Local Universe samples studied in this work, an optimized
procedure is necessary for dealing with the large number of objects.
Nusser \& Branchini (\cite{nusbran}) developed an optimized version of
Peebles' original LAP formalism, the {\it Fast Action Minimization}
method. The various optimization aspects of the FAM implementation
proved to be crucial for our purposes. The relevant optimization
hinges on {\bf four} major aspects of the FAM scheme.

The first FAM improvement involves the choice of time basis functions
$q_n(D)$. Its convenient choice of time basis functions yields a
simple expression for the action of the system and for its derivatives
with respect to ${\bf C}_{i,n}$. Both quantities relate to the
internal gravity term ${\bf g}_{int}$ of the system. Once the action
and its derivatives are evaluated numerically, the minimum of the
action is determined by means of the conjugate gradient method (Press
{\it et al.}  \cite{press}). The orbits of the system are then fully
specified through the set of parameters ${\bf C}_{i,n}$ found in
correspondence to the minimum.

Closely related to the first aspect is that of tuning the choice of
the time basis functions $q_n(D)$ such that only a limited number
$N_f$ of basis functions is needed to successfully parameterize the
orbits of the system. This is in particular beneficial to the the
physical configuration we are studying here, involving Megaparsec
scale dynamics characterized by quasi-linear or mildly nonlinear
motions.

Note that using the growth factor $D$ as time variable makes the
equations of motions almost independent of the value of $\Omega_m$
(Nusser and Colberg 1998).  As a consequence FAM orbits and peculiar
velocities in a generic $\Omega_m$ universe can be obtained by
appropriate scaling those assuming a flat cosmology.

A final major aspect of the FAM implementation involves the efficient
computation of the internal (self-consistent) gravity ${\bf g}_{int}$
from the particle distribution in the sample. To this end, the
gravitational forces acting on the particles at the different epochs
are computed by means of the TREECODE technique (Bouchet \& Hernquist
\cite{bouchet}). By proceeding in this fashion, the FAM method is able
to reconstruct the orbits of $\approx 10^4-10^5$ mass tracing objects
back in time. This makes FAM numerically fast enough to perform a
large number of orbit reconstructions, essential for performing the
intended statistical analysis presented in the following sections.

In this work we use $N_f=6$ basis functions to parameterize the
orbits, choosing a tolerance parameter $tol=10^{-4}$ to search for the
minimum of the action $S$ and setting a softening parameter of $0.27
\hmpc$ to smooth the gravitational force in the TREECODE. 
Orbit searching in dynamically relaxed systems is a difficult
exercise since one has to choose among the many solutions found at the
extrema of the action.  However, since the purpose of FAM is to
investigate large scale dynamics dominated by coherent flows rather
than virial motions, our evaluations translates into an orbit search
restricted to solutions which do not deviate too much from the Hubble
flow i.e. to the simplest orbits that represents the minima of the
action.  Therefore, we set the initial guess for ${\bf C}_{i,n}$
according to linear theory prescription and search for the minimum of
the action to avoid multiple solutions found a stationary points which
typically describe passing orbits (Peebles \cite{peebles94}).  We have
checked that this choice of parameters is optimal in the sense that
decreasing $tol$, increasing $N_f$ or changing the input set of ${\bf
C}_{i,n}$ does not modify the final results appreciably.

Distortions in the resulting FAM-predicted peculiar velocities mainly
arise from two systematic artifacts (Branchini, Eldar \& Nusser
\cite{branchini02}). One is the discrete sampling of the mass
distribution within $V_{int}$. The second, and overriding one, is the
failure of FAM in reproducing the virial motions within high-density
regions that is a direct consequence of having considered
solutions that represents perturbations to the Hubble flow.  This
deficiency of the FAM reconstructions is clearly illustrated by the
residual velocity vector maps (see eq.~\ref{eq:velres}) in
Fig.~\ref{fig:map30nb} and Fig.~\ref{fig:map100nb} (bottom row). These
show the velocity vector differences between the ``real'' measured,
i.e. $N$-body, velocities and the corresponding FAM reconstructions
(here based on either the galaxy distribution in a $30\hmpc$ central
region or the extended $100\hmpc$ region). The maps show how the
largest residuals are the ones found in the high density regions:
although the FAM$_{30}$ and FAM$_{100}$ velocity fields do show
pronounced velocities near these regions they are not the proper
``real'' virialized velocities they should have been. The residual
fields thus underline the fact that FAM's inaptitude to deal with
regions characterized by large virial motions. Instead, in those
situations it will lead to a false prediction of coherent inward
streaming velocities, an effect pointed out by Nusser \& Branchini
(\cite{nusbran}) and which can be also noted in our images when
carefully studying them.

Finally, for practical reasons, since we are merely interested in
measuring the effect of external gravity fields we make a further
simplifying hypotheses. We ignore redshift distortion effects by
working in real space In this respect, we should point out that
extensions of the action principle method allowing a direct processing
of redshift space information have been proposed and shown to work
(Phelps~\cite{phelps2000}, Phelps~\cite{phelps2002}, also see Sharpe
et al.~\cite{sharpe}).


\subsection{The role of biasing}

In this work we perform orbit reconstructions by assuming that all the
mass of the systems is associated to point mass objects. More
explicitly, we are making two different hypotheses.  The first one is
that we are able to identify a set of objects that trace the
underlying mass density field in an unbiased way.  The second one is
that the internal structure of these objects is irrelevant for our
reconstruction purposes.

The first assumption hardly applies to real galaxies that are most
likely to be biased tracers of the mass distribution, as indicated by
the relative bias between galaxies with different luminosities, colors
and morphological type (Loveday {\it et al.} 1995).  However, if
galaxies and mass particles share the same velocity field so that the
biasing relation remains constant along the streamlines, then the
problem can be easily circumvented by specifying the biasing scheme at
the present epoch (Nusser and Branchini 2000).

Within the standard lore of galaxies embedded in a virialized halo of
dark matter that grow through hierarchical merging of smaller systems,
neglecting the internal structure of objects is an assumption that is
best justified {\it a posteriori} by showing how well Numerical Action
methods can reproduce N-body velocities. Although the goodness of this
assumption has been quantified by a number of numerical tests
(e.g. Nusser and Branchini (2000) and Branchini Eldar \& Nusser
(2002)) little effort has been devoted to understand why numerical
action methods can accurately reconstruct the velocity field of a
large N-body simulation.

One of the reason for this success is that only $\sim 5 \%$ of the
points used in our reconstructions, that were randomly selected from
the N-body simulation, belong to virialized regions where FAM
reconstruction fails.  Fortunately, the locality of this ``virial
effect'' allows us to partially circumvent this problem by applying a
modest spherical tophat smoothing of $2 \hmpc$ to the FAM-predicted
velocities. This tophat filter has been specifically important for the
quantitative aspects of our study, where such systematic problems may
sort distorting conclusions. This smoothing has been invoked in
quantitative comparisons between FAM and $N$-body velocities presented
in this work, in particular in the regression analyses.

Little is known about the ability of numerical action methods to
reconstruct the orbits of virialized systems. Indeed, when applied to
extended objects rather than point masses, numerical action methods
follow a single center of mass point per virialized objects,
completely neglecting its merging history.  Some argument can be given
to back our choice of neglecting the internal structure of virialized
objects.  First of all, after tracing back the merging history of
virialized halos in N-body experiments a simple visual inspection
reveals that particles ending up in the same halo at z=0 are contained
within regions with simple boundaries at high redshifts. As a
consequence, high order terms in the gravity potential about the halo
center of mass are probably rather small.  This probably minimize the
role of major mergers whose rate for galaxy-size halos peaks in the
redshift range 2-4 (Volonteri, Haardt and Madau 2003) while peculiar
motions mostly develop at $z<2$ (Branchini and Carlberg 1994).  These
qualitative arguments clearly need to be confirmed by appropriate
numerical analyses similar to that of Branchini and Carlberg (1994)
but extending out to scales of cosmological interest.


\subsection{Ordered reconstructions}

To obtain an idea of the level of improvement obtained through the use
of successively higher order FAM evaluations, Figure~\ref{fig:orbits}
depicts 2D projections of the corresponding FAM particle orbit
reconstructions within a local spherical volume of $\approx 30 \hmpc$.
The black dots indicate the positions for each object in the sample,
while the lines emanating from each dot represent the computed
trajectories followed by these objects as they moved towards their
present location. The illustrated configuration is taken from one of
constructed mock catalogs, and resembles that of the Local Universe
(see section~\ref{sec:mkNBG}). Each successive FAM reconstruction is
based on the same (present-day) particle distribution. The four frames
correspond to successively higher order FAM approximations, involving
an increasing number $N_f$ of basis functions $q_n(D)$. The top-left
panel shows FAM reconstructed orbits with $N_f=1$, which in fact
corresponds to the conventional first order Zel'dovich approximation
and thus represent the orbits that would have been obtained by the
PIZA method (Croft \& Gazta{\~n}aga~\cite{croftgazt}). These are
followed by panels with $N_f=2$ (top right), $N_f=3$ (bottom left) and
$N_f=6$ (bottom right). They show a clear and steady improvement
towards the $N_f=6$ FAM evaluation. Testing proved that even higher
order computations did not yield improvements significant enough to
warrant the extra computational effort.

In summary, the galaxy orbits in our Local Universe environment are
found at a minimum of the action which is not too far, yet different,
from linear theory predictions.  The FAM technique thus yields a
significant modification of the recovered galaxy orbits and peculiar
velocities for configurations that evolved well beyond the linear
regime (see e.g. Figure~\ref{fig:orbits}). Potentially its ability to
deal with nonlinear circumstances might even prove of benefit to
recover sets of cosmological initial conditions satisfying nonlinear
observational constraints at the present day, which indeed has been
suggested by Goldberg \& Spergel~\cite{goldsperg}.


\subsection{LAP and External forces}

The original cosmological Least Action Principle formulation by
Peebles (~\cite{peebles89}) considered a fully self-consistent, i.e.
isolated, system of point masses.  For practical reasons, the original
implementation had to be restricted to systems of at most a few dozen
objects. Almost exclusively, the Local Group of galaxies formed the
focus of these LAP studies
(Peebles~\cite{peebles89},~\cite{peebles90},~\cite{peebles94}, Dunn \&
Laflamme~\cite{dunnlaf93}).

While these studies did indeed yield a substantial amount of new
insight into the dynamical evolution of the Local Group, the issue of
incorporating the dynamical influence exerted by external mass
concentrations remained a major unsettled question. External forces do
represent a significant component of the dynamics of the Local Group,
as had been shown by Raychaudhury \& Lynden-Bell (~\cite{raychlynd}).
They established beyond doubt that the Local Group cannot be
considered a tidally isolated entity, and demonstrated that the Local
Group is acted upon by an appreciable quadrupolar tidal force. The
resulting tidal torque appears to be responsible for the large angular
momentum of the Local Group as a whole, as Dunn \& Laflamme
(\cite{dunnlaf93}) showed in an elegant and pioneering analysis using
orbits computed by the LAP variational method. They confirmed that the
tidal influence of the external matter distribution is indeed
essential to explain its present angular momentum.

In the course of time various strategies emerged to include external
dynamical influences.  The nature of these methods are mainly set by
the character of the physical system under consideration, and to some
extent was dependent on the available computational resources. Three
strategies are outlined below.

\subsubsection{Directly Including External Masses}

To incorporate the external tidal influence within the LAP analysis
the work by Peebles (\cite{peebles89}, \cite{peebles90},
\cite{peebles94}), Peebles et al.~\cite{peebles2001} and Dunn \&
Laflamme (\cite{dunnlaf93}) attempted to identify a few principal
external mass concentrations which would be responsible for the major
share of the external gravitational influence. While in his first LAP
study Peebles (\cite{peebles89}) considered the Local Group mainly as
an isolated system, sequel studies (Peebles \cite{peebles90}, Peebles
et al.~\cite{peebles2001}) attempted to assess the possible external
influence by neighbouring matter concentrations.  In
Peebles~\cite{peebles90} he attempted to condense the external tidal
force into two nearby mass concentrations, the Sculptor and Maffei
group, each modeled as a single mass. Both were incorporated as 2
extra particles, with properly scaled masses, within the action $S$ in
order to take them along in a fully self-consistent variational
evaluation. A similar approach was followed by Dunn \& Laflamme
(\cite{dunnlaf93}), be it that they included five galaxies/groups in
the local cosmic neighbourhood which arguably contribute a significant
torque on the Local Group. Also in a later application (Peebles et
al.~\cite{peebles2001}) this approach was followed, be it with an
extensive outer region between $4h^{-1}_{75}$Mpc and
$20h^{-1}_{75}$Mpc whose mass distribution was condensed into a coarse
sample of some 14 major external objects.

This ``self-consistent'' strategy is feasible to pursue within the
context of the original, computationally intensive, LAP
implementation. This approach may therefore be followed in LG
resembling situations in which a few objects suffice to represent the
main aspects of a system's dynamical evolution. On the other hand,
cosmic systems of a considerably larger scale than the Local Group
would in general be too demanding for. Supercluster sized regions,
with scales of up to a few tens of Megaparsec, count many more
individual objects than a galaxy group. These systems have also not
yet reached a formation stage so advanced that they have already
largely decoupled from the global Hubble expansion, so the resulting
external gravitational influence is usually the shared responsibility
of a large number of external matter concentrations. Accounting for
the large-scale tidal field would quickly become prohibitively
expensive in terms of the computational effort for conventional LAP
analyses.

\subsubsection{Inserting External Tidal Potential}
\label{sec:exttidpot} 

An alternative strategy is to incorporate the external gravity in the
LAP scheme via an approximate expression for the external
contribution. This may be most directly achieved by inserting an extra
external tidal potential term $\Phi_{tidal}(t)$ in the action $S$
(eqn.~\ref{lapS}). As on sufficiently large, Megaparsec, scales we may
expect this term to evolve according to linear gravitational
instability perturbation growth,
\begin{equation}
  {\displaystyle \Phi_{tidal}(t)}\ =\ D_{\phi}(t)\ {\displaystyle \Phi_{tidal}
    (t_0)}\ =\ {D(t) \over a(t)}\ {\displaystyle \Phi_{tidal}(t_0)}, 
\end{equation}
in which $\Phi_{tidal}(t_0)$ is the tidal term at the present cosmic
epoch and $D_{\phi}=(D/a)$ the linear growth term for the
gravitational potential (the growth factor $D_{\phi}$ and cosmic
expansion factor $a(t)$ are set to be equal to unity at the present
epoch, $D(t_0)=a(t_0)=1.$). Thus, instead of evolving it
self-consistently along with the considered system, the external field
is determined at one epoch -- usually the present one -- and then
incorporated as an extra linearly growing gravity field term
$\Phi_{tidal}(t)$ in the action $S$:
\begin{eqnarray}
  S\ =\ \int_0^{t_0} dt&\ \Bigl\lbrack &\sum_i\,{\displaystyle m_i a^2 \over
    \displaystyle 2} \left({\displaystyle d {\bf x}_i \over \displaystyle d t}
  \right)^2 +\ {\displaystyle G \over \displaystyle a} \sum_{i \not= j} 
  {\displaystyle m_i m_j \over \displaystyle 
    |{\bf x}_i - {\bf x}_j|} +\nonumber\\ 
  &&\ +\ {\displaystyle 2 \over \displaystyle 3}\pi G \rho_b a^2 
  \sum m_i {\bf x}_i^2\ - \Phi_{tidal}(t)\,\Bigr\rbrack \,.
  \label{eqn:sfamtot}
\end{eqnarray}
There are various possibilities to compute the tidal potential term
$\Phi_{tidal}$, usually from the current mass distribution. One option
is to compute it directly from a sample of $M_{ext}$ external objects
which is deemed responsible and representative for the major share of
the external tidal force field,
\begin{equation}
  \Phi_{tidal}(t_0)\ =\ \sum_{i}\ m_i\ \left\{{\displaystyle G \over 
      \displaystyle a}\,\sum_{j=1}^{M_{ext}} 
    {\displaystyle m_j \over \displaystyle |{\bf x}_i - {\bf y}_j|}\right\}\ .
  \label{phiextdir} 
\end{equation} 
Note that none of these external objects ($j=1,\cdots,M_{ext}$) is
taken into account as far as the action of the system and the
computation of their orbits is concerned, except for their ``passive''
role in determining $\Phi_{tidal}$. An alternative approach is to
insert an approximate analytical expression for $\Phi_{tidal}$, in
particular one including the dipolar and quadrupolar contributions,
${\bf d}$ and ${\bf T}$, to the tidal potential,
\begin{equation}
  \Phi_{tidal}(t_0)\ =\ \sum_{i}\ {\bf d}\cdot {\bf x}_i\ +\ {1 \over 2}\,
  {\bf x}_i\cdot{\bf T}\cdot{\bf x}_i\ . 
  \label{phiextsmth}
\end{equation}
Equivalently, one may chose to insert the corresponding expressions
directly into the expression for the derivative of the action with
respect to an expansion coefficient, $\partial S/\partial
C^{\alpha}_{i,n}=0$, evidently equal to zero within this variational
approach.

The first, ``direct'', procedure (eqn.~\ref{phiextdir}) was followed
by Shaya, Peebles \& Tully (\cite{shaya}), who for the purpose of
studying the velocity field within the surrounding $30\hmpc$ modelled
the relevant external mass distribution after the distribution of rich
Abell clusters from Lauer \& Postman (\cite{laupost}). To some extent,
Sharpe {\it et al.} (\cite{sharpe}) operated along the same lines, be
it that they added the resulting tidal term directly to the
reconstructed velocities produced by the LAP procedure.  However,
while in principle exact, such a concentrated and static mass
distribution may involve considerable uncertainties and can be highly
sensitive to the uncertainties in the location of a few dominant point
masses. As this spatial point distribution is supposed to form a
suitable model for the underlying large scale matter distribution this
may be even more worrisome.

Potentially more elegant may therefore be the modelling of a smooth
tidal field along the line of the second procedure
(eqn.~\ref{phiextsmth}), as suggested by Schmoldt \& Saha
(\cite{schmoldtsaha}). The corresponding dipolar and quadrupolar term
may then be based on the best available determinations of these
parameters.  On the other hand, when the LAP volume is comparatively
large, the analytical approximation may represent an
oversimplification of the force field, neglecting potentially
important local variations within the external force field.

\subsubsection{Selfconsistent and Direct FAM approach}

The indirect ``potential'' approach which we described above
(eqn.~\ref{phiextdir} or eqn.~\ref{phiextsmth}) may not properly
account for the temporal evolution of the external field in the case
of nonlinearly evolving systems. The formalism assumes a static,
merely linearly evolving, gravitational potential. However, the matter
concentrations which generate the external tidal forces will
themselves get displaced as the cosmos evolves. These displacements
may be relatively minor for distant masses, but for the more nearby
entities this may be entirely different. A detailed treatment of the
external mass distribution will be necessary when the influence of the
nearby external objects on the evolution of small ``interior'' regions
is comparable to or even dominant over the selfgravity of the region.
It will be equally crucial to follow the detailed whereabouts of
nearby matter concentrations in the case of a large ``interior''
region in which a marked contrast between the central regions and the
outer realms may result in a significantly different dynamic
evolution.

This prompted us to follow an alternative and direct approach, a fully
self-consistent strategy in which also the external matter
concentrations are accounted for in the computation of the system of
evolving particle orbits. Alongside that in the ``local'' region for
which we seek to reconstruct the velocity field, also the system of
objects in the exterior regions ($30 \hmpc < r < 100 \hmpc$) are
considered. Non-uniform manifestations of the external influence can
only be included by pursuing such a direct and systematic approach. It
is only through the availability of the FAM technology that we were
enabled to do so for a Megaparsec system consisting of a large number
of objects.


\begin{table*}
  \centering 
  \caption{$N$-body simulation parameters. 
    Column 1: cosmological model. 
    Column 2: $\Omega_0$ mass density parameter.
    Column 3: $\Lambda$, cosmological constant parameter.
    Column 4: $\Gamma$, power spectrum shape parameter.
    Column 5: $\sigma_8$, density perturbation amplitude spectral normalization.
    Column 6: size of the computational box in $\hmpc$.
    Column 7: number of particles in the simulations.}
  \begin{tabular}{lcccccc}\hline 
    \\
    Cosmology & $\Omega_{0}$ & $\Lambda$ & $\Gamma$ &
    $\sigma_{8}$ & boxlength & N$_{obj}$ \\
    \\
    \hline
    \\
    $\Lambda$CDM& 0.3 & 0.7 & 0.25 & 1.13 & 345.6 & 192$^3$\\
    \\
    $\tau$CDM   & 1.0 & 0.0 & 0.25 & 0.55 & 345.6 & 192$^3$\\
    \\
    \hline
    \label{table:parameters}
  \end{tabular}
\end{table*}


\section{Cosmological scenarios}
\label{sec:setupc} 

The mock catalogs on which we apply our Fast Action Minimization
analysis are extracted from $N$-body simulations in two different
cosmological settings. Their characteristics, in terms of their
relevant parameters, are listed in 
Table~\ref{table:parameters}. The
table also lists the simulation specifications. The first scenario is
a flat $\Lambda$CDM model with a cosmological constant term
$\Omega_{\Lambda,0}~=~0.7$ ($\Omega_0=0.3,\,
\Omega_0~+~\Omega_{\Lambda,0}=1.0,\,\Gamma=0.25,\,n=1$). The second
model is a $\tau$CDM Einstein-de Sitter
($\Omega_0=1.0,\,\Omega_{\Lambda,0}=0,\Gamma=0.25,n=1$) model,
motivated by the decaying particle model proposed by Bond \&
Efstathiou (\cite{bondef91}). Both scenarios were chosen to be viable
with respect to the current observational constraints, implying
similarities in many overall properties and appearances, yet with some
significant differences with respect to their dynamical repercussions.
This may provide indications on whether the galaxy motions in our
local cosmic neigbourhood do contain information on the structure
formation scenario.

\begin{figure*}[t]
  \centering
  \vskip -5.5cm
  \mbox{\hskip -0.7truecm\includegraphics[width=7.in]{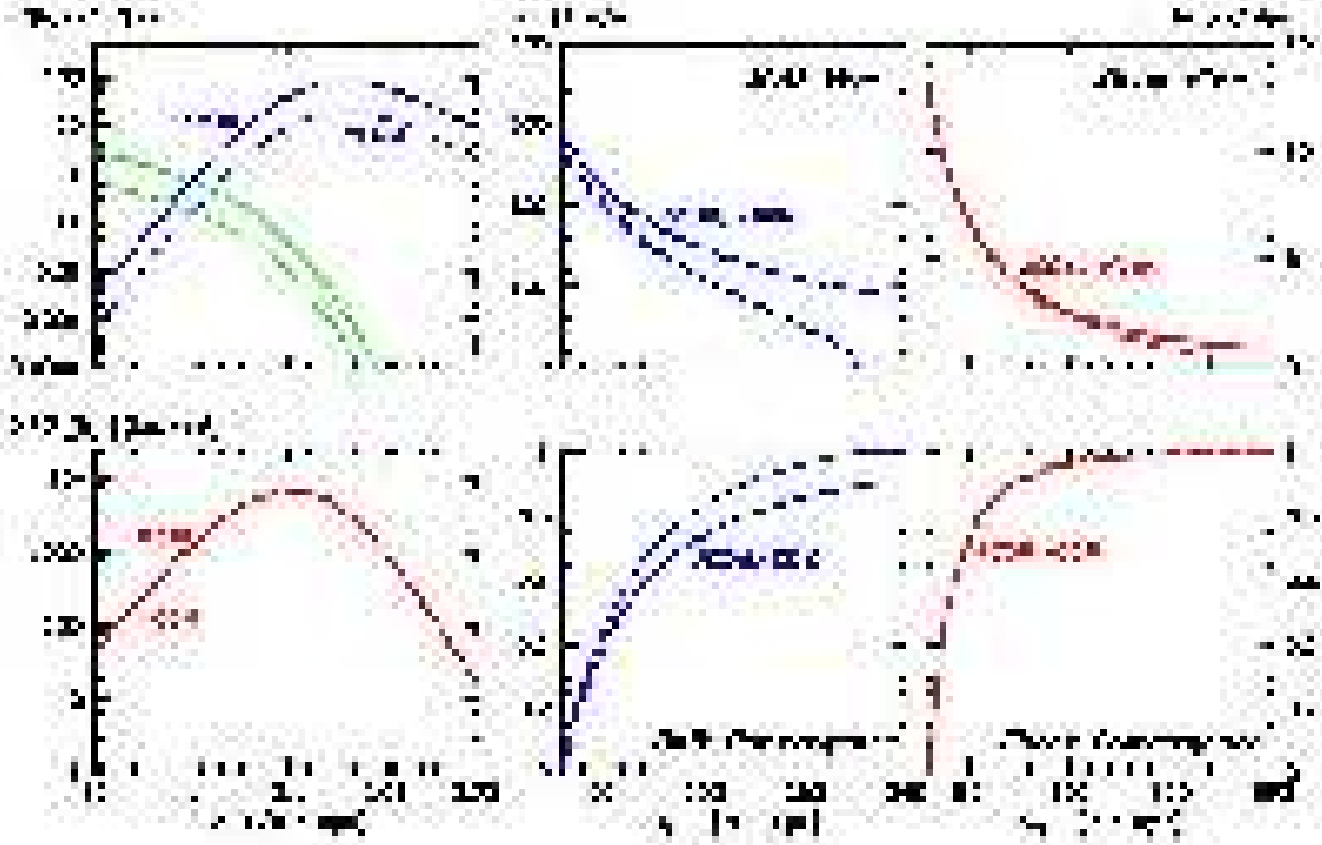}}
  \vskip -1.2truecm
  \caption{Spectral characteristics for the two studied cosmological
scenarios. Upper left: Power spectrum $P(k)$ (green lines) and
$k^3P(k)$ (blue lines), for the $\Lambda$CDM (continuous lines)
and $\tau$CDM (dotted lines) cosmological scenarios used in this
study.  Lower left: velocity power $k^3 P_v(k)$, the amount of
velocity power per logarithmic waveband interval.  Upper centre:
expected bulk flow velocities ($\kms$) over a spherical (tophat)
volume of radius $R_{TH}$. The solid line is the real theoretical
prediction, the dashed one concerns the restricted power spectrum
represented in the simulation box. Centre low: the convergence of the
bulk flow, defined according to eqn.~\ref{eq:fbulk}. Upper
right: Expected shear flow $(\kmsmpc)$ over a spherical (tophat)
volume of radius $R_{TH}$. Similar to bulk flow, the solid line
concerns the full theoretical power spectrum while the dashed
corresponds to the restricted power spectrum represented in the
simulation box. Lower right: convergence of the shear flow, defined
according to eqn.~\ref{eq:fbulk}.  Note that for the righthand frames
the corresponding ordinate axis are the ones on the righthand side of
the plotbox, while for the central frames it is the one on its
lefthand side.}
  \label{fig:pksig}
\end{figure*}

\begin{figure*}[t]
  \centering
  \vskip -0.75truecm
  \includegraphics[width=7.0in]{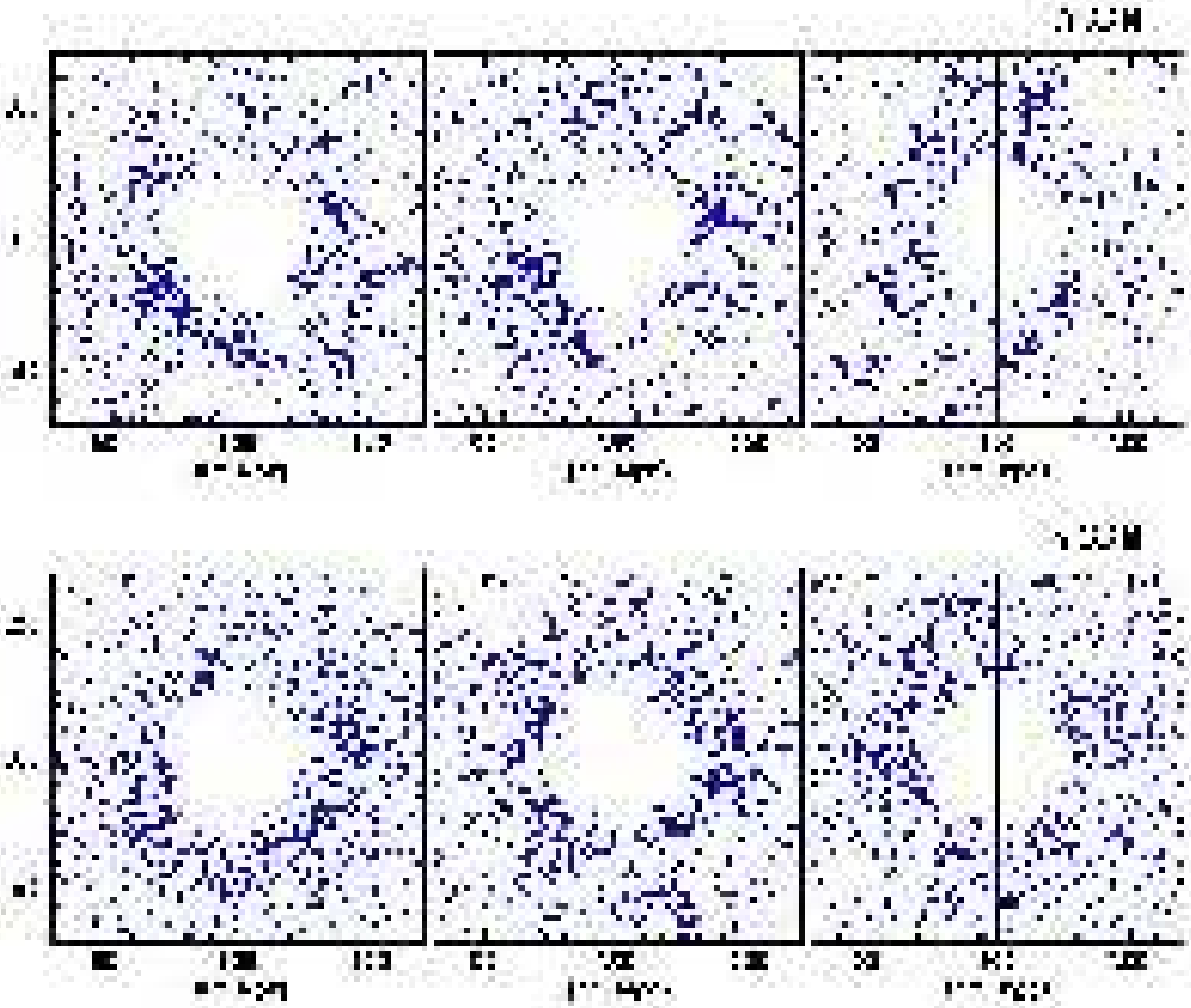}
  \vskip -2.7truecm
  \caption{Central slices through a mock galaxy catalog realization. 
    Depicted is the galaxy distribution in the external ($r>30~\hmpc$)
    volume.  For each catalog, we show the $x-y$ (left), $x-z$
    (centre) and $y-z$ (right) plane.  Top row: $\Lambda$CDM mock
    catalog; Bottom row: $\tau$CDM catalog.}
  \label{fig:psczlt}
\end{figure*}

In both cases the amplitude of density fluctuations is normalized on
the basis of the observed abundance of rich galaxy clusters in the
local universe. This abundance depends on the magnitude of the matter
field fluctuations on the mass scale characteristic for galaxy
clusters. This translates into a dependence on the amplitude of
density fluctuations on cluster scales modulated by the mean global
matter density. A variety of studies (e.g. White, Efstathiou \& Frenk
\cite{wef93}, also see Eke, Cole \& Frenk \cite{eke}) found that in
order to yield the present-day cluster abundance the amplitude of
density fluctuations in spheres of radius $8\hmpc$, $\sigma_8$, and
$\Omega_0$ are related by
\begin{equation}
  \sigma_{8}=0.55 \Omega_{0}^{-0.6}\,, 
  \label{eq:pkclust}  
\end{equation}
The resulting power spectra are depicted in Figure~\ref{fig:pksig} 
(top left). On all scales, the density fluctuations in the $\tau$CDM
scenario, represented by the dotted lines (for both $P(k)$ (green 
lines) and $k^3P(k)$ (blue lines)), are less pronounced than 
those of the $\Lambda$CDM scenario:
the two power spectra have a similar shape and differ by a simple
scaling factor over the entire wavelength range. Visually, this is
immediately reflected in the stark differences between the spatial
galaxy distribution in the resulting mock catalogs.
Figure~\ref{fig:psczlt} provides such a visual comparison. It shows
the ``external'' PSC$z$ catalog mimicking galaxy distribution in three
mutually perpendicular central slices in the case of the $\Lambda$CDM
scenario (top row), together with the same set of frames for a
$\tau$CDM mock galaxy catalog (bottom row). On all scales, the
$\tau$CDM galaxy distribution looks considerably more uniform than
that in the $\Lambda$CDM Universe. Not only is the clustering of
galaxies in the $\Lambda$CDM scenario more pronounced, it also
delineates considerably larger structures, a manifestation of the
power spectrum's amplitude at the corresponding large wavelengths.

Because the higher average matter density in the $\Omega_0=1$
$\tau$CDM Universe does almost fully compensate for the lower
amplitude of the density fluctuations the resulting gravity and
velocity perturbation fields in the $\Lambda$CDM and $\tau$CDM
scenarios are very similar. The velocity power spectra $k^3 P_v(k)$
are shown in the bottom lefthand panel of Figure~\ref{fig:pksig}:
their functional dependence is the same over the entire wavelength
range. The larger mass corresponding to a given density excess in the
$\tau$CDM Universe evidently effects a stronger gravitational force.
The resulting large scale motions scale as $f(\Omega_0) \propto
\Omega_0^{0.6}$.  This happens to be almost exactly the inverse of the
average density perturbation amplitude scaling (eq.~\ref{eq:pkclust}),
which is proportional to $\Omega_0^{-0.6}$ (eqn.~\ref{eq:pkclust}).
While this is exactly the factor involved in the normalization of the
power spectrum, in terms of $\sigma_8$, the lower level of density
fluctuations gets precisely compensated by the higher amount of mass
involved with them. This can be directly observed from the velocity
power spectra $P_v(k)$ for the two scenarios (Fig.~\ref{fig:pksig},
lower lefthand frame). The velocity power spectra for both scenarios
are exactly equal over the entire wavelength range, both in functional
dependence as well as in amplitude. Note that also the gravity
perturbations in the $\Lambda$CDM scenario are substantially stronger
than those in the $\tau$CDM cosmology: because they scale with
${\scriptscriptstyle 3 \over \scriptscriptstyle 2} \Omega H_0^2$ and
the amplitude of the density perturbations, which according to
eq.~\ref{eq:pkclust} is $\propto \Omega^{0.6}$, the average peculiar
gravitational acceleration is proportional to $\Omega^{0.4} H_0^2$.

The comparison between $k^3 P(k)$ (Fig.~\ref{fig:pksig}, top panel,
blue lines) and $k^3 P_v(k)$ (Fig.~\ref{fig:pksig}, bottom
panel) in the same figure shows the shift of the velocity
perturbations, with respect to the density perturbations, towards a
more large-scale dominated behaviour. This follows directly from the
continuity equation, connecting the velocity and density perturbations
such that the velocity power spectrum relates to $P(k)$ through
$P_v(k) \propto P(k)\,k^{-2}$.

The large-scale behaviour of the (linear) velocity perturbation field
immediately illuminates the difficulty in tracing the full array of
matter inhomogeneities responsible for the cosmic motions within a
specific cosmic region. To account for all noticeable contributions it
is necessary to probe out to large depth. This is manifestly evident
for the first order component in the externally induced flow, the
``bulk flow'' $v_{bulk}$. A measure for the expected bulk flow within
a (tophat) spherical region of size $R_{TH}$,
\begin{equation}
  {\bf v}_{bulk}({\bf x})\,\equiv\, \int_{V} {\rm d}{\bf x}'\,{\bf v}
  ({\bf x}')\,
  W_{TH}({\bf x}-{\bf x}',R_{TH})\,,
\end{equation}
is represented by the (root square) average value $\sigma_v$, whose
value may be inferred from the Fourier integral
\begin{eqnarray}
  \sigma_v(R_{\rm TH})&\,=\,&H_0 f(\Omega_0)\ \sigma_{-1}\\
  &\,\equiv\,&H_0 f(\Omega_0)\ \sqrt{\int\d3k\,P({\bf k})\,{\hat W}_{TH}^2
    ({\bf k})\,k^{-2}}\,.\nonumber
  \label{eq:vsigma}
\end{eqnarray}
In these relations, $W_{TH}({\bf x},R_{TH})$ and ${\hat W}_{TH}({\bf
  k})$ are the expressions for the tophat window filter, spatially and
in Fourier space, and ${\bf \sigma_{-1}}$ is the spectral moment $\sigma_j$
for $j=-1$ (see Bardeen et al.~\cite{bbks}, henceforth BBKS).

How substantial the large scale origin of the bulk flow is may be
readily appreciated from figure~\ref{fig:pksig} (top centre). Because
the linear character of fluctuations on large scales, the spectral
$\sigma_v$ (eq.~\ref{eq:vsigma}) does provide a reasonable
order-of-magnitude estimate of the magnitude of the large-scale bulk
motions.  The figure shows the estimated bulk flow amplitudes,
$\approx \sigma_v$, as a function of the (tophat) window radius: the
bulk flow is clearly a large scale phenomenon, converging only very
slowly towards large spatial scales. In both the $\Lambda$CDM scenario
and the $\tau$CDM scenario the externally induced bulk flow on a scale
of $30 \hmpc$ will be in the order of $200 \kms$. Of this overall bulk
flow, more than $100 \kms$ has to be ascribed to inhomogeneities on
scales exceeding $200 \hmpc$ ! When assessing the motions in a local
volume of $30\hmpc$ radius, in terms of relative external
contributions, inhomogeneities on a scale larger than $100\hmpc$ still
contribute more than $25\%$ of the total while the ones larger than
$200\hmpc$ are still responsible for more than $10\%$ (see
Fig.~\ref{fig:pksig}, centre bottom). We should therefore expect to
find substantial external contributions in the $\Lambda$CDM and
$\tau$CDM simulations.  Note that this relative contribution to the
bulk flow, the ``bulk convergence'', is defined as the relative
contribution by matter perturbations within a radius $R_{TH}$ to the
externally induced bulk flow on a scale of $30\hmpc$ (the size of the
NBG volume):
\begin{equation}
  {\cal F}_{\rm bulk}\,\equiv\,1\,-\,{\displaystyle \sigma_v(R_{TH}) \over 
    \displaystyle \sigma_v(30\hmpc)}\,.
  \label{eq:fbulk}
\end{equation}

The second order aspect of the velocity field which we seek to study
is the induced velocity shear $s_{ij}$,
\begin{equation}
  s_{ij}\,\equiv\,{\displaystyle 1 \over \displaystyle 2} 
  \left\{ {\partial v_i \over \partial x_j}+{\partial v_j \over \partial x_i}
  \right\} - {\displaystyle 1 \over \displaystyle 3} (\nabla \cdot {\bf v})\,
  \delta_{ij}\,. 
\end{equation}
Also the velocity shear reveals interesting and distinguishing
differences between the $\Lambda$CDM and the $\tau$CDM scenario. In
the linear regime the expected magnitude of the shear tensor $s_{ij}$,
on a tophat scale $R_{TH}$, may be evaluated through its direct
proportionality to the tidal shear ${\cal T}_{ij}$. Quantifying
$s_{ij}$ by means of its (root square ) average $\sigma_s$ (van de
Weygaert \& Bertschinger~\cite{rienbert}), we find
\begin{equation}
  \sigma_s\,=\,H_0 f(\Omega_0)\,\sigma_0(R_{TH})\,\sqrt{\displaystyle 1 - 
    \gamma^2 \over \displaystyle 15}\, 
\end{equation}
in which the (dimensionless) spectral parameter $\gamma \equiv
\sigma_1^2/\sigma_0\,\sigma_2$ is defined through the $0^{th}$,
$1^{st}$ and $2^{nd}$ spectral moments $\sigma_j$ (see
BBKS~\cite{bbks}).  The predictions for the two cosmological scenarios
are shown in topright frame of Fig.~\ref{fig:pksig}. With respect to
the bulk flow there is a marked difference in coherence scale: the
major contributors to the tidal shear are located at considerably
closer distances than the sources of the bulk flow
(fig.~\ref{fig:pksig}, cf. lower right with lower centre). Most of the
shear inducing matter inhomogeneities are found within a radius of
$100\hmpc$, accounting for more than $95\%$ of its value
(fig.~\ref{fig:pksig}, lower right). On a scale of $30\hmpc$ we expect
an external tidal shear of $\approx 7 \kmsmpc$ for both the
$\Lambda$CDM scenario and the $\tau$CDM model.

\section{Mock catalog Construction and Analysis}


\subsection{$N$-body simulations}

The two $N$-body simulations used in this work were carried out by
Cole {\it et al.}  (\cite{cole}) within the context of an extensive
study of PSCz catalogue resembling galaxy mock samples in a large
variety of cosmological structure formation scenarios. They consist of
$192^3$ particles in a computational box of $345.6 ~\hmpc$. They are
dynamically evolved using an AP$3$M code in which the force is
smoothed with a softening parameter of $0.27~\hmpc$.

The purpose of this study is a demanding task for truely
representative $N$-body simulations.  The $N$-body simulations should
provide an optimal compromise between a high mass resolution on the
small scale side and, on the large-scale side, a cosmic volume large
enough to be dynamically representative. The large dynamic range
requirement involves a mass resolution refined enough to resolve mass
entities comparable to galaxies. This translates into an average
inter-particle separation that needs to be smaller or comparable to
that of galaxies in real observational catalogues.  On the other hand,
the simulations have to extend over a cosmic volume which is large
enough to incorporate the major share of the gravitational influence
exerted by the inhomogeneous cosmic matter distribution. Given the
slow convergence of the bulk flow and its large coherence scale this
is particularly challenging, and will be in the order of several
hundreds of Megaparsec (see discussion in the previous section and
fig.~\ref{fig:pksig}). Although hardly any current $N$-body
simulations would fully fulfill the dynamic range requirements, the
used $N$-body simulations do appear sufficiently adequate for a
meaningful investigation of the relevant systematic trends and
effects. This remains true in a qualitative sense, even though on the
basis theoretical arguments (see e.g. Fig.~\ref{fig:pksig}) and
observational indications (e.g. Hudson et al.~\cite{hudsv}) we know
there may be substantial bulk flow contributions stemming from even
larger spatial scales.

In this respect it is important to note is that the mock catalog
realizations in this work are constrained by the finite size of the
simulation box. The practical repercussions of being confined to a
limited simulation volume may be inferred from the dashed curves in
Fig.~\ref{fig:pksig}.  They show the corrections to the expected bulk
flow and velocity shear predictions (solid curves) when only the
inhomogeneities in the restricted volume of the $345.6~\hmpc$
simulation box are incorporated. Because perturbations on scales
exceeding the fundamental scale of the box are absent, the realized
power spectrum has a rather sharp and artificial large-scale cutoff:
the limited boxsize $L_{box}$ implies a cutoff in the power spectrum
at low wavenumber $k_{box}=2\pi/L_{box}$.  From Fig.~\ref{fig:pksig}
we can conclude that this correction is particularly apt for the bulk
flow, predictions for the velocity shear seem hardly affected. As a
consequence, on scales over $\approx 100~\hmpc$ the bulk flows in the
realized $N$-body simulations will be severely repressed and far from
representative. Although large-scale mode adding procedures have been
proposed to partially remedy this situation (Tormen \&
Bertschinger~\cite{torbert} and Cole~\cite{cole97}), our $\tau$CDM and
$\Lambda$CDM simulations did not include such MAP (mode adding
procedure) extensions. Conclusions with respect to the convergence of
the FAM reconstructed velocity flows should therefore be referred to
with respect to the suppressed velocity power spectrum indigenous to
our $N$-body simulations (notice that this dynamic range issue is
truely cumbersome to nearly any study attempting to assess velocity
flows in computer simulations).


\subsection{Mock Catalog Construction}
\label{sec:mk} 

From the full $N$-body simulations we extract mock catalogs made to
resemble the local Universe. The $\Lambda$CDM and $\tau$CDM $N$-body
simulations are processed through specified observational masks to
imprint the required characteristics on the resulting mock catalogs.
We distinguish two types of mock catalogs. From each $N$-body
simulation we extract ten different ``local'' mock catalogs mimicking
the NBG catalogue and, with these ``local'' samples representing their
interior, ten different ``extended'' samples resembling the PSC$z$
catalogue.

The ``local'' class of mock samples is meant to sample the mass
distribution within a $30~\hmpc$ region in and immediately around the
Local Supercluster. These catalogs constitute {\it volume-limited}
galaxy samples mimicking the Nearby Galaxy Catalog of Tully
(\cite{tullycat}).  Mock catalogs of the second type are designed to
account for the mass distribution out to distances of $100~\hmpc$.
These ``extended'' samples represent {\it flux-limited} samples, for
which we take the IRAS PSC$z$ galaxy catalog (Saunders {\it et al.}
\cite{saunders}) as template. The PSCz sample is not only ideal for
our purposes in that it covers one of the largest volumes of the
Universe amongst the available galaxy redshift surveys, but also in
that it concerns a survey covering a large fraction of the sky and
involves a well-defined uniformity of selection. Assuming that on
large linear scales IRAS galaxies define an unbiased tracer of the
underlying dark matter, Hamilton etal~\cite{hamilton2000} found that
its real-space power spectrum is consistent with that of a
COBE-normalized, untilted, flat $\Lambda$CDM model with $\Omega_m=0.3$
and $\Omega_{\Lambda}=0.7$.

In both {\it flux-limited} and {\it volume limited samples}
the mass of mock galaxies have been rescaled to the value of  
$\Omega_m=0$ of the parent N-body simulations, listed in 
Table~\ref{table:parameters}.

In Table~\ref{table:catalogs} we have listed the main characteristics
of all mock catalogs used in this work. .

\begin{table}[h]
  \centering 
  \caption{Characteristics of the mock catalogs. Column 1: label of the 
    catalog.
    Column 2: cosmological model. Column 3: for each mock catalog type:
    number of generated catalogs.
    Column 4: average number of particles $\langle N_{obj} \rangle$ in 
    each catalog.
    Column 5: external radius of the catalog in units of $\hmpc$.}
  \label{table:catalogs}
  \begin{tabular}{llrrr}\hline
    \\
    Cosmology & Set &  Number of & $\langle N_{obj} \rangle$ &  Maximum \\
    &           & catalogs  &             &  radius  \\
    \\
    \hline\hline
    \\
    $\Lambda$CDM& NBG&   10     &   2740     &    30    \\
                & PSC$z$&  10   &   10900     &   100    \\
    \\
    \hline
    \\
    $\tau$CDM   & NBG&   10     &   2803     &    30    \\     
                & PSC$z$&  10   &   11207     &   100    \\     
    \\
    \hline\hline
  \end{tabular}
\end{table}

In constructing the mock samples galaxies were identified
with N-body particles selected randomly, exclusively according to the
catalog selection criteria.  Therefore, we did not attempt to
include {\it bias} descriptions to model possibly relevant differences
in the spatial distribution of dark matter and galaxies. This is
different from the original use of the simulations (Cole {\it et
al.}~\cite{cole}), in which various bias descriptions were invoked to
construct artificial galaxy samples whose two-point correlation
function and large-scale power spectrum largely matched that of the
APM survey (Maddox et al.~\cite{apmmaddox}).  The analysis of the
small-scale nonlinear power spectrum of the PSCz by Hamilton \&
Tegmark (~\cite{hamilton2002}) even implies the bias on small scales
to be very complex, involving a scale-dependent galaxy-to-mass bias.
We, however, prefer not to include an extra level of modelling
prescriptions. Our interest concerns the kinematics and dynamics of
the matter distribution in the Local Universe, and the velocities of
galaxies are thought to reflect these almost perfectly: they are mere
probes moving along with the underlying dark matter flows,
irrespective of their particular bias relation with respect to the
dark matter distribution. The sole strict assumption is therefore that
of having no velocity bias (Carlberg {\it et al.}~\cite{carl90}),
which on the large-scale Megaparsec scales at hand should be a more
than reasonable approximation.

\subsubsection{Mock NBG catalogs} 
\label{sec:mkNBG} 

The mock NBG catalogs are obtained by extracting spherical volumes of
$30 \hmpc$ from the $N$-body simulation particle distribution.  The
positions of the spheres in the parent simulations are not random but
chosen to mimic as close as possible the characteristics of a Local
Group look-alike region.  Therefore, each mock catalog is centered on
a particle moving at a speed of $625 \pm 25$ km s$^{-1}$, residing in
a region in which the shear within $5\hmpc$ is smaller than 200 km
s$^{-1}$, where the fractional overdensity measured within the same
region ranges between -0.2 and 1.0.

The velocity vector of the central particles defines a Galactic
coordinate system and a Zone of Avoidance.  Particles within the Zone
of Avoidance are removed and substituted with a population of
synthetic objects distributed using a random-cloning technique
(Branchini {\it et al.}~\cite{branchini99}).  The Zone of Avoidance
[ZA] in the mock samples is designed to mimic that of the PSC$z$
catalog (Saunders {\it et al.}~\cite{saunders}) and is smaller than
the one of the real NBG catalog (Tully~\cite{tullycat}).

Each spherical region contains on average $2\times10^4$ particles.
This set of particles is randomly resampled in order to produce an
unbiased catalog of around 2800 objects, a number that matches that
of the galaxies in the real NBG catalog (Table~\ref{table:catalogs}).  This procedure
preserves, within shot noise errors, density fluctuations and thus
does not alter the orbit reconstruction.

\begin{figure}[h]
  \centering
\mbox{\hskip -0.1truecm\includegraphics[width=3.6in]{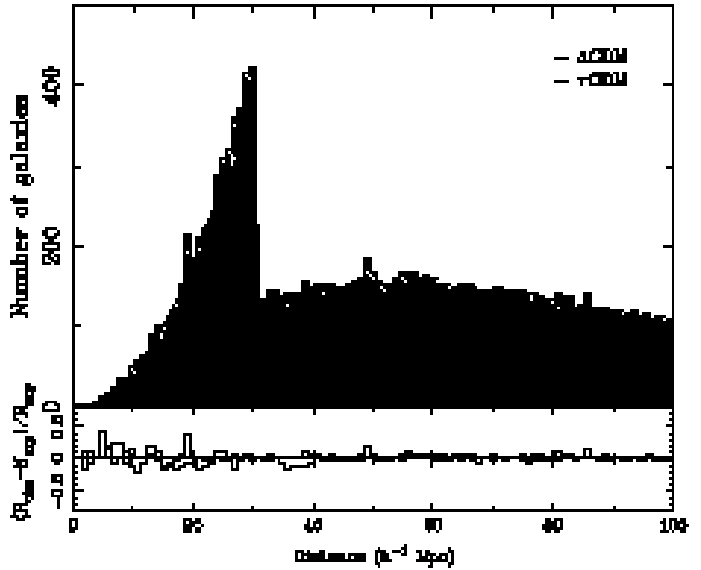}}
  \caption{Object counts as a function of distance in the PSC$z$
    mock catalogs. Upper panel: the two histograms show the average
    counts in the mock samples for the $\tau$CDM and $\Lambda$CDM
    cosmological models.  The continuous line show the expected
    counts. Lower panel: the histograms show the fractional difference
    between the observed and expected counts averaged over all PSC$z$
    mock samples in the two cosmological models explored.}
  \label{fig:sf}
\end{figure}

These NBG mimicking mock catalogs define {\it volume-limited} galaxy
samples, so that the number of objects within a distance $x$ therefore
increases as $x^3$. This is indeed what the resulting realizations
yield, as may be discerned from the central part of the corresponding
histogram in Figure~\ref{fig:sf} ($x \leq 30 \hmpc$).

\subsubsection{Mock PSC$z$ catalogs}
\label{sec:mkPSCz}

The second set of mock catalogs was obtained by carving out spherical
regions of radius $100 \hmpc$ from the $N$-body simulations.  Each of
these new mock samples is centered on the same central position as
that of corresponding NBG mock catalogs, with which they share the
objects within the central $30 \hmpc$.

While the central $30\hmpc$ region coincides with the NBG mock sample,
the particle distribution in the external region ($30\hmpc < x <
100\hmpc$) is supposed to mimic that of galaxies in the {\it
  flux-limited} IRAS PSC$z$ catalog. To achieve this the objects
beyond $30 \hmpc$ were selected from the $N$-body particle samples
according to the PSC$z$ selection function used by Branchini {\it et
  al.} (\cite{branchini99}):
\begin{equation}
  \psi(x)=Ax^{-2 \alpha} \left(1+ \frac{x^{2}}{x_{\ast}^{2}} 
  \right)^{- \beta} \quad \textrm{if } x > x_{s}\ .
  \label{eq:sel} 
\end{equation}
In this expression $x$ is the distance of the galaxy in~$\hmpc$, 
while the parameters have the values $\alpha=0.53$, $\beta=1.8$,
$x_{s}=10.9\hmpc$, and $x_{\ast}=84\hmpc$ (Branchini
\cite{branchini99}). The selection function $\psi(x)$ defines the
relative {\it number} density of the galaxy sample with respect to the
{\it number} density of the $N$-body particle simulation. While the
number density of PSC$z$ galaxies is equal to that of the particles in
the simulation at $10.9 \hmpc$, the amplitude $A$ of $\psi(x)$ is
normalized such that $\psi(10.9)=1$.

\begin{figure*}[t]
  \centering
  \vskip -3.0truecm
  \mbox{\hskip -0.5truecm \includegraphics[width=7.5in]{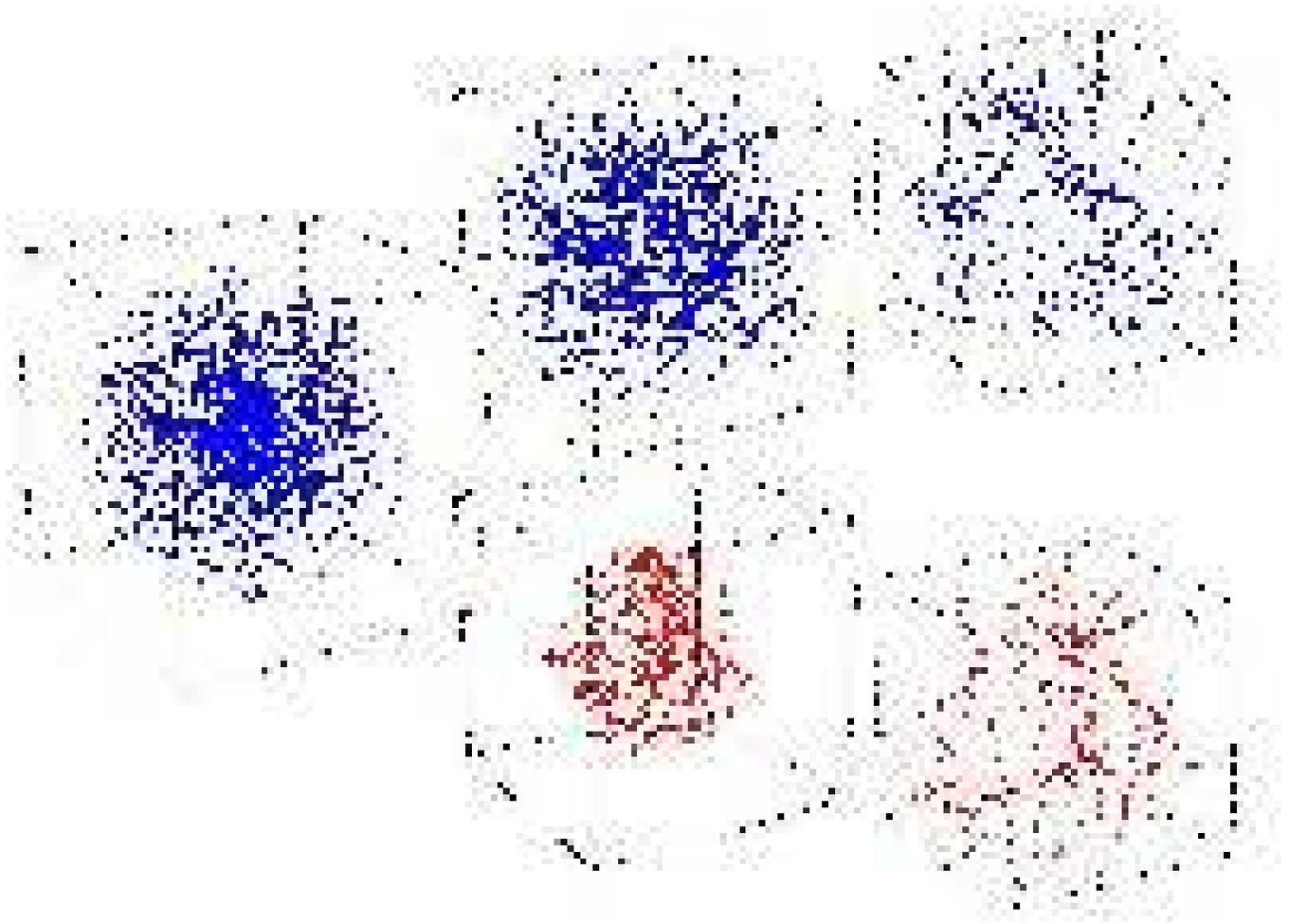}}
  \vskip -2.5truecm
  \caption{Compilation of cubic subregions and various mock subsamples of 
    the object distribution in one of the $\Lambda$CDM PSC$z$ mock
    catalogs. The full mock catalog of objects, the PSCz look-alike
    volume and the NBG look-alike volume out to a radius of $100
    \hmpc$, is shown in the large lefthand cubic box (with size $200
    \hmpc$). The boxes along the top row and those along the bottom
    row represent mutually exclusive object samples. The lower ones
    represents the inner cosmic region ($x < 30\hmpc)$, the top row
    represent a corresponding part of the exterior PSCz region ($30
    \hmpc < x < 100\hmpc$).  Emanating from the complete cubic volume
    and running along the top row are two successive boxes which zoom
    in on the external PSCz mock catalog objects, with cubic sizes of
    $160 \hmpc$ (middle right cube) and $80 \hmpc$ (lower right cube)
    respectively. The lower row of two cubic volumes the central
    (inner) NBG mimicking object sample. The full inner sample,
    comprising a sphere of radius $30 \hmpc$ is shown in the lower
    central cubic volume, whose size of $80 \hmpc$ is equal to the
    central PSCz box, while the righthand cube focuses in on the $40
    \hmpc$ size central region.}
  \label{fig:5cube}
\end{figure*} 

Two additional steps concern the treatment of ``Zone of Avoidance''
objects and the evening of the matter density throughout the full
external sample volume.  A first step is the processing of sampled
objects in the Zone of Avoidance such that the resulting sample
conforms to a reality resembling situation. The ZA
``removal$+$substitution'' is implemented in the same way as in the
case of the NBG mock catalog construction, with the replacement
achieved with the same random-cloning technique.  Finally, in order to
guarantee a uniform average {\it mass} density throughout the volume,
the mass of the objects in the {\it flux-limited} external object
sample ($30\hmpc < x < 100\hmpc$) has been scaled by the inverse of
the selection function $\phi(x)$.

\begin{figure*}
  \centering
  \vskip 0.5truecm
  \includegraphics[width=7.0in]{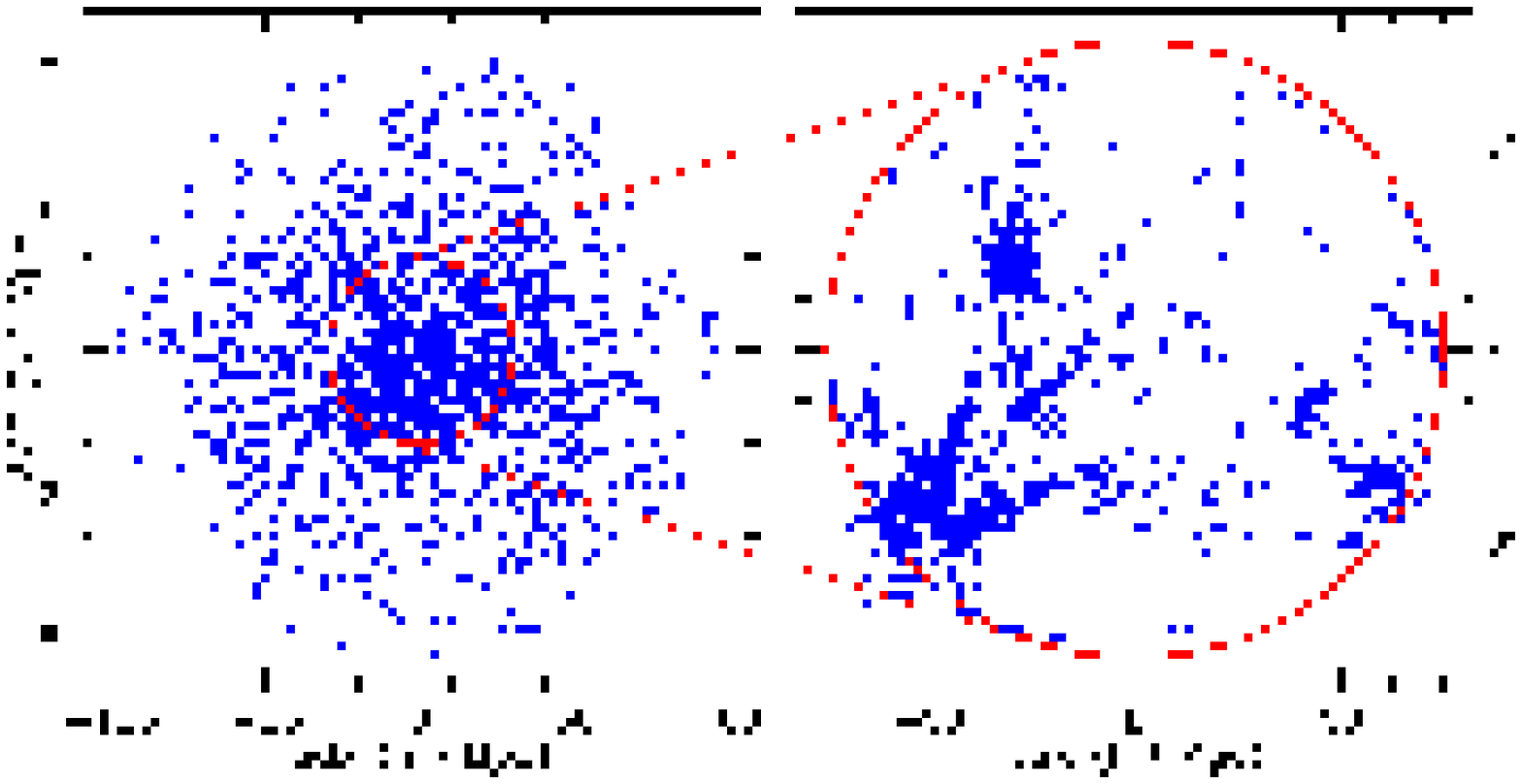}
  \caption{2D projection of the particle distribution in one of the 
    $\Lambda$CDM PSC$z$ mock catalogs (left panel).  The inner circle
    divides the NBG look-alike volume limited region from the PSC$z$
    flux limited one.  The right panel show the 2D projected
    velocities in a slice of 10$\hmpc$ cut through the inner part of
    the catalog.}
  \label{fig:mocks}
\end{figure*}

\subsubsection{Mock catalog realizations}
\label{sec:mkcat} 

From both figure~\ref{fig:5cube} and figure~\ref{fig:mocks} one can
obtain an impression of the spatial context of the local NBG mock
sample within the wider environment of the surrounding $100\hmpc$
PSC$z$ sample. To visually appreciate the selection criteria of the
catalogs, and their interrelationship, figure~\ref{fig:5cube} shows a
three-dimensional view of one set of the $\Lambda$CDM mock catalogs,
extracted from the particle distribution in the $N$-body simulation of
structure formation in a $\Lambda$CDM scenario.  Emanating from the
full PSC$z$ + NBG mimicking galaxy samples in the top righthand cube
is a row of two cubes showing the content of the external PSC$z$ mock
catalog (righthand) and the content of the central NBG mock catalog
(lefthand).  Figure~\ref{fig:mocks} elaborates on this, and shows the
projected particle distribution in the same PSC$z$ + NBG mock catalog
(left panel) while focusing in on the central region (right panel).
The circle (left panel) indicates the boundary of the volume-limited
region comprised by the mock NBG galaxy sample, which in the right
panel has been enlarged to show the corresponding velocity field
within this NBG region. Velocities of objects within a 10$\hmpc$ thick
slice are shown by means of arrows whose size is proportional to the
amplitude of the galaxy velocity components within this slice.

As a matter of test, we checked the distance distribution of the
resulting mock galaxy samples. The histograms of the resulting mock
catalog distributions are shown in Figure~\ref{fig:sf}. The upper part
of Figure~\ref{fig:sf} shows the number of galaxies -- averaged over
all PSC$z$ mock catalogs for both two cosmological models -- as a
function of distance $x$ over the full range $x \leq 100\hmpc$.
Clearly visible is the discontinuity at $30 \hmpc$, marking the
transition from volume-limited NBG-like region to the PSC$z$-like
flux-limited outer region. For comparison, the solid line shows the
theoretically expected counts (Eq.~\ref{eq:sel}).  The generated mock
samples appear to match the expected distance distribution rather
well. This is further underlined by the fractional difference between
observed and expected counts, $ {\rm N}_{\rm obs}/{\rm N}_{\rm exp} -1
$, shown in the lower part of Figure~\ref{fig:sf}. The fractional
difference between mock samples displays a perfect featureless scatter
pattern: Poisson noise free of systematic effects.


\subsection{Mock Catalog Analysis} 

\subsubsection{FAM velocity field reconstructions}
\label{sec:famvelrec} 

On the basis of the FAM reconstructions of the galaxy velocities and
the comparison with the true velocities -- i.e. those in the original
$N$-body simulation -- we assess to what extent the matter
distribution within the confines of each different mock galaxy sample
does contribute to the total velocity of the galaxies. In these
idealized circumstances of the $N$-body world the galaxy positions and
velocities are known to perfect accuracy, thus circumventing the need
to investigate the effects of measurement errors and deceptive
systematic biases in the galaxy peculiar velocities. This should
provide us with a better understanding of the nature and magnitude of
genuine physical influences.

Three different velocities are accorded to each galaxy located within
the ``local'' spherically shaped $30\hmpc$ NBG region. The first
velocity is that of the ``true'', $N$-body velocity. For each of the
in total 20 NBG mimicking galaxy mock catalogs, the FAM
reconstructions produce two additional velocity estimates. One FAM
velocity results from the application of the FAM analysis to the
restricted inner $30 \hmpc$ NBG-like region itself. The second FAM
based velocity is obtained on the basis of the FAM analysis on the
extended, ``full'', $100\hmpc$ PSC$z$ survey resembling sample (in
which the ``local'' NBG sample occupies the interior central region).
In the following, we will indicate these FAM velocities by the names
of FAM$_{30}$ and FAM$_{100}$ velocities.

\subsubsection{FAM$_{30}$ versus FAM$_{100}$ reconstructions}

The mutual comparison between each of the three different galaxy
velocities -- the FAM$_{30}$, the FAM$_{100}$ and the full $N$-body
velocities -- is expected to yield abundant information on the
dynamics and development of the structure in the interior $30\hmpc$
region:

The FAM$_{30}$ velocities are the galaxy velocities which would have
been the product of the combined gravitational interaction of --
solely -- the matter concentrations within the central $30\hmpc$
volume. Any deficiency with respect to the ``real'' $N$-body velocity
of each galaxy has to be ascribed to the gravitational impact of
matter inhomogeneities outside the local NBG region.

By tracing the mass distribution further out to a distance of
$100\hmpc$, invoking the matter distribution in the complete PSC$z$
mimicking mock samples, we will then evaluate the extent to which
matter inhomogeneities within a 100$\hmpc$ scale are able to account
for the motions within the local 30$\hmpc$ region. From this we can
infer in how far the external influence over the local region can be
ascribed to matter fluctuations situated between a radius of 30$\hmpc$
and 100$\hmpc$.

In this study we also have to take into account the fact that a single
30$\hmpc$ region cannot be considered representative for the whole
Universe, and generic conclusions on the basis of the kinematics
within a single 30$\hmpc$ volume cannot be drawn. This is also true
for the the NBG mock samples in this work, even though they were
selected according to some strict criteria (see section 5.2). Analysis
and conclusions will therefore be based on a straightforward average
over the 10 different $30\hmpc$ mock samples which were constructed
for each cosmological scenario. The dispersion in the extracted
parameter values will provide a reasonable estimate for their
significance.

\subsubsection{Analysis of Reconstructions}

The basic product of the FAM reconstructions are velocity maps, in
essence a velocity vector at the location of each galaxy in the
sample.  Our analysis consists of three different but complementary
tracks.  The first and most straightforward one is the visual
inspection of the resulting velocity vector maps. It provides a direct
impression of the extent to which a FAM reconstructed field reproduces
the true velocities. Also, it will provide a direct impression of a
spatial coherence in the differences between true and reconstructed
field, which is an incisive way to uncover systematic contributions
like e.g.  a bulk flow component.
 
The second examination is a strictly local analysis, a pure
point-to-point comparison between the velocities predicted by the FAM
reconstructions on the one hand and the ``true'' $N$-body velocity of
the same object on the other hand. To some extent, the analysis by
means of scatter plots is the most direct and objective quantitative
comparison between two fields. Various velocity related quantities
will be assessed in this fashion. Note that these localized
comparisons cannot address the presence of spatial coherence in the
cosmic flows (even though they may uncover systematic effects caused
by external influences).

Finally, the third track is targeted towards a factual description of
the spatial coherence within the velocity fields or, rather, in the
residual fields between the ``true'' velocities and the reconstructed
velocities. Systematic trends in these residual fields are interpreted
as manifestations of external forces. Of these we shall determine the
first-order -- bulk flow -- and second order -- velocity shear --
components.


\section{FAM velocity vector maps}

For reasons of consistency and to achieve optimal transparency the
illustrated velocity vector maps in the following discussion all
concern the same mock sample of NBG calculations. For the illustration
of the FAM$_{30}$ (Fig.~\ref{fig:map30nb}) and the FAM$_{100}$
(Fig.~\ref{fig:map100nb}) reconstructions we use one of the
$\Lambda$CDM $30\hmpc$ NBG mock catalogs. It is the same galaxy sample
that was shown in 3-D in Fig.~\ref{fig:5cube} and in projection along
the ``x-y'' plane in Fig.~\ref{fig:mocks}.

The vector maps in Figure~\ref{fig:map30nb} and
Fig.~\ref{fig:map100nb} depict the projections of the raw unsmoothed
galaxy velocities, for galaxies within a central slice of $10 \hmpc$.
The size of the arrows is proportional to the amplitude of the
peculiar velocity component within this slice, each arrow starting at
the location of the galaxy. Both figures consist of three successive
rows. The velocity maps in the first row correspond to the ``real''
world of the $N$-body simulation. The second row depicts the velocity
maps for the FAM reconstructions, the FAM$_{30}$ reconstruction in
Fig.~\ref{fig:map30nb} and the FAM$_{100}$ reconstruction in
Fig.~\ref{fig:map100nb}. The last row shows the resulting residual
velocity vector fields,
\begin{equation}
  {\bf v}_{\scriptstyle{\rm res}}\equiv {\bf v}_{\scriptstyle{\rm FAM}}-
  {\bf v}_{\scriptstyle{\rm Nbody}}\,, 
  \label{eq:velres}
\end{equation}
the vector difference between the $N$-body velocities and the
corresponding FAM velocity reconstructions, [$N$-body - FAM$_{30}$]
and [$N$-body - FAM$_{100}$].

Each row has three panels, containing the vector maps in the three
mutually perpendicular ``central'' slices. Each plane is identified by
means of the index combination ``x-y'', ``x-z'' or ``y-z'' (top
figure), the index pair identifying the horizontal and vertical axis
along which the panel is seen. Imagining these three planes passing
through the centre of the $30 \hmpc$ NBG volume provides a spatial
impression of the full 3-D velocity field. Note that here the choice
of Cartesian coordinate system does not have any special significance,
arbitrarily set by the axes of the total $345.6\hmpc$ simulation box
(the ``fundamental'' box) from which the mock catalogs were distilled.
This is unlike vector maps (e.g. Fig.~\ref{fig:nbfam30b}) in some
later sections.


\subsection{$N$-body sample: the ``observed'' velocities}

\begin{figure*}[t]
   \mbox{\hskip -0.65truecm\includegraphics[width=7.5in]{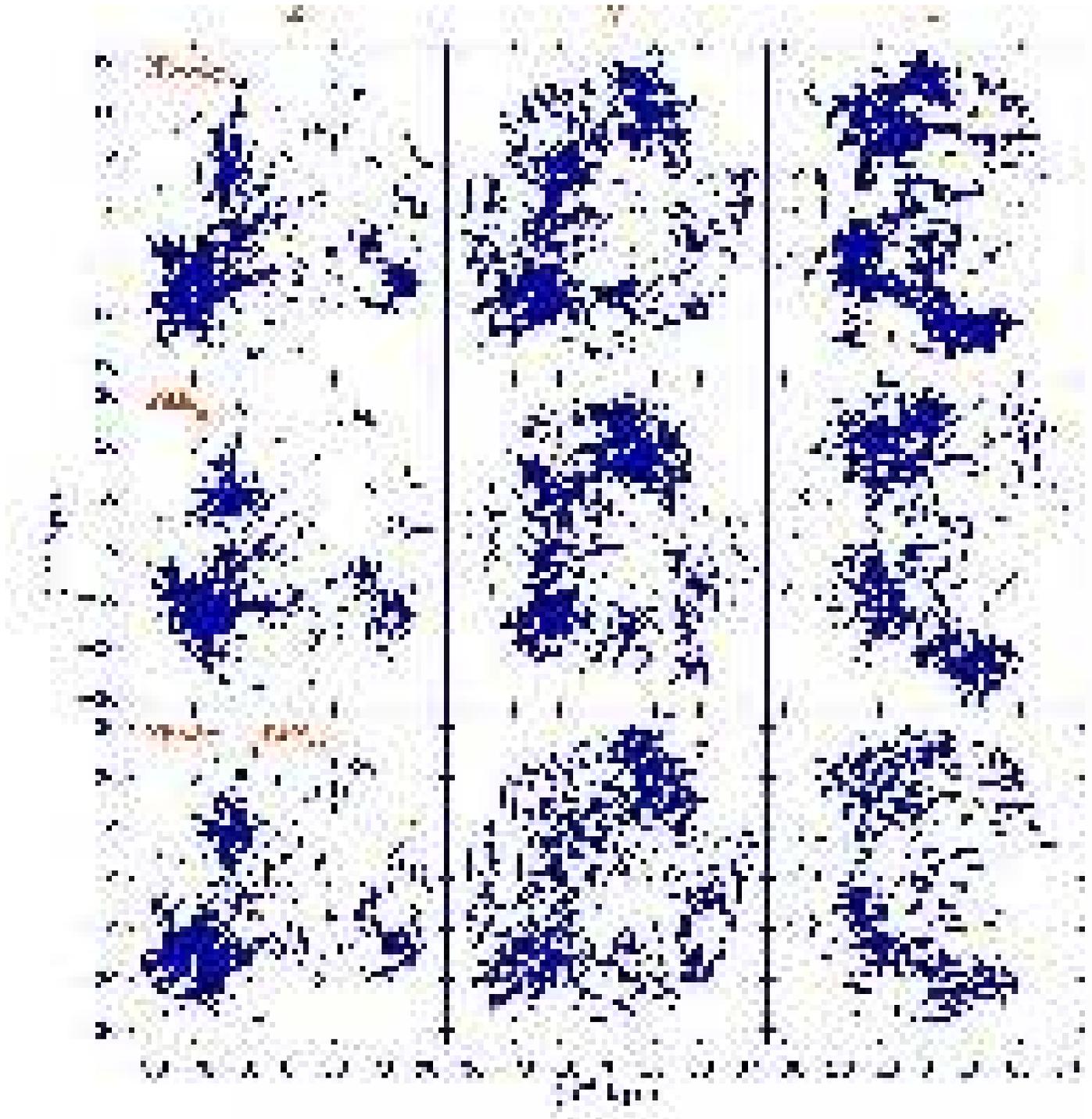}}
   \caption{2D projected, unsmoothed velocities at each particles position 
     in one of the NBG-$\Lambda$CDM mock catalogs. Each row contains
     three frames, corresponding to the central $y-z$, $x-z$ and $x-y$
     plane through the point sample. Bottom row: projected $N$-body
     velocities. Central row: projected FAM$_{30}$ velocities. Bottom
     row: residual $N$-body-FAM$_{30}$ velocities.}
  \label{fig:map30nb}
\end{figure*}

\begin{figure*}[t]
  \centering
  \mbox{\hskip -0.65truecm\includegraphics[width=7.5in]{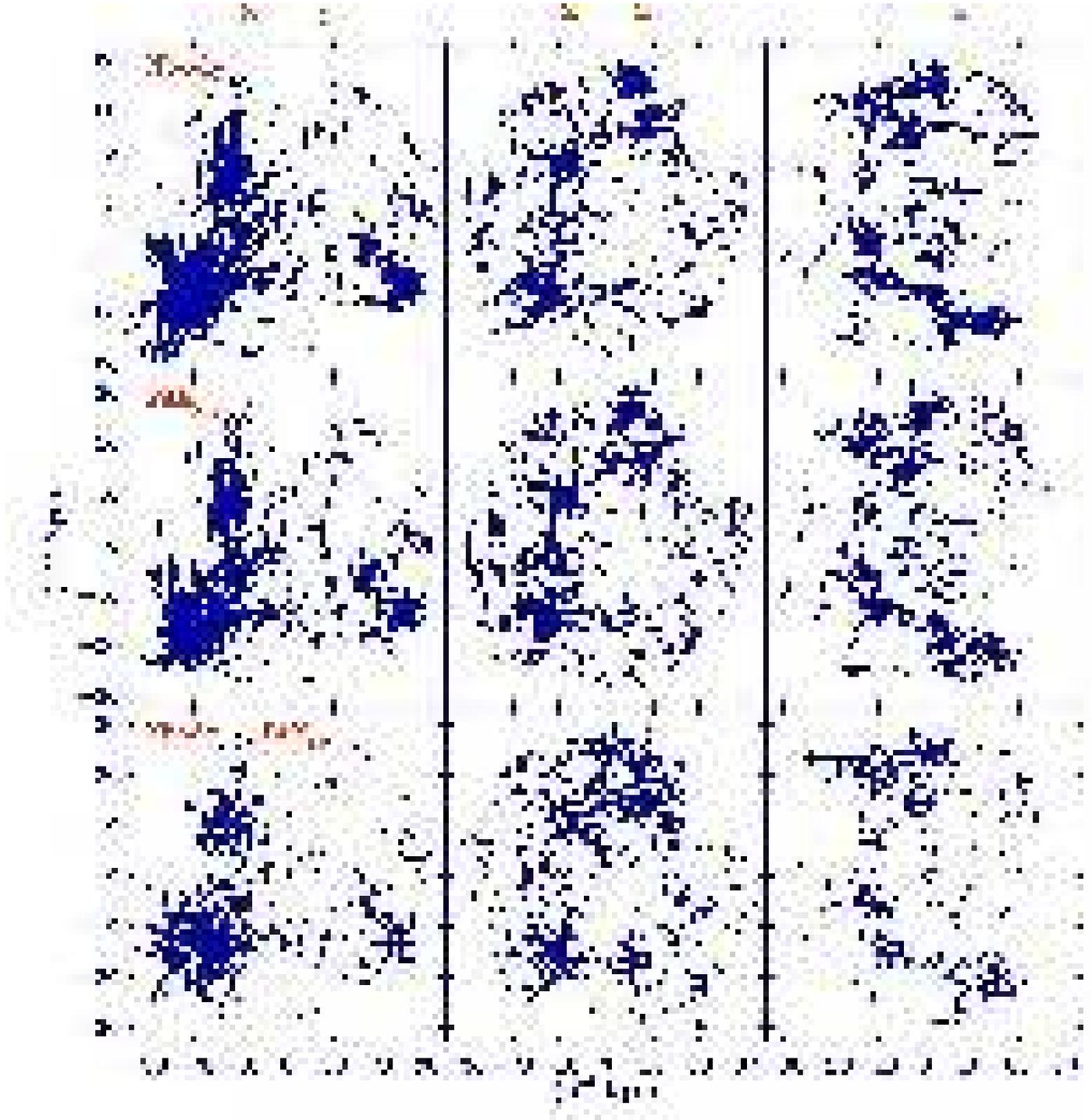}}
  \caption{Same as Fig.~\ref{fig:map30nb}, for the PSCz based FAM$_{100}$ 
    reconstructions. Shown are the 2D projected, unsmoothed velocities
    at each particles position in the central NBG mimicking volume of
    the full PSCz-$\Lambda$CDM mock catalogs, the same region as
    depicted in Fig.~\ref{fig:map30nb}. Each row contains three
    frames, corresponding to the central $y-z$, $x-z$ and $x-y$ plane
    through the point sample. Bottom row: projected $N$-body
    velocities. Central row: projected FAM$_{100}$ velocities. Bottom
    row: residual $N$-body-FAM$_{100}$ velocities.}
  \label{fig:map100nb}
\end{figure*}

The velocity vectors in the top row vector maps depict the ``real''
$N$-body velocities of the ``galaxies'' located within the three
``central'' slices (the same for Fig.~\ref{fig:map30nb} and
Fig.~\ref{fig:map100nb}). The galaxy distribution is characterized by
a few dense, massive and virialized clumps, visible as high
concentrations of large and randomly directed velocity vectors. The
truely massive concentration visible in the lower left of the $x-y$
panel is part of a superstructure extending beyond the boundaries of
the NBG region. It represents a major and dominant source for the
motions in this area. This may be appreciated from the observed
velocity flow towards this clump and the overall distortion of the
flow in its vicinity. The large configuration visible in the ``y-z''
slice contains several dense compact regions embedded in a ridge-like
structure running curvedly from the lower righthand corner to a
location slightly left from the centre. At least partially related to
this mass concentration in and around the ridge is the bulk flow along
the right-to-left direction.

Overall, the ``x-y'' and ``y-z'' vector maps indicate the presence of
a dominant coherent ``bulk flow'' pattern which can be traced
throughout the whole NBG volume. By coincidence, the orientation of
the coordinate axes is such that the direction of the bulk flow is
almost perfectly aligned along the ``y''-axis: for this particular
mock sample the ``y''-axis does represent a physically significant
direction defined by the streaming pattern itself.  The bulk flow
seems to be directed towards some (fictitious) point outside the local
$30\hmpc$ region.

A dominant and conspicuous coherent flow pattern also characterizes
the $x-z$ velocity vector map. While the flow in the two other planes
seems to be almost exclusively dominated by a bulk flow, here the
pattern has a more complex geometry, readily recognizable as a typical
``velocity shear'' pattern. The specific shearing motion in this plane
consists of a compressional component along the top lefthand to lower
righthand direction, in combination with a dilational stretch along
the perpendicular direction from the lower lefthand towards upper
righthand corner.

\subsection{NBG samples: FAM$_{30}$ velocity vector maps}
\label{sec:velmapnbg}

The role of the local cosmic matter distribution on the motions in the
local Universe is assessed on the based of the ``FAM$_{30}$
velocities''. They are the peculiar velocities computed by FAM on the
basis of the local matter distribution, supposedly reflected by the
galaxies within the NBG catalogs. The corresponding reconstructed
velocities are shown in the second panel row of
Figure~\ref{fig:map30nb}. With their final position as boundary
condition, each velocity vector is located at the same galaxy position
as in the $N$-body maps (top row). Note that the vector maps in
Figure~\ref{fig:map30nb} and Figure~\ref{fig:map100nb}, and also the
later ones in Fig~\ref{fig:vmapmultp}, Fig~\ref{fig:nbfam30b} and
Fig~\ref{fig:nbfam30bs}, show the pure unsmoothed velocity vectors (and
do not ``correct'' for the virialized regions).

The FAM$_{30}$ velocity maps are distinctly different from the
corresponding $N$-body velocity maps (top row): a coherent flow
pattern is almost entirely absent. The FAM$_{30}$ reconstructions
obviously did not recover the strong bulk flow observed in the
$N$-body velocity maps, nor the striking shear pattern in the $x-z$
plane. Because the FAM$_{30}$ velocity field reconstructions solely
relate to the matter distribution within the inner 30$\hmpc$ NBG
region, this indicates that the major share of coherent bulk flow and
the velocity shear are due to the matter distribution outside the
central 30$\hmpc$. This is most readily apparent in the velocity
residual maps [$N$-body - FAM$_{30}$], the difference between the
$N$-body and the FAM$_{30}$ velocity vector fields (bottom row of
Fig.~\ref{fig:map30nb}). In the residual field [$N$-body - FAM$_{30}$]
we recognize the same characteristic flow patterns, strong spatial
correlation, long-range coherence and overall morphology as in the
full $N$-body velocity field. This represents convincing evidence for
the external origin of the large-scale ``bulk'' and ``shear''
component in the local velocity flow.

Prominently visible in the residual velocity field is the strong bulk
flow along the ``y''-axis.  Overall, the spatial pattern of the
residual bulk flow appears to reproduce that of the $N$-body flow
field. However, some minor yet significant differences between the
residual and the full $N$-body bulk flow can be discerned. The
amplitude of the corresponding velocities in the residual map is
somewhat smaller than the equivalent $N$-body velocities: apparently
part of the bulk flow is induced by the local NBG matter distribution.
This does not seem to be true for the velocity shear: the shear
patterns in the ``x-z'' plane of the residual and $N$-body velocity
fields are almost identical (except for the virialized motions in
high-density clumps). Apparently, the velocity shear component is
almost exclusively due to external matter distribution. As a locally
flattened matter configuration would induce an internal shear flow,
this appears to imply a local matter distribution whose geometry is
hardly flattened or elongated.

Closer inspection of the FAM$_{30}$ velocity field provides a more
detailed view of the small-scale flow pattern mentioned above. In the
``x-y'' plane the large-scale ($N$-body) bulk flow has virtually
completely disappeared. Instead, the dominant motion in the ``x-y''
plane is a streaming flow towards a prominent matter concentration
within this region (lower left). On the other hand, in the ``y-z''
slice a trace of the $N$-body bulk flow along the ``y'' axis remains,
be it that the corresponding velocities have considerably smaller
amplitudes than their $N$-body counterparts. These local motions
appear to be effected by the matter located along the lower ridge,
supporting the impression that this feature is a local extension or
outlier of the large-scale matter configurations responsible for the
full bulk flow. Examination of the panels in Fig.~\ref{fig:5cube} and
Fig~\ref{fig:mocks} indeed seems to suggest that the density ridge in
the lower half of the ``y-z'' plane is indeed connected to structures
just outside the NBG volume, while this perhaps may be true for the
massive matter clump in the ``x-y'' plane too. This may not come as a
surprise: the local matter distribution will to some extent be
correlated with the external matter configuration so that the locally
induced bulk flow is expected to reflect at least partially the full
$N$-body bulk flow.

In summary, the inability of the FAM$_{30}$ reconstruction to recover
the large-scale bulk flow and velocity shear is a consequence of the
fact that they are a result of the action of the mass distribution on
scales larger than the internal $30 \hmpc$ size region while the
FAM$_{30}$ velocities are entirely and self-consistently determined by
the mass distribution within this interior region. The residual
[$N$-body - FAM$_{30}$] maps, which are a model for the possible
findings of a real-world observational campaign, provide the most
elucidating illustration of their ``external'' origin. Even though
they do provide convincing evidence for their external nature, they do
not provide sufficient information to infer the identity and nature of
the main source of the flow patterns.  In principle, however, we may
deduce a substantial amount of information on the basis of a careful
quantitative analysis: the work by Lilje, Yahil \& Jones~(\cite{lilje})
still sets a prime example. To this end, we will investigate the
external matter distribution in the PSC$z$ $100\hmpc$ sized regions.

As a final note, we point to the rather artificial nature of velocity
vectors in the vicinity of the massive clump in the ``x-y'' slice as
indicative for the self-consistent nature of the FAM reconstructions.
Its location near the edge of the NBG volume even appears to have
generated the rather contrived infall motions along the rim of the NBG
sphere.


\subsection{PSC$z$ samples: FAM$_{100}$ velocity vector maps} 
\label{sec:pscz}

The contribution by the relatively nearby external matter
agglomerations, within a distance of $100\hmpc$, to the motions in the
local Universe is investigated on the basis of the ``FAM$_{100}$
velocities''. FAM produces these peculiar galaxy velocities on the
basis of the galaxy sample in the full mock PSC$z$ galaxy sample,
extending out to $100\hmpc$ around the center of our local region. The
corresponding reconstructed velocities are shown in the second panel
row of Figure~\ref{fig:map100nb}. It is the analogy for the
``FAM$_{100}$ velocities'' of Figure~\ref{fig:map30nb}, and concerns
the same $30\hmpc$ central region (the NBG region is the central
subregion of the PSC$z$ mimicking catalog). The FAM$_{30}$ maps showed
the dominant influence of externally induced forces on the motions in
the local 30$\hmpc$ NBG region: on the basis of the FAM$_{100}$ maps
we seek to assess whether the major share of the responsible external
matter agglomerations may be identified within the realm of a PSC$z$
like volume.

Comparison of the first and the second row of panels in
Fig.~\ref{fig:map100nb} shows the large degree of similarity between
the FAM$_{100}$ velocities (panels 2$^{nd}$ row) and the $N$-body
velocities (panels top row). Unlike the FAM$_{30}$ maps in
Fig.~\ref{fig:map30nb} we find that the FAM$_{100}$ maps successfully
reproduce most of the large-scale behaviour and most of the finer
details of the $N$-body velocity field. The degree of similarity is
particularly evident in the corresponding residual velocity field
[$N$-body - FAM$_{30}$] (bottom row panels). With the exception of the
high-density virialized regions the residual velocities are very small
and mostly randomly oriented: no significant spatial correlations and
spatial coherence can be detected.

The detailed similarity between the $N$-body and the FAM$_{100}$ maps
shows that it is sufficient to take account of the mass distribution
out to $100\hmpc$ for explaining, in considerable detail, the velocity
flows in the local NBG volume. Moreover, the detailed rendering of the
velocity field by FAM is a convincing demonstration of the capacity of
the FAM technique to accurately describe the dynamics implied by the
observed local galaxy distribution. The quantitative comparisons in
the following sections will provide ample support to this claim.

Of course, the above conclusion is partially related to the
realizations of the cosmological scenarios we have studied. The
behaviour of the power spectrum $P(k)$ on large scales will
considerably influence the generality of our findings. A power
spectrum with more power on large scales would modify our findings:
potentially it may be so that we need a representation of the matter
distribution out to larger radii than $100\hmpc$. In this respect it
is important to note that the used $N$-body velocity fields do not
have any contributions from wavelengths larger than $\approx 175
\hmpc$ (both for $\tau$CDM as well as $\Lambda$CDM simulations, see
Fig.~\ref{fig:pksig}). This merely for the technical reason of the
simulation box imposing an upper limit to the scale on which we can
represent $P(k)$. The extent to which this may influence our
conclusions may be readily appreciated from Figure~\ref{fig:pksig}
(right column, top and bottom panel: compare solid lines with dashed
ones). The velocity field perturbations of $\tau$CDM and $\Lambda$CDM
carry out considerably further than the fundamental scale of the
simulation box, in particularly affecting the resulting bulk flows.

\begin{figure*}[t]
   \vskip -1.0truecm
   \centering
   \mbox{\hskip -1.65truecm\includegraphics[width=14.5cm,angle=270.0]
     {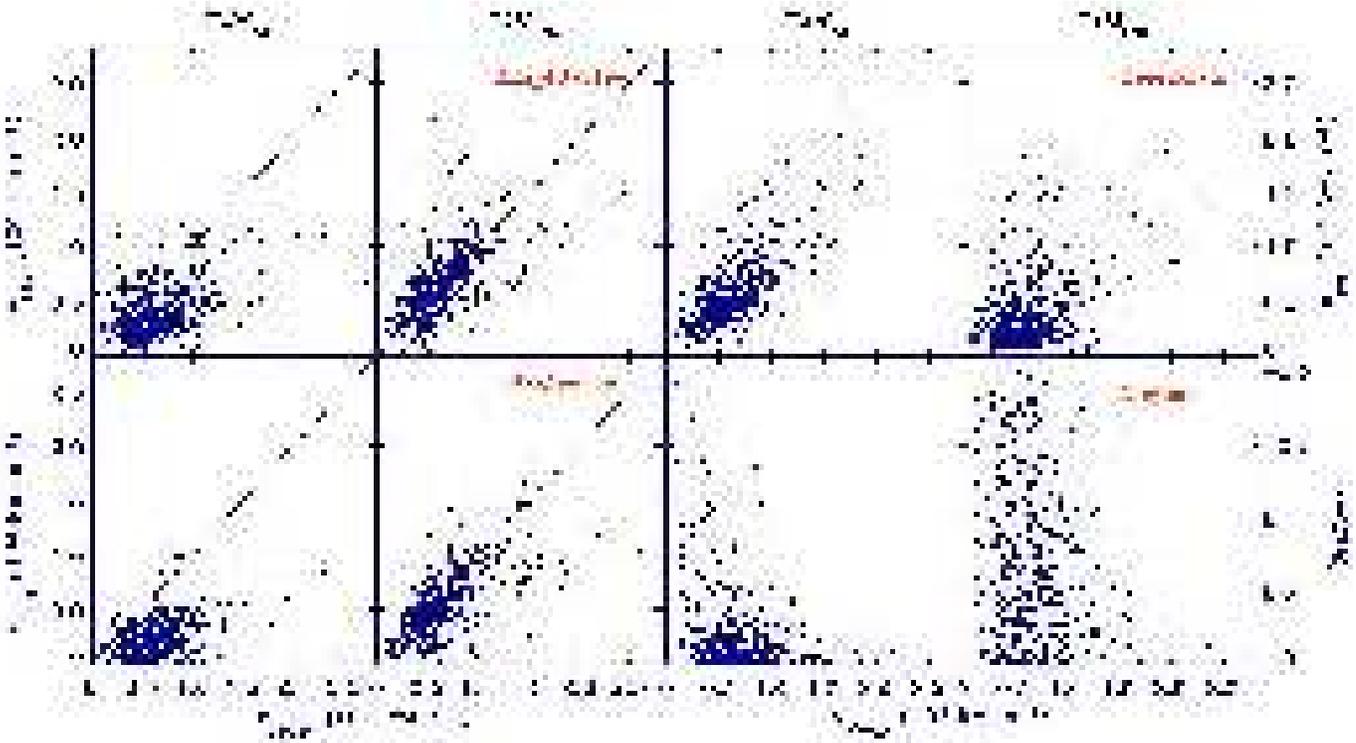}}
   \vskip -2.5truecm
   \caption{Point-to-point comparison (scatter plot) of 
     four quantities related to FAM reconstructed velocities and the
     corresponding ``real'' object velocity in the underlying $N$-body
     samples.  The $N$-body realization, and the mock sample, concern
     the $\Lambda$CDM scenario. For each quantity two panels are
     shown: the lefthand one is the one for FAM$_{30}$ reconstruction,
     the righthand one for the FAM$_{100}$ reconstruction.  Top
     lefthand: the amplitude of the velocities. Top righthand:
     residual velocity amplitudes FAM-$N$-body. Bottom lefthand: FAM
     velocity component projected along the $N$-body velocity.  Bottom
     righthand: angle $\theta$ between FAM velocity and $N$-body
     velocity, in terms of $\mu=\cos(\theta)$.}
  \label{fig:vvscat}
\end{figure*} 



\section{Point-to-point comparison}
\label{sec:pntpntnbg}

Scatter diagrams are used to assess the point-to-point comparisons
between quantitative aspects of the ``real'' galaxy velocities in the
original $N$-body samples and the computed velocities in the
FAM$_{30}$ and FAM$_{100}$ reconstructions. This analysis is meant to
be a direct, in principal local, assessment of systematic trends in
the velocity flows in volume of the NBG sample. The comparisons
involve a component of the ``true'' $N$-body velocity (abscissa)
versus the equivalent quantity for either the FAM$_{30}$ or
FAM$_{100}$ velocities, or of the corresponding residuals (ordinate).
Since the objects had been artificially added to the Zone of
Avoidance, any particles inside this region (see
sec.~\ref{sec:mkPSCz}) are excluded from these diagrams.


\subsection{Systematics}

If we neglect the small-scale sources in the deficiencies of FAM
reconstructions, the differences between FAM$_{30}$ and FAM$_{100}$
scatter plots are mainly to be ascribed to the corresponding
differences in the external gravitational influence acting over the
two corresponding sample volumes. In the external gravitational
influence the the corresponding leading velocity terms are the bulk
flow ${\rm v}_{\scriptstyle{\rm bulk}}$ and the velocity shear
$s_{ij}$,
\begin{equation}
  v_{\scriptstyle{\rm Nbody},i}\,\approx\,v_{\scriptstyle{\rm FAM},i}\,+\,
  v_{\scriptstyle{\rm bulk},i}\,+\,\sum_{j=1}^3\,{s}_{ij}\, x \hat{x}_j\,+\,
  \ldots
  \label{eq:vfamvnbd}
\end{equation} 
In the above $i,j$ denotes the Cartesian component indices. The
vectors $\hat{x}_{j}$ represent the vector components along the
Cartesian $j$ direction of the spatial unity vector oriented along the
object position vector ${\bf x}$. Following this definition, $x
\hat{x}_{j}$ is the $j$-component of the position vector ${\bf x}$,
with $x$ the distance of the object.

Systematic differences in FAM velocity-$N$-body velocity scatter
diagrams are therefore to be attributed to differences in bulk flow,
shear and possibly higher order contributions. Because each of these
large-scale phenomena will manifest themselves in distinctly different
ways, we seek to identify them from the scatter diagrams. An
horizontal offset in the scatter diagram would be the trademark for a
bulk flow component. Velocity shear would manifest itself as a
distinctly characteristic correlation between residuals and
velocities, although the prominence of this signal will be dictated by
shear magnitude, configuration, and orientation with respect to the
reference system (as is true for the bulk flow). In reality, the
situation will be more intricate. Subtle correlations between
small-scale and large-scale contributions will bring about a change in
the slope of the scatter diagram of FAM reconstructed velocity
components against their full $N$-body values
(see~\ref{sec:veldecomp}).


\subsection{Velocity Scatter Diagrams Analysis}

\begin{figure*}[t]
   \vskip 0.5truecm
   \centering
   \mbox{\hskip -0.5truecm\includegraphics[width=5.0in,angle=90.0]
     {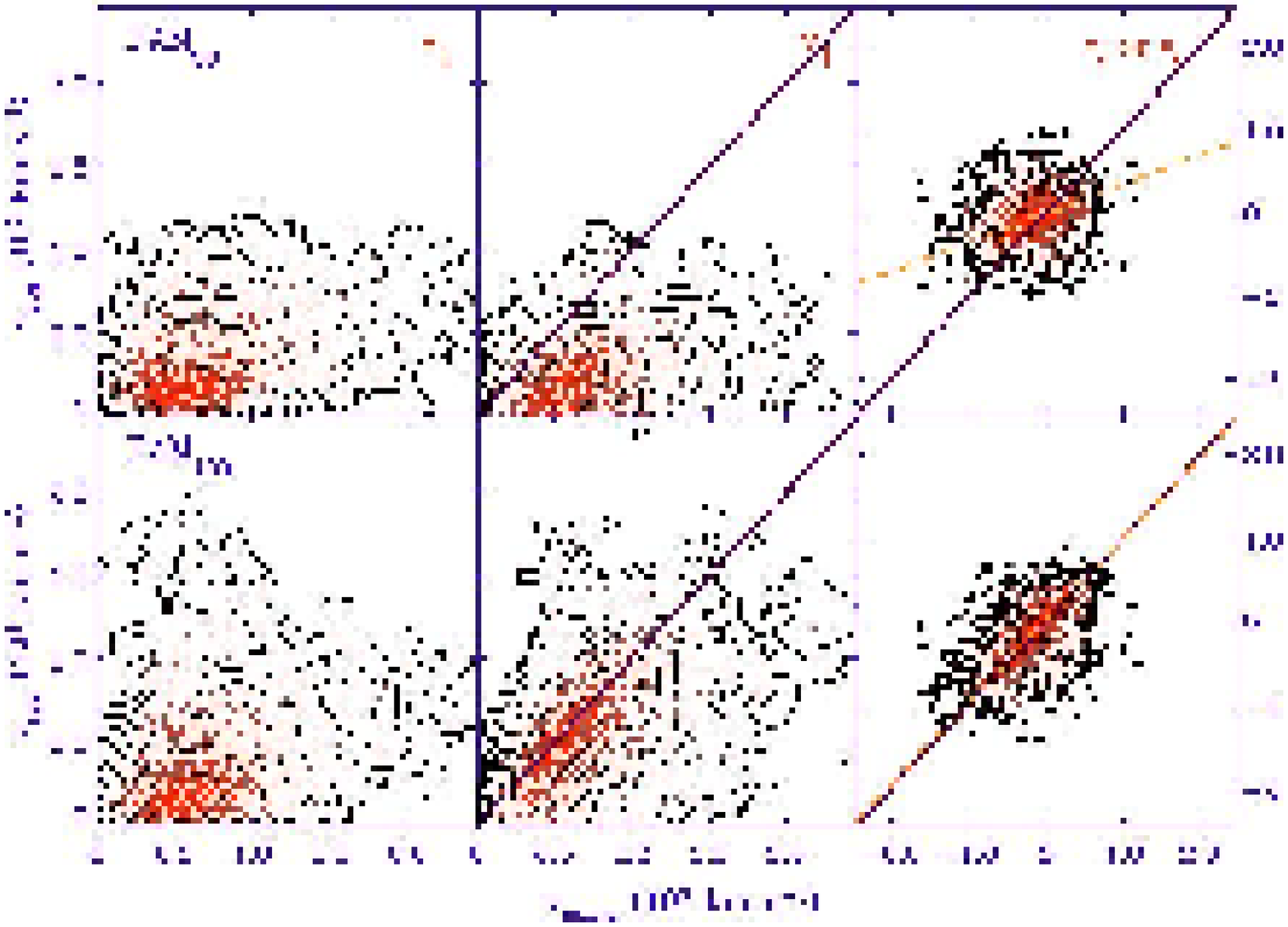}}
   \caption{Point-to-point comparison (scatter plot) of 
     FAM velocity decompositions wrt. the corresponding $N$-body
     velocity of the involved $\Lambda$CDM mock sample. Both the
     parallel projection $v_{\parallel}$ (centre frames) as well as
     the perpendicular projection $v_{\perp}$ along the $N$-body
     velocity direction are analyzed. Top row: FAM$_{30}$
     reconstruction against $N$-body velocity equivalent.  Bottom row:
     FAM$_{100}$ reconstruction against $N$-body velocity equivalent.
     Lefthand column: $v_{\parallel}$. Central column: $v_{\perp}$.
     Righthand column: The $x$ component of the velocity, $v_x$. The
     solid line in the righthand column frames indicates the identity
     line $v_{x,FAM}=v_{x,N-body}$. The dashed line indicates the best
     linear fit. Evidently, the FAM$_{30}$ deviate systematically from
     the identity line. The point scatter density is characterized by
     the contours, indicating iso-density surfaces. See
     Table~\ref{table:linreg}.}
 \label{fig:vvscatregr}
\end{figure*}

Scatter diagrams are presented in three successive figures. The
depicted scatter diagrams all relate to a $\Lambda$CDM mock catalogue,
and each of these point-to-point analyses relates to a different
aspect of the velocity field reconstructions. Figure~\ref{fig:vvscat}
contains four different panels, of which each contains two scatter
diagrams: FAM$_{30}$ versus $N$-body quantity (left) and the
equivalent FAM$_{100}$ versus $N$-body quantity (right). The diagrams
in Fig.~\ref{fig:vvscatregr} focus on the correlations between these
quantities and the scatter around regression relations. The figure
addresses three velocity-related quantities, each taking one column of
each 2 panels: the top one for the comparison of the FAM$_{30}$
components with their $N$-body counterparts, and the same for the
FAM$_{100}$ components in the bottom frame.

\begin{table*} 
  \centering 
  \caption{Results of the linear regressions in the tests with the
    mock catalogs. 
    Col. 1: Cosmological model. 
    Col. 2: Component
    Col. 3: label of the set.
    Col. 4: the average Spearman (non-parametric) correlation coefficient 
    Col. 5: average linear correlation index of the best fitting line and
            its $1\sigma$ scatter. 
    Col. 6: average zero point of the best fitting line and its 
            $1\sigma$ scatter. 
    Col. 7: average slope of the best fitting line and its $1\sigma$
    scatter. 
    Col. 8: average dispersion around the fit and its $1\sigma$ scatter }
  \label{table:vscatlrg}
  \begin{tabular}{lclccccc}
    \hline
    \\
    Cosmology & Component & Set &    $R_{Spear}$   &    $R_{lin}$    &$a_0$        &$a_{lrg}$ &
$\sigma_{lrg}$ \\
              &           &     &  &         &(km s$^{-1}$)&       & (km
s$^{-1}$)  \\
\\
    \hline\hline
\\
              &$\mid v
\mid$&FAM$_{30}$&0.54\,$\pm$\,0.11&0.56\,\,$\pm$\,0.11&\hfill -29.43\,\,$\pm$\,\ 84.28&
                                     0.66\,\,$\pm$\,0.21&198\,\,$\pm$\,\ 73\\
              &
&FAM$_{100}$&0.68\,$\pm$\,0.12&0.62\,\,$\pm$\,0.16&\hfill 83.38\,\,$\pm$\,156.39&
                                     0.85\,\,$\pm$\,0.26&212\,\,$\pm$\,\ 89\\
\\
  $\Lambda$CDM&$v_{\parallel}$&FAM$_{30}$&0.43\,$\pm$\,0.14&0.37\,\,$\pm$\,0.20&
              \hfill -198.21\,\,$\pm$\,173.86&0.65\,\,$\pm$\,0.30&251\,\,$\pm$\,114\\
              &
&FAM$_{100}$&0.59\,$\pm$\,0.23&0.52\,\,$\pm$\,0.29&\hfill -135.98\,\,$\pm$\,242.78&
                                     0.99\,\,$\pm$\,0.40&281\,\,$\pm$\,158\\
\\
              &  $v_i$
&FAM$_{30}$&0.54\,$\pm$\,0.20&0.48\,\,$\pm$\,0.29&\hfill 19.35\,\,$\pm$\,140.90&
                                     0.66\,\,$\pm$\,0.25&240\,\,$\pm$\,130\\
              &          &FAM$_{100}$&0.71\,$\pm$\,0.27&0.61\,\,$\pm$\,0.31&\hfill 5.96\,\,$\pm$\,
\ 78.94&
                                      0.86\,\,$\pm$\,0.23&233\,\,$\pm$\,107\\
\\
              &  $v_{res,i}$
&FAM$_{30}$&0.65\,$\pm$\,0.11&0.66\,\,$\pm$\,0.11&\\
              &          &FAM$_{100}$&0.40\,$\pm$\,0.14&0.47\,\,$\pm$\,0.19&\\
\\
              \hline
\\
              &$\mid v \mid$
&FAM$_{30}$&0.56\,$\pm$\,0.05&0.59\,\,$\pm$\,0.05&\hfill -54.10\,\,$\pm$\,\ 67.09&
                                     0.84\,\,$\pm$\,0.18&203\,\,$\pm$\,\ 57\\
              &
&FAM$_{100}$&0.66\,$\pm$\,0.10&0.59\,\,$\pm$\,0.13&\hfill -146.58\,\,$\pm$\,140.40&
                                     1.34\,\,$\pm$\,0.26&289\,\,$\pm$\,\ 80\\
\\
    $\tau$CDM&$v_{\parallel}$ &FAM$_{30}$&0.52\,$\pm$\,0.09&0.48\,\,$\pm$\,0.15&
              \hfill -313.16\,\,$\pm$\,134.10&1.04\,\,$\pm$\,0.24&272\,\,$\pm$\,100\\
              &
&FAM$_{100}$&0.63\,$\pm$\,0.14&0.55\,\,$\pm$\,0.19&\hfill -430.59\,\,$\pm$\,302.29&
                                     1.60\,\,$\pm$\,0.46&360\,\,$\pm$\,171\\
\\
              &  $v_i$
&FAM$_{30}$&0.68\,$\pm$\,0.13&0.63\,\,$\pm$\,0.15&\hfill -4.75\,\,$\pm$\,107.95&
                                     0.80\,\,$\pm$\,0.19&221\,\,$\pm$\,\ 81\\
              &
&FAM$_{100}$&0.79\,$\pm$\,0.17&0.69\,\,$\pm$\,0.13&\hfill 27.60\,\,$\pm$\,\ 70.70 &
                                     1.12\,\,$\pm$\,0.11&255\,\,$\pm$\,\ 87\\
\\
              &  $v_{res,i}$
&FAM$_{30}$&0.55\,$\pm$\,0.15&0.57\,\,$\pm$\,0.13&\\
              &          &FAM$_{100}$&0.22\,$\pm$\,0.13&0.27\,\,$\pm$\,0.14&\\
\\
    \hline\hline
  \end{tabular}
\label{table:linreg}
\end{table*}

A straightforward comparison is that between the Cartesian velocity
components $v_i$ of the FAM reconstructed velocities and the $N$-body
velocities (Fig~\ref{fig:vvscatregr}, righthand column). 
Complementary regressions involve coordinate system independent
aspects of galaxy velocities. These involve the velocity amplitude
$|v_{\scriptstyle{\rm FAM}}|$ (Fig.~\ref{fig:vvscat}, top lefthand
panel), the component of each FAM velocity parallel to the
corresponding $N$-body velocity, $v_{\parallel}$ ($\equiv$
v$_{\scriptstyle{\rm proj}}$), and the additional perpendicular
component $v_{\perp}$ (Fig~\ref{fig:vvscatregr}, first column).
Misalignments between the real $N$-body velocity and the FAM velocity
reconstructions should indicate in how far a reconstruction has been
failing to take into account all relevant gravitational forces along
the path of a particle. Systematic misalignments reveal themselves in
the scatter diagram of the angle $\theta$ between the FAM velocity and
the galaxies' $N$-body velocity ${\bf v}_{\scriptstyle{\rm Nbody}}$
(in Fig.~\ref{fig:vvscat} we plot $\mu \equiv \cos(\theta)$, bottom
righthand panel). In terms of the character and systematics of the
underlying physics and dynamics the residual velocities, [$N$-body -
FAM], represent highly informative aspects in our analysis. They are
assessed in (Fig.~\ref{fig:vvscat}, top righthand panel) and
Figure~\ref{fig:vvresscat}.

The significance and strength of correlations between the $N$-body and
FAM velocity components in the scatter diagrams are analyzed by means
of a linear regression and correlation analysis. To circumvent
excessive pollution of the computed parameters by the virialized
motions in high-density regions (see sect.~\ref{sec:fam}), the galaxy
velocity components in these regression analyses involve $2 \hmpc$
tophat filtered velocity fields. The resulting numerical values of the
correlation parameters are listed in Table~\ref{table:vscatlrg}.
Table~\ref{table:vscatlrg} is organized in two separate sections, one
for the regression analysis results of the $\Lambda$CDM mock samples
(top section) $\tau$CDM samples (bottom section). For
both the $\Lambda$CDM and the $\tau$CDM section we list the results
for four velocity related quantities, each separately for the
FAM$_{30}$ and the FAM$_{100}$ reconstructions.

The presence of significant correlations between FAM reconstructions
and their $N$-body counterparts is evaluated on the basis of the
nonparametric Spearman correlation coefficient $R_{Spear}$. The linear
regression parameter $R_{lin}$ quantifies the linearity of the
relation. Prevailing in most situations, the linear regression
parameters are used to characterize the relation between reconstructed
and real $N$-body velocities: the zero-point (offset) $a_0$, the slope
$a_{lrg}$ and the dispersion $\sigma_{lrg}$ around the linear
regression relation. We assume equal errors in FAM and $N$-body
velocities, as both are affected by similar shot noise errors (while
$2 \hmpc$ top hat smoothing significantly reduces the impact of virial
motions on FAM velocity predictions). In addition, we also list the
rms scatter of the parameters, estimated on the basis of the results
for the 10 different mock catalogs (for each of the four different
configurations).


\subsection{Inventory}

\subsubsection{Velocity Amplitude} 

In the top lefthand frame of Fig.~\ref{fig:vvscat} the FAM$_{30}$ and
FAM$_{100}$ velocity amplitudes are compared with their $N$-body
counterpart $|v_{\scriptstyle{\rm Nbody}}|$.  The FAM$_{30}$ diagram
differs considerably from the FAM$_{100}$ diagram: the FAM$_{30}$
velocities are systematically smaller than their FAM$_{100}$
counterparts. Also, while the latter have a strong one-to-one
correlation to the $N$-body velocities, the FAM$_{30}$ diagram shows a
systematic offset with respect to this relation (the solid line) and a
somewhat larger scatter.

While the FAM$_{100}$ diagram tapers out to higher velocities and even
shows a few points with $|v_{\scriptstyle{\rm FAM}}| > 2000 \kms$,
there is a firm ceiling of $|v_{\scriptstyle{\rm FAM}}|\approx 1300
\kms$ for the FAM$_{30}$ velocities. It is a direct reflection of the
FAM$_{30}$ reconstructions {\it missing out on the gravitational force
  contributions by the external mass distribution}. The asymmetric
nature of the scatter in both diagrams is due to particles in high
density regions. 

To a good approximation, the correlation between the FAM$_{100}$
velocity amplitudes and $|v_{\scriptstyle{\rm Nbody}}|$ is that of a
linear identity relation: the solid line, $|v_{\scriptstyle{\rm
Nbody}}|=|v_{\scriptstyle{\rm FAM}}|$, forms a good fit to the scatter
diagram (see Table~\ref{table:vscatlrg}: $a_{lrg} \rightarrow 1$).
The significantly higher value of Spearman's correlation coefficient
(Table~\ref{table:vscatlrg}: $R_{Spear}\approx 0.68$ vs.
$R_{Spear}\approx 0.54$ for FAM$_{30}$) indicates and confirms the
visual impression of Fig.~\ref{fig:map100nb}) of the tight
correspondence between the FAM$_{100}$ and $N$-body vector velocity
fields.  The FAM$_{30}$ results stand in marked contrast: the majority
of the FAM$_{30}$ velocities have a systematically lower amplitude
than their $N$-body counterparts. It results in a relation with a
significantly shallower slope than that of the identity relation
$|v_{\scriptstyle{\rm Nbody}}|=|v_{\scriptstyle{\rm FAM}}|$ (also see
table~\ref{table:vscatlrg}): objects with a higher velocity have a
larger discrepancy. The contribution by the missing large-scale
velocity component ${\bf v}_{\rm lss}$ to the amplitude of the FAM
velocity includes a cross-term ($-{\bf v}_{\rm Nbody}\cdot{\bf v}_{\rm
lss}$), a term dependent on the velocity ${\bf v}_{\rm Nbody}$ of the
galaxy. Most of the missing large-scale velocity component ${\bf
v}_{\rm lss}$ is due to the absence of a bulk flow term in the
FAM$_{30}$ reconstructions. Subtle and/or higher order external
gravitational effects play an additional role: the velocity vector
diagrams did already reveal that the presence of shear should be one
of the main contributors (cf. eq.~\ref{eq:vfamvnbd}).

When comparing the $\Lambda$CDM FAM velocity amplitudes with those of
the $\tau$CDM reconstructions it is evident that in the case of the
FAM$_{30}$ reconstructions the latter adhere considerably better to
the corresponding $N$-body values. The linear fitting slope $a_{lrg}$
(see table~\ref{table:vscatlrg}) is considerably closer to unity for
the $\tau$CDM samples than for the $\Lambda$CDM samples. Over a
30$\hmpc$ volume the external density inhomogeneities in the
$\Lambda$CDM cosmology will induce considerably higher bulk flows than
the more moderate $\tau$CDM perturbations, which is entirely in line
with the theoretical expectation (fig~\ref{fig:pksig}). In the case of
the FAM$_{100}$ reconstructions the qualitative differences are far
less prominent.  On the scale of $\sim 100\hmpc$ the mass distribution
in both the $\tau$CDM and $\Lambda$CDM simulation volumes have
converged to homogeneity and no major bulk flows are to be expected.

\subsubsection{Velocity Decompositions} 
\label{sec:veldecomp}

In the two righthand frames of Fig.~\ref{fig:vvscatregr} we show the
scatter diagrams for the x-component of the FAM$_{30}$ (top) and
FAM$_{100}$ velocities (bottom). Although with a significant level of
scatter, the FAM$_{100}$ diagram can be fitted quite well by a
straight line with a slope close to unity (linear regression line:
dashed,unity line: solid). That the equivalent FAM$_{30}$ diagram may
also be fitted by a straight line, be it with a slope significantly
smaller than unity is not entirely straightforward. It stems from an
intricate interplay between the small scale velocity field and its
larger scale contributions, which in most circumstances are not
uncorrelated (Berlind, Narayanan \& Weinberg \cite{bernarw2000}).

Table~\ref{table:vscatlrg} lists the linear regression parameters.
Although the average best fitting slopes for the FAM$_{100}$
velocities are either larger ($\tau$CDM) or smaller ($\Lambda$CDM)
than unity, the deviation from unity is considerably smaller than that
for the FAM$_{30}$ velocities, well within the $\sim 1\sigma$
uncertainty interval. In all, these regression results do adhere to
the expected and noted trend of FAM$_{100}$ velocities accounting for
practically all contributions to the local velocity field and
FAM$_{30}$ velocities systematically neglecting significant external
contributions. Notice that the scatter around the regression lines for
the FAM$_{30}$ and FAM$_{100}$ reconstructions is of comparable
magnitude (as may be inferred from the superposed number density
contours). 

While the choice of any Cartesian coordinate system is an arbitrary
one we have also addressed the decomposition of the particle velocities
in one defined by the system itself. The FAM velocities are decomposed
in a component projected along the corresponding $N$-body velocity,
$v_{proj}$ (or $v_{\parallel}$) and the complementary perpendicular
component, $v_{\perp}$. The second column of
figure~\ref{fig:vvscatregr} contains the scatter diagrams for the
parallel component of the FAM$_{30}$ (top) and FAM$_{100}$ (low)
velocities. Qualitatively, the behaviour of both diagrams resembles
that of the velocity amplitude scatter diagrams in
fig.~\ref{fig:vvscat}. A $1-1$ relation between FAM$_{100}$ velocities
and $N$-body velocity amplitude represents a reasonable fit (solid
line: slope $a_{lrg}\approx 0.99$). The FAM$_{30}$ diagram not only
appears to deviate strongly from such a $1-1$ relation, it may even
fail to fit any linear relation. Also, none of the projected
FAM$_{30}$ velocity components appears to supersede a value of
$\approx 1200 \kms$. Given the fact that the equivalent FAM$_{100}$
component even surpasses values of $\approx 2000 \kms$, this confirms
the systematic deficiency of the gravitational field in the FAM$_{30}$
evaluations.

From the scatter diagrams for the perpendicular FAM velocity
components, $v_{\perp}$, one can infer that almost all systematic
effects are confined to the parallel components $v_{\parallel}$. For
both FAM$_{30}$ and FAM$_{100}$ the complementary perpendicular
component lacks a systematic correlation with the $N$-body velocity.
It mainly represents unrelated scatter, with a magnitude concentrated
around values of $\sim 200-250 \kms$.  The only difference between the
FAM$_{30}$ and FAM$_{100}$ reconstructions is that for the latter
$v_{\perp}$ involves considerably higher values, reflecting the higher
amplitude of the FAM$_{100}$ velocities. FAM$_{30}$ velocities, on the
other hand, involve stronger misalignments
(section~\ref{sec:velalign}).

\subsubsection{Velocity Alignments} 
\label{sec:velalign}

Misalignments between the reconstructed FAM velocities and the
$N$-body velocity vectors are the result of a few effects.  A major
source is that of localized small-scale effects. These are not
expected to lead to systematic offsets: they will have a {\it noisy}
character and reflect random motions in highly nonlinear environments,
in particular those of dense virialized regions.  Because these have
no preferred direction, they behave like randomly oriented
``residual'' velocities wrt. to the real $N$-body velocities of
galaxies.  Of an entirely different nature are misalignments stemming
from the systematic neglect of the external gravitational forces.
Because the resulting residual velocity vectors comprise systematic
components along one or a few preferred directions, a distinctly
anisotropic distribution is the result. This reflect itself as a
systematic trend for total $N$-body galaxy velocities to be aligned
along the residual velocity components.

For both the FAM$_{30}$ and FAM$_{100}$ reconstructions we computed
the angles between the $N$-body velocity vector and the FAM
velocities. The lower righthand panel in Figure~\ref{fig:vvscat}
confirms that the alignments of ``residuals'' and total velocity is
indeed considerably stronger for the FAM$_{30}$ reconstructions than
the FAM$_{100}$ ones. The figure plots, for each galaxy in the sample,
the misalignment angle $\theta$ (or, rather, $\mu=\cos(\theta)$)
versus the $N$-body velocity magnitude $v_{\scriptstyle{\rm Nbody}}$.
For the FAM$_{100}$ velocities we see a near isotropic distribution of
angles. With the exception of a minor concentration near perfect
alignment, $\cos(\theta)=1$, the distribution is sweeping out nearly
uniformly over the full range of $\mu \equiv \cos(\theta) = 1
\rightarrow -1$. If at all there is a trend in velocity amplitude, it
appears to be the weak tendency for large velocities to be better
aligned.

The above results reflect the observation that FAM$_{100}$ residual
velocities do mainly consist of small-scale random effects. The
FAM$_{30}$ residuals form a telling contrast. They are heavily aligned
along the full $N$-body velocities, with a very strong concentration
near $\theta=0$.  Although occasionally there are serious
misalignments, their occurrence diminishes rapidly towards large
$\theta$. When they occur it almost exclusively concerns small
velocities, mostly corresponding to serious misalignments between the
locally induced velocity and the added external velocity component.

\begin{figure*}[t]
   \centering
   \mbox{\hskip -0.5truecm\includegraphics[width=6.0in]{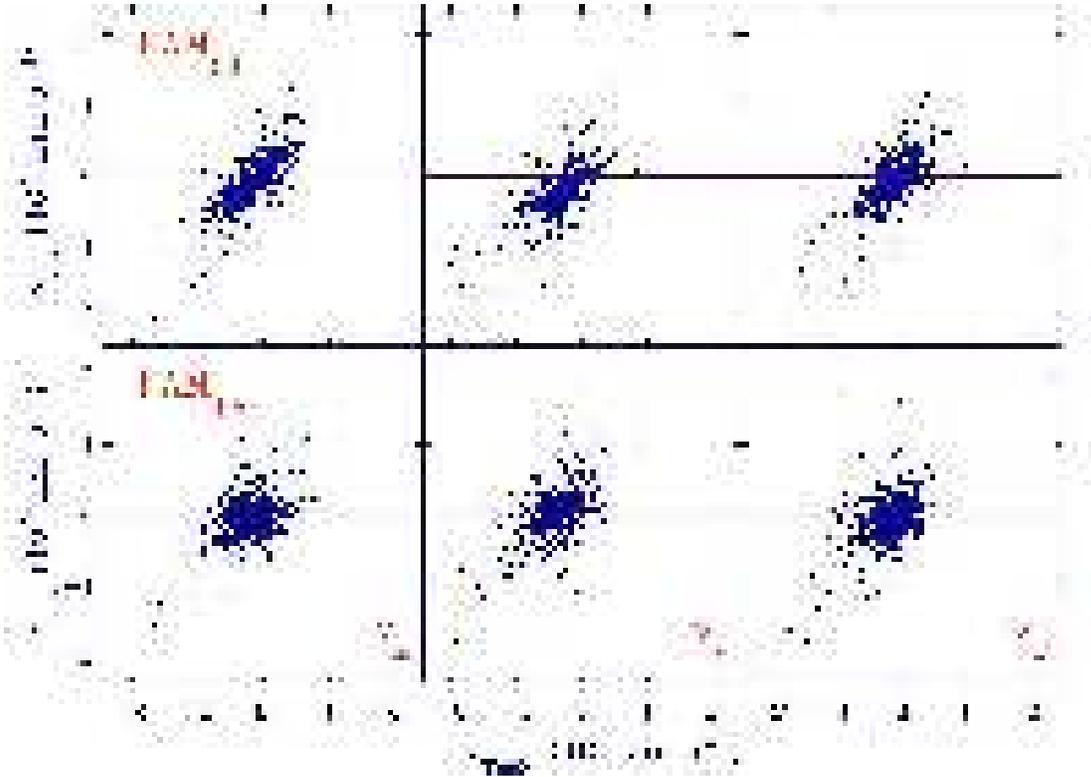}}
   \caption{point-to-point comparison (scatter plot) of the Cartesian 
     components of the residual velocity ${\bf v}_{FAM}-{\bf
       v}_{Nbody}$ components against corresponding $N$-body velocity
     ${\bf v}_{Nbody}$.  From left to right: $v_x$, $v_y$ and $v_z$.
     Top row: FAM$_{30}$ reconstructed velocity. Bottom row:
     FAM$_{100}$ reconstructed velocity.}
 \label{fig:vvresscat}
\end{figure*} 

\subsubsection{Velocity Residuals}

The residuals accumulate all systematic physical effects as well as
random artifacts. They are therefore an excellent source of information 
on the dynamical
role of matter concentrations in the various galaxy sample volumes. If
there are large external contributions to the galaxies' velocity these
will constitute a major part of the residuals. On the other hand, if
most of those influences are contained within the sample volume
treated by FAM, the residuals may mainly reflect localized
nonlinearities and artifacts of the FAM method. Scatter diagrams
involving the residual velocities will indicate systematic trends and
are well suited for elucidating the character and underlying dynamics
of external influences.

Figure~\ref{fig:vvresscat} elaborates on this observation. In two
successive rows, the top one for the FAM$_{30}$ reconstructions and
the bottom one for the FAM$_{100}$ reconstructions, it displays the
residuals ${\bf v}_{res}$ for each of the three Cartesian velocity
components, $v_x$, $v_y$ and $v_z$. Each panel plots the velocity
component residual as a function of the corresponding $N$-body
velocity component.

The mark of a bulk velocity is a constant offset of the scatter
diagram, a translation of all FAM velocities by a constant term. This
is indeed what is observed in the FAM$_{30}$ $v_y$ scatter diagram:
the vast majority of points is located beneath the
$v_y=0~\kmsmpc$ line. It is a telling confirmation of the
impression yielded by the corresponding velocity vector fields in
Fig.~\ref{fig:map30nb}. The velocity vector field revealed the
presence of strong bulk flow oriented almost perfectly along the
$y$-axis: clearly visible in the $N$-body velocity field, hardly
present in the corresponding FAM$_{30}$ velocity field, and
representing a major component of the residual field [$N$-body -
FAM$_{30}$].  When turning to the equivalent FAM$_{100}$ diagram, the
indicative offset for a bulk flow has almost completely disappeared.
This implies that the source(s) for the bulk flow should be found
within the region between $30\hmpc$ and $100\hmpc$. The equivalent
$v_x$ and $v_z$ FAM$_{30}$ residual scatter diagrams do confirm the
visual impression of there hardly being a bulk flow contribution along
the $x$- and $y$-directions.

Additional systematic behaviour is readily apparent in
Fig~\ref{fig:vvresscat}: the diagrams show an almost linear increase
of residual velocity with $N$-body velocity. Also, we find that the
$v_x$ scatter is skewed towards negative $v_{res}$ values while the
$v_z$ diagram is skewed towards positive $v_{res}$ values.  In the
equivalent FAM$_{100}$ scatter diagrams the linear increase of
$v_{res}$ and the asymmetry in the $v_x$ and $v_z$ diagram has almost
disappeared: the dense core of points has turned into a compact and
nearly horizontal bar symmetrically distributed around the $v_{res}=0
\kms$ line. To a large extent this is explained by the much smaller
contribution of external tidal shear to the flows over $100\hmpc$
volumes (cf. Fig.~\ref{fig:pksig}).

The mark for external shear is a near linear increase of residual
velocities as a function of their $N$-body (or measured) velocity.
Depending on the location ${\bf x}$ of a galaxy within the sample
volume and with respect to the shear configuration its participation
in a shear flow will involve a velocity component $v_s\equiv\sum
s_{ij} x_j$. This may involve a negative or a positive contribution.
With such shear contributions representing a non negligible component
to the total velocity, its systematic contribution to a largely random
local residual signal reshuffles the velocities such that on average
the largest velocity involves the largest residual contribution.

With prominent large-scale bulk and shear motions at large, the
FAM$_{100}$ residual scatter diagram has largely transformed into a
featureless and purely random point distribution. The residuals mainly
involve uncorrelated small-scale effects and are nearly independent of
the amplitude of the $N$-body velocity. Some additional artifacts are
seen upon closer inspection: the presence of diffuse ``S''-shaped
point clouds in both the FAM$_{30}$ and FAM$_{100}$ residual diagrams,
tapering off towards a steep tail at both the negative and positive
side of the plots. 

The corresponding scatter diagrams for the velocity residual
amplitudes $|{\bf v}_{\scriptstyle{\rm res}}| = |{\bf
  v}_{\scriptstyle{\rm FAM}}-{\bf v}_{\scriptstyle{\rm Nbody}}|$
represents a summary of the systematic trends
(Fig.~\ref{fig:vvscatregr}). The FAM$_{30}$ velocity residuals show a
near linear increase as a function of the $N$-body velocity, starting
with an offset, indicative of the ingredients of bulk and shear flow
in the residuals. The lack of any clear correlation between $|{\bf
  v}_{\scriptstyle{\rm res}}|$ and $|{\bf v}_{\scriptstyle{\rm
    Nbody}}|$ in the case of the FAM$_{100}$ residuals confirms the
absence of such systematic components. More clearly than in the case
of the individual Cartesian components, the presence of local
nonlinear motions may be discerned from the extensive surrounding
clouds of outliers.


\subsection{Power Spectrum Dependence} 

The contrast between FAM$_{30}$ and FAM$_{100}$ scatter diagrams is
more pronounced in the case of the $\Lambda$CDM mock catalogs than in
those assembled for the $\tau$CDM universes. This clearly reflects the
fact that within the $\Lambda$CDM scenario cosmic structure is
characterized by a larger coherence scale. It implies the presence of
larger and more coherent structures whose size exceeds $30 \hmpc$.
Their combined gravitation impact will yield a stronger systematic
impact in the velocity-velocity comparisons. On the other hand, the
dispersions listed in table 3 also show that it would hardly be
possible to infer information on the cosmological scenario on the
basis of one individual realization. The large dispersion around the
average slopes, in particular in the case of the $\Lambda$CDM
Universe, show that the magnitude of the external dynamical effects
may vary appreciably as a function of the location of the (mock) NBG
sample within the simulation box. Local measurements will therefore be
unable to separate cosmological effects from those stemming from local
variations.


\subsection{Nonlinearities} 

The point-to-point diagrams discussed above all contain a substantial
level of scatter around the inferred regression relations. With a few
exceptions the scatter of velocity quantities is in the order of $\sim
200-250 \kms$, for both the $\Lambda$CDM as well as the $\tau$CDM FAM
reconstructions. The main source for this scatter are the virial
motions in the high density and mildly nonlinear environments. Also
shot noise provides a substantial additional contribution. In the case
of small filter radii, another source of scatter is formed by spurious
very close pairs of points in the parent $N$-body catalog which for
artificial reasons failed to collapse into a single object (Branchini,
Eldar \& Nusser \cite{branchini02}). Scatter may also be due to higher
order multipole components in the external gravity field. An
inspection of the particle configurations and the velocity vector maps
does unmistakably show significant systematic variations on top of
dipolar and quadrupolar components.  However, tests restricting the
analysis to points in the central regions of the sample produced no
substantial decrease in level of scatter. This seems to argue for a
minor role of such contributions.


\section{Bulk Flow and Tidal Shear:\\
  \hskip 0.5truecm Velocity Flow Multipole Components}
\label{sec:nbg+bs}

In the previous sections we have found that in order to obtain a good
representation of the local cosmic velocity field it is necessary to
take into account the external gravitational influence. This was
accomplished through the incorporation of the fully detailed external
mass distribution contained in the (flux-limited) galaxy catalogs.
This involved the galaxy distribution out to distances of $100\hmpc$.
The reconstructions showed that modelling of velocity fields by FAM
with the inclusion of matter concentrations on such large scales is
indeed rewarding.

In nearly all situations where the local volume $V_{int}$ is suitably
large, the small-scale details of the external mass configuration are
rather irrelevant for constructing an appropriate model of the flows
in the local Universe. An appropriate approximate expression for the
the gravitational potential $\Phi_{ext}({\bf r})$ inside the internal
volume $V_{int}$ due to the surrounding external matter distribution
follows from its expansion in multipole contributions. Assuming a
spherical local volume with radius $R_{int}$, the potential
$\Phi_{ext}$ may be written in terms of a multipole expansion of
spherical harmonics $Y_{lm}(\theta,\phi)$ (see e.g.
Jackson~\cite{jackson})
\begin{eqnarray}
  \Phi_{ext}({\bf r})&\,=\,&-\,\int_{R_{int}}^{\infty}\,{\displaystyle G 
    \rho({\bf x}') \over 
    \displaystyle |{\bf x}-{\bf x}'|}\,d{\bf x}' \nonumber\\
  \\
  &\,=\,&-\,\sum_{l=0}^{\infty}\,\sum_{m=-l}^{m=l}\,{\displaystyle 4 \pi G 
    \over \displaystyle 2l+1}\,{\cal Q}_{lm}\,Y_{lm}(\theta,\phi)\,r^l
  \,,\nonumber
  \label{eq:phiextmulpl}
\end{eqnarray}
in which the {\it multipole moments} ${\cal Q}_{lm}$ relate to the
external density field $\rho({\bf x}')$ as
\begin{equation}
  {\cal Q}_{lm}\,=\,\int_r^{\infty}\,\rho({\bf x}')\,r'^{l-3}\,Y_{lm}^*
  (\theta',\phi')\,d{\bf x}'\,.\\
  \label{eq:multmom}
\end{equation}
Most contributions to the external gravity ${\bf g}_{ext}$ will be
confined to these dipole and quadrupole components, induced by the
corresponding large-scale constellations in which the surrounding
matter concentrations have grouped themselves. Here we will assess the
approximation in which the potential expansion
(eqn.~\ref{eq:phiextmulpl}) is restricted to the monopole term $l=0$,
the dipole term $l=1$ and the quadrupole term $l=2$,
\begin{equation}
  \Phi_{ext}({\bf r})\,\approx\,\Phi_0({\bf r})\,+\,\Phi_1({\bf r})\,+\,
  \Phi_2({\bf r})\nonumber\\
\end{equation}
with
\begin{equation}
  \Phi_l({\bf r})\,=\,-\,\sum_{m=-l}^{m=l}\,{\displaystyle 4 \pi G \over 
    \displaystyle 2l+1}\,{\cal Q}_{lm}\,Y_{lm}(\theta,\phi)\,r^l\,.
\end{equation}
To explore the nature of the external component in the total 
gravitational field in the Local Universe we proceed by probing it
through the resulting peculiar velocity field. The amplitude of higher
order terms may be assumed to be so small that one cannot expect to
deduce any significant value, given the sizeable errors in the
available galaxy peculiar velocity datasets, we may expect this to be
a reasonable approximation. We investigate the velocity field by
decomposing the residual velocity field -- i.e. the component in the
velocity field which could not be accounted for in the FAM$_{30}$
reconstruction and supposedly induced by external influences -- into
its multipole components. Once we have determined the bulk flow
component and shear tensor components in the tidal velocity field, we
will assess whether we can indeed relate this to the external
gravitational (``tidal'') influence within the local Universe.

Restricting the description of the external gravitational influence to
the first few orders of its multipole expansion has several
advantages. The large-scale external dipole and quadrupole gravity
perturbations retain a largely linear character, simplifying the
velocity field analysis and thus retaining the direct linear relation
between gravity and velocity field.  Also, by discarding its
small-scale fluctuating contributions a physically more transparent
image of the velocity field is obtained. This allows a straightforward
relation and translation towards the corresponding large-scale pattern
of the surrounding mass distribution. A final practical issue of some
importance is the fact that the dipole and quadrupole characterization
is particularly suited for an implementation in FAM.  Restricting the
external force field to these moments alleviates the need to take into
account a large sample of external galaxies. Not only is the latter
computationally expensive, in practice it is even not always feasible.


\subsection{Velocity Field Multipole Decomposition}
\label{sec:vdec}  

In the multipole analysis we restrict ourselves to the externally
induced velocity components, ${\bf v}_{ext}$, which in the following
we frequently designate by the term {\it ``tidal''}\footnote{in the
  following we regularly use the word {\it ``tidal''} to shortly
  indicate the externally induced component of a gravity or velocity
  field. As it includes a dipolar contribution, strictly speaking this
  is not an appropriate term. Also see sect.~\ref{sec:exttidpot}.}.
For each object, the {\it ``tidal''} velocity vector is determined by
subtracting the internally induced velocity field, ${\bf v}_{int}$,
from the object's full velocity. The latter is usually the $N$-body
velocity of the mock galaxy, although we will assess the possibility
of using the FAM$_{100}$ velocity as a reasonable alternative. The
internal velocity ${\bf v}_{int}$ is deduced by evaluating, through
our FAM computations, the impact of the internal matter distribution
within the internal catalog volume $V_{int}$. The resulting (residual)
peculiar velocity vector ${\bf v}_{ext}$ field may then be expressed
in terms of a Taylor series description as function of spatial
position ${\bf x}$.

For the practical implementation, we follow the general scheme
described by Kaiser (\cite{kaiser}). The velocity field Taylor
expansion is truncated at the quadratic term and is restricted to the
dipole and quadrupole moments (and a minor monopole term). The {\it
  tidal} velocity field ${\bf v}_{ext}$, is then modeled by the the
first two components, a bulk flow vector, ${\tilde u}_{i}$, and a
quadratic shear tensor contribution, ${\tilde s}_{ij}$,
\begin{equation}
  v_{ext,i}= {\tilde u}_{i}+{\tilde s}_{ij}x\hat{x}_{j}, \qquad 
  {\rm where} \qquad i,j = \{1,2,3\}\ , 
  \label{veldecomp}
\end{equation}
in which $i,j$ denotes the Cartesian component indices. As in
eqn.~\ref{eq:vfamvnbd}, the vectors $\hat{x}_{j}$ represent the vector
components along the Cartesian $j$ direction of the spatial unity
vector oriented along the object position vector ${\bf x}$. Using
these notations, we can easily reconfigure eqn.~\ref{veldecomp} and
express the $i$-component of the velocity of object $n$ into a product
of the vectors $F_{n,I}$ and $V_{Ii}$,
\begin{eqnarray*}
  v_{n,i}= \sum_{I=1}^4 F_{n,I}({\bf x})V_{Ii} \nonumber
\end{eqnarray*}

\begin{table*}
    \centering 
    \begin{tabular}{llcccccl}\hline\hline
      \\
      Field& Cosmology & $v_{bulk}$ & $s$ & $s_1$ &$s_2$ & $s_3$ &\\
      Configuration&        & (km s$^{-1}$) \\
      \\
      \hline\hline
      \\
      &$\Lambda$CDM&313\,\,$\pm$\,65& 13.3\,\,$\pm$\,3.8&8.8\,\,$\pm$\,3.6&
      0.6\,\,$\pm$\,2.5&\,\ -9.4\,\,$\pm$\,2.3&
      (km s$^{-1}$ Mpc$^{-1}$) \\
      &$\tau$CDM   &216\,\,$\pm$\,57&13.6\,\,$\pm$\,4.6&8.9\,\,$\pm$\,3.1&
      1.1\,\,$\pm$\,2.0&-10.0\,\,$\pm$\,3.7& \\
       $N-$body -- FAM$_{30}$\\
      &$\Lambda$CDM&                &399\,\,$\pm$\,115&264\,\,$\pm$\,107&
      19\,\,$\pm$\,\,\,76&-283\,\,$\pm$\,\,\,70& 
      (km s$^{-1}$) \\
      &$\tau$CDM&                &409\,\,$\pm$\,137&267\,\,$\pm$\,\,\,92&
      33\,\,$\pm$\,\,\,60&-300\,\,$\pm$\,110&\\
      \\
      \hline
      \\
      &$\Lambda$CDM&81\,\,$\pm$\,51& 3.7\,\,$\pm$\,1.2&2.6\,\,$\pm$\,0.9&
      -0.2\,\,$\pm$\,0.8&\,\ -2.4\,\,$\pm$\,1.0& 
      (km s$^{-1}$ Mpc$^{-1}$) \\
      &$\tau$CDM   &121\,\,$\pm$\,54&4.8\,\,$\pm$\,1.1&3.4\,\,$\pm$\,1.0&
      \,0.0\,\,$\pm$\,0.8&\,\ -3.3\,\,$\pm$\,0.8& \\
      $N-$body -- FAM$_{100}$\\
      &$\Lambda$CDM&                &110\,\,$\pm$\,36&78\,\,$\pm$\,\,\,27&
      \,\,-6\,\,$\pm$\,25&\ -72\,\,$\pm$\,\,\,30& 
      (km s$^{-1}$) \\
      &$\tau$CDM&                &145\,\,$\pm$\,33&101\,\,$\pm$\,\,30&
      \,\,-1\,\,$\pm$\,23&-100\,\,$\pm$\,\,23&\\
      \\
      \hline
      \\
      &$\Lambda$CDM&317\,\,$\pm$\,89& 14.0\,\,$\pm$\,4.6&9.4\,\,$\pm$\,4.0&
      0.5\,\,$\pm$\,2.9&\,\ -9.9\,\,$\pm$\,3.0& 
      (km s$^{-1}$ Mpc$^{-1}$) \\
      &$\tau$CDM   &245\,\,$\pm$\,60&15.3\,\,$\pm$\,4.3&9.8\,\,$\pm$\,2.5&
      1.4\,\,$\pm$\,2.9&-11.2\,\,$\pm$\,3.8& \\
      FAM$_{100}$ -- FAM$_{30}$\\
      &$\Lambda$CDM&                &421\,\,$\pm$\,138&281\,\,$\pm$\,120&
      15\,\,$\pm$\,\,86&-296\,\,$\pm$\,\,\,91& 
      (km s$^{-1}$) \\
      &$\tau$CDM&                &458\,\,$\pm$\,128&293\,\,$\pm$\,\,\,75&
      42\,\,$\pm$\,\,87&-335\,\,$\pm$\,115&\\
      \\
      \hline
      \\
      &$\Lambda$CDM&81\,\,$\pm$\,51& 2.4\,\,$\pm$\,1.3&-0.6\,\,$\pm$\,1.7&
      -0.2\,\,$\pm$\,1.4&\,\ 0.9\,\,$\pm$\,1.3& 
      (km s$^{-1}$ Mpc$^{-1}$) \\
      $N-$body - FAM$_{mpl}$ \\
      &$\tau$CDM   &121\,\,$\pm$\,54&2.9\,\,$\pm$\,0.8&-0.6\,\,$\pm$\,1.6&
      -0.5\,\,$\pm$\,1.6&0.9\,\,$\pm$\,1.7& \\
      \\
      \hline\hline
    \end{tabular}
    \vskip 0.3truecm
    \caption{Average results for the tidal bulk amplitude and  
      shear eigenvalues. The errors represent the $1\sigma$ scatter 
      around the average value.
      Col. 1: Tidal field region where the velocity components were computed. 
      The first two sections refer to the tidal field computed from the $N$-body 
      -- FAM models. In the third section the external tidal influences are determined
      by the large region modeled on FAM$_{100}$ instead of the $N-$body samples.
      The fourth region refers to the residual field between $N-$body and FAM$_{mpl}$  after the external contributions have been accounted for in the FAM modeling: FAM$_{30}$ + tidal bulk + tidal shear (see text sec.~8.2 and sec.~8.5 for details).
      Col. 2: Cosmological model. 
      Col: 3: bulk flow, $v_{bulk}$.
      Col: 4: shear eigenvalue amplitude, $|s|$.
      Col: 5: $s_1$ eigenvalue (stretching).
      Col. 6: $s_2$ eigenvalue (middle).
      Col. 7: $s_3$ eigenvalue (compressional).}
      \label{table:multplbs}
    \end{table*}

\begin{figure*}[!t]
  \vskip 0.5truecm
  \centering
  \mbox{\hskip -0.5truecm\includegraphics[width=7.5in]{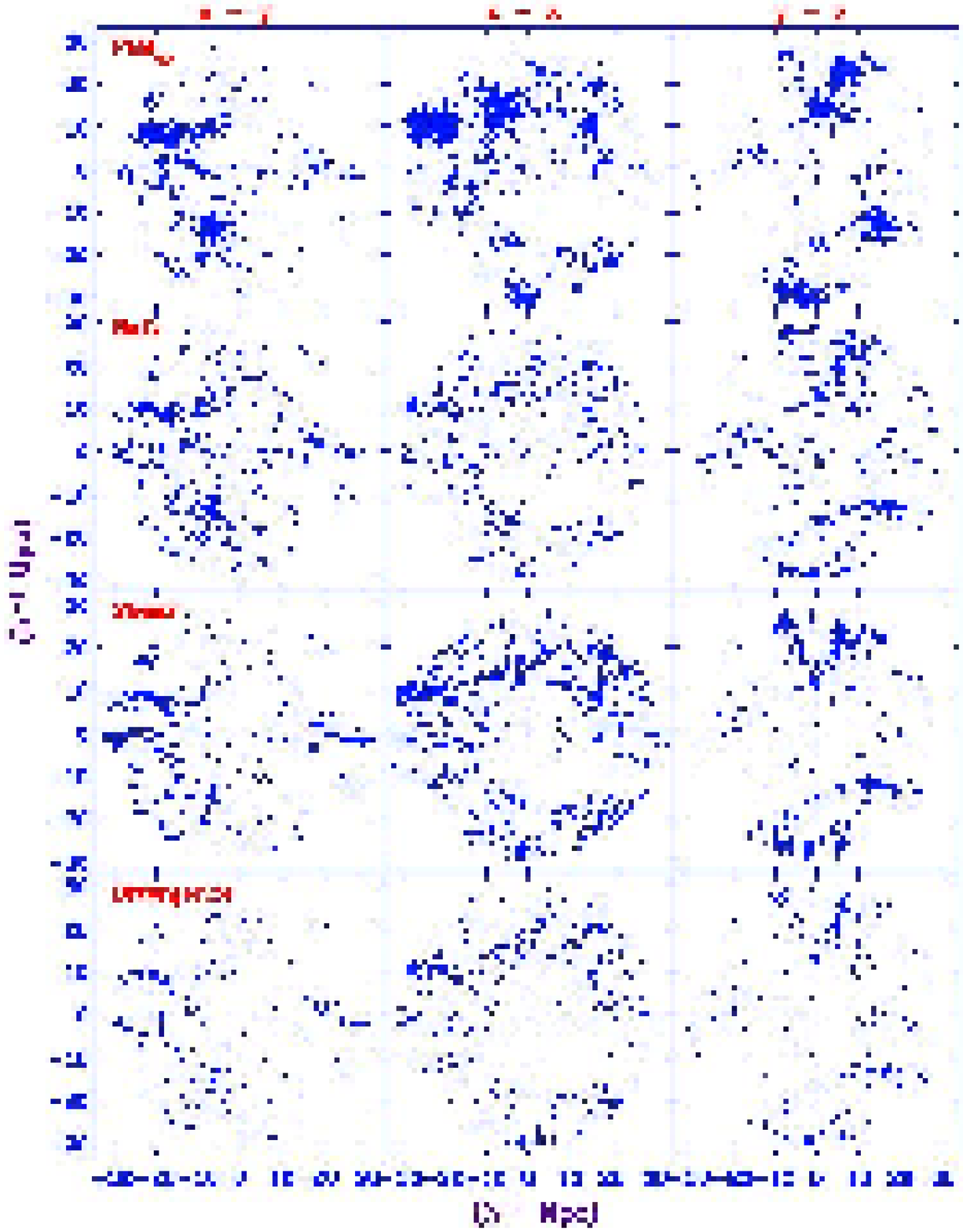}}
\end{figure*}
\begin{figure*}[!t]
  \caption{Velocity Field Multipole Decomposition: the total $N$-body 
    velocity field, involving the same one as in
    Fig.~\ref{fig:nbfam30b} and Fig.~\ref{fig:nbfam30bs}, decomposed
    into its four different components. The coordinate system is that
    defined by the tidal shear tensor, see Fig.~\ref{fig:nbfam30b}. At
    each row we depict the velocities in the $(x-y)$, $(x-z)$ and
    $(y-z)$ plane. Top: the locally induced velocity, approximated by
    the FAM$_{30}$ realization. Second row: the monopole component, a
    result of the slight local expansion due to its underdensity wrt.
    the global Universe. Third row: the bulk flow, of which most is
    concentrated in the $(x-y)$ and $(y-z)$ plane. Fourth row: the
    shear flow component.  }
  \label{fig:vmapmultp}
\end{figure*}

\noindent 
in which the data 4-vector $F_{I}$ and the velocity field 
component 4-vector $V_{Ii}$ are defined as 
\begin{eqnarray}
  F_{n,I} & = & \{1,x\hat{x}_{1},x\hat{x}_{2},x\hat{x}_{3}\} \nonumber \\
  V_{Ii} & = & \{{\tilde u}_{i},{\tilde s}_{i1},{\tilde s}_{i2},
  {\tilde s}_{i3} \}\ .
  \label{vectors} 
\end{eqnarray}
The ``dipolar'' bulk flow components ${\tilde u}_i$ and 
``quadrupolar'' velocity shear components ${\tilde s}_{ij}$ can then
be obtained by solving for the vectors $V_{Ii}$ on the basis of a
fitting analysis (to be precise, ${\tilde s}_{ij}$ also includes a
minor residual ``monopole'' expansion/contraction term). We accomplish
this by computing for each Cartesian component $i$ the values for the
multipole elements ${\tilde u}_i$ and ${\tilde s}_{ij}$ which minimize
$\chi^2$
\begin{equation}
  \chi^{2}=\sum_{n=1}^{N_{objs}}\Big(v_{n,ext,i}-\sum_{I}F_{n,I}({\bf x})
  V_{Ii}\Big)^{2}\ ,
  \label{chi2}
\end{equation}
to be evaluated on the basis of the data sample of $N_{obj}$ 
objects at locations ${\bf x}_n$ and with inferred ``external''
velocities ${\bf v}_{n,ext}$.  The bulk flow and velocity shear in the
externally induced velocity component ${\bf v}_{ext}$, along with a
residual expansion term,
\begin{equation}
  v_{ext,i}\,\approx\,v_{exp,i}+v_{bulk,i}\,+\,v_{shear,i}
  \label{eq:vextexpblkshr}
\end{equation}
will follow directly from the inferred values of the 4-vector
components $V_{Ii}$:
\begin{eqnarray}
  v_{exp,i}&\,=\,&{1 \over 3}\,Tr({\tilde s})\,x \hat{x}_i\nonumber\\
  v_{bulk,i}&\,=\,&{\tilde u}_i\,;\qquad\qquad v_{shear,i}\,=\,\sum_{j=1}^3\,
  {\tilde s'}_{ij} x \hat{x}_j
  \label{eq:vexpblksh}
\end{eqnarray}
in which $Tr(s)$ is the trace of the tensor $s_{ij}$ and 
${\tilde s'}_{ij}$ the traceless shear tensor
\begin{eqnarray}
  Tr({\tilde s})&\,\equiv\,&{\tilde s}_{11}+{\tilde s}_{22}+{\tilde s}_{33}
  \nonumber\\
  {\tilde s'}_{ij}&\,=\,&{\tilde s}_{ij}-{1 \over 3}Tr({\tilde s})\,
  \delta_{ij}\,.
\end{eqnarray}


\subsection{Velocity Multipole Analysis: results}
\label{sec:velmultres}

The results of our analysis are summarized in
Table~\ref{table:multplbs}. It lists the average quantities for the
tidal bulk flow and shear components for the two cosmological
scenarios discussed in this work. The table has been organized in four
(horizontal) sections. Each corresponds to another ``differential''
velocity field, the difference between two differently processed
velocity fields.

For both the $\Lambda$CDM and the $\tau$CDM model each of the quoted
values in Table~\ref{table:multplbs} involve the average and standard
deviation determined on the basis of ten different realizations. This
adds up to 8 configurations, two cosmologies per section. For each of
the 8 configurations, in the third column the table lists the dipole
component of the external velocity field, the bulk flow $v_{bulk}$.
Subsequently, the velocity shear is specified in terms of the three
eigenvalues $s_1$, $s_2$ and $s_3$ of the traceless shear tensor. This
is preceded in the fourth column by the amplitude $s$ of the shear.
Note that shear is quoted in two units. First, 
units of $\kmsmpc$, followed by the equivalent velocity differential
in $\kms$ over a volume of $30\hmpc$ radius. The intention of the
latter is to offer a directly appreciable comparison between the
relative importance of bulk flow and shear contributions.

Each of the four sections specifies the values of the computed
dipole and quadrupole moments of the velocity field of the
corresponding sample. The first section relates to a multipole
analysis of the differential velocity field between the full $N$-body
velocity field and the FAM$_{30}$ velocity reconstructions of the
inner $30\hmpc$ region, $N$-body - FAM$_{30}$.
The resulting residual velocity field has been generated by the mass
distribution beyond a radius of $30\hmpc$. On these linear scales the
inferred dipole and quadrupole components of the velocity field may be
directly related to the moments of the surrounding mass
distribution. 

The second section of Table~\ref{table:multplbs} does the same for the
larger $100\hmpc$ region. The outcome of similar analyses are
presented in the third and fourth section.  The third section repeats
the analysis of the first section, except that the external tidal
influences are determined on the basis of the difference between the
FAM velocity reconstructions within the large $100\hmpc$ region and
the inner $30\hmpc$ region. Earlier, in Section~\ref{sec:pscz}, we
have found that the major share of the origin of the external tidal
field is confined to this region and that it therefore may well be
determined from the residuals between FAM$_{100}$ and FAM$_{30}$. The
comparison between the inferred multipole moments of the velocity
differences between FAM$_{100}$-FAM$_{30}$ in the third section and
those in the first section are therefore expected to be rather
similar, any systematic differences originating in tidal effects
generated beyond a radius of $100\hmpc$. 
The fourth section in Table~\ref{table:multplbs} refers to
the values of the residual tidal velocity field between $N$-body --
FAM$_{mpl}$, the FAM sample after having accounted for the missing external tidal contributions, FAM$_{30}$ + tidal bulk + tidal shear (see sec.~\ref{sec:mulvelmodel}).
If indeed all significant contributions can be characterized by their dipolar and
quadrupolar contributions, the multipole values in this section are
expected to be negligible.


\subsection{Velocity Multipole Contributions: Maps} 

\begin{figure*}[!t]
  \centering \vskip 0.5truecm
  \mbox{\hskip -0.5truecm\includegraphics[width=7.5in]{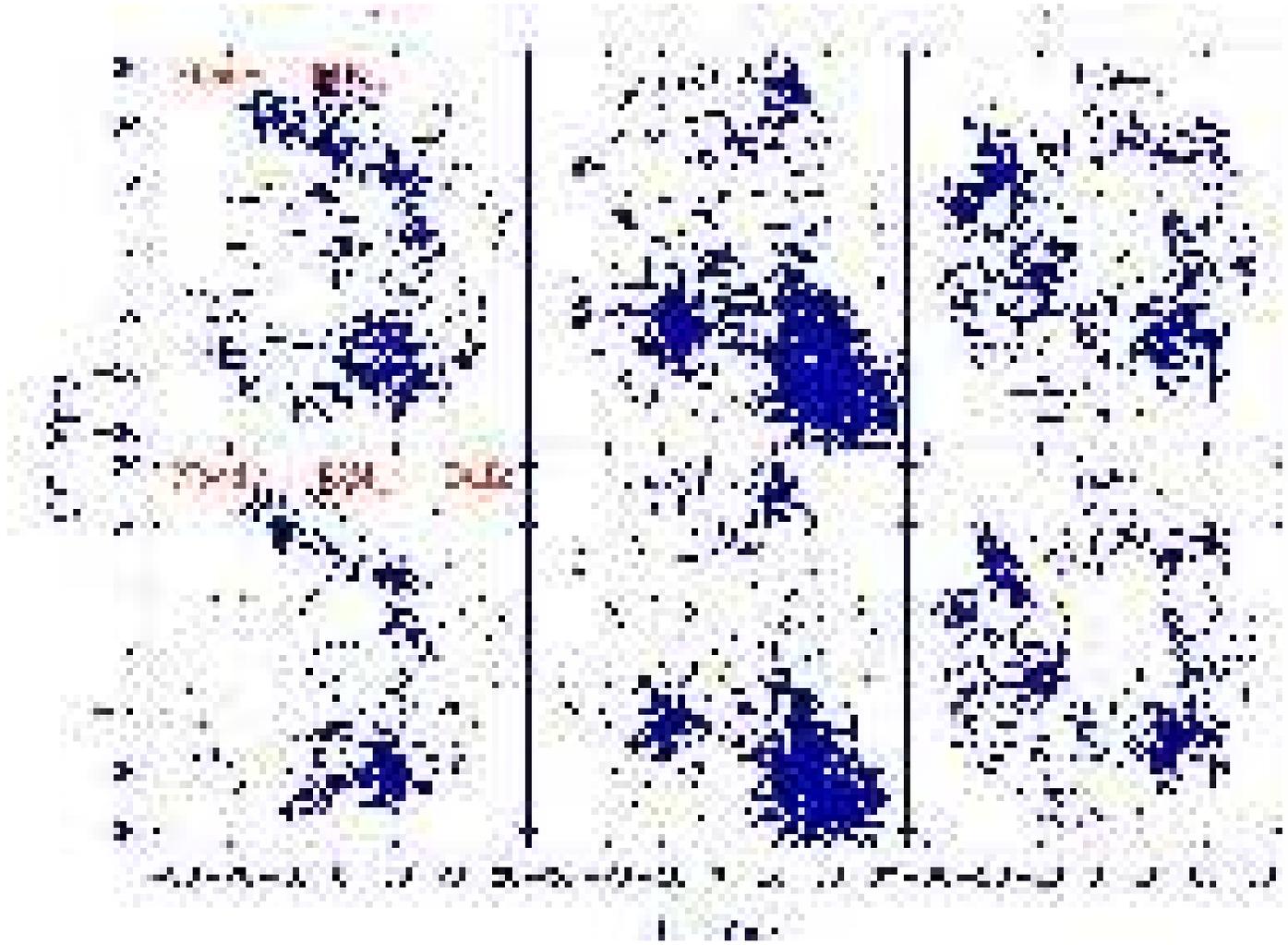}}
  \caption{2D projected peculiar (residual) velocities for the same mock 
    catalog as in Fig.~\ref{fig:map30nb} and Fig.~\ref{fig:map100nb}
    in three perpendicular central planes of $10\hmpc$ width. The
    coordinate frame is rotated such that the bulk flow velocity is
    oriented along the $x$-direction. Within the plane perpendicular
    to the $x$-axis, the $y$ and $z$ axes are chosen arbitrarily.
    First row: the residual velocity (ie. tidal velocity) $N-$body -
    FAM$_{30}$.  Second row: residual velocities after subtraction
    bulk flow component. The resulting residual field is clearly
    dominated by a shear pattern, most notably in the $y-z$ plane.}
  \label{fig:nbfam30b}
\end{figure*}

For a direct visual appreciation of the various multipole
contributions to the tidal velocity field we assess the ``tidal''
velocity field $N$-body - FAM$_{30}$, the velocity field generated by
the mass distribution beyond a radius of $30\hmpc$, for one of the
$\Lambda$CDM catalogs. The presented maps concern the same
$\Lambda$CDM catalog as those presented in the maps of
Figures~\ref{fig:map30nb}. The map of the projection of this ``tidal''
velocity flow onto three central planes is shown in the top row of
Figure~\ref{fig:nbfam30b}.

\begin{figure*}[!t]
  \centering
  \vskip 0.5truecm
  \mbox{\hskip -0.5truecm\includegraphics[width=7.5in]{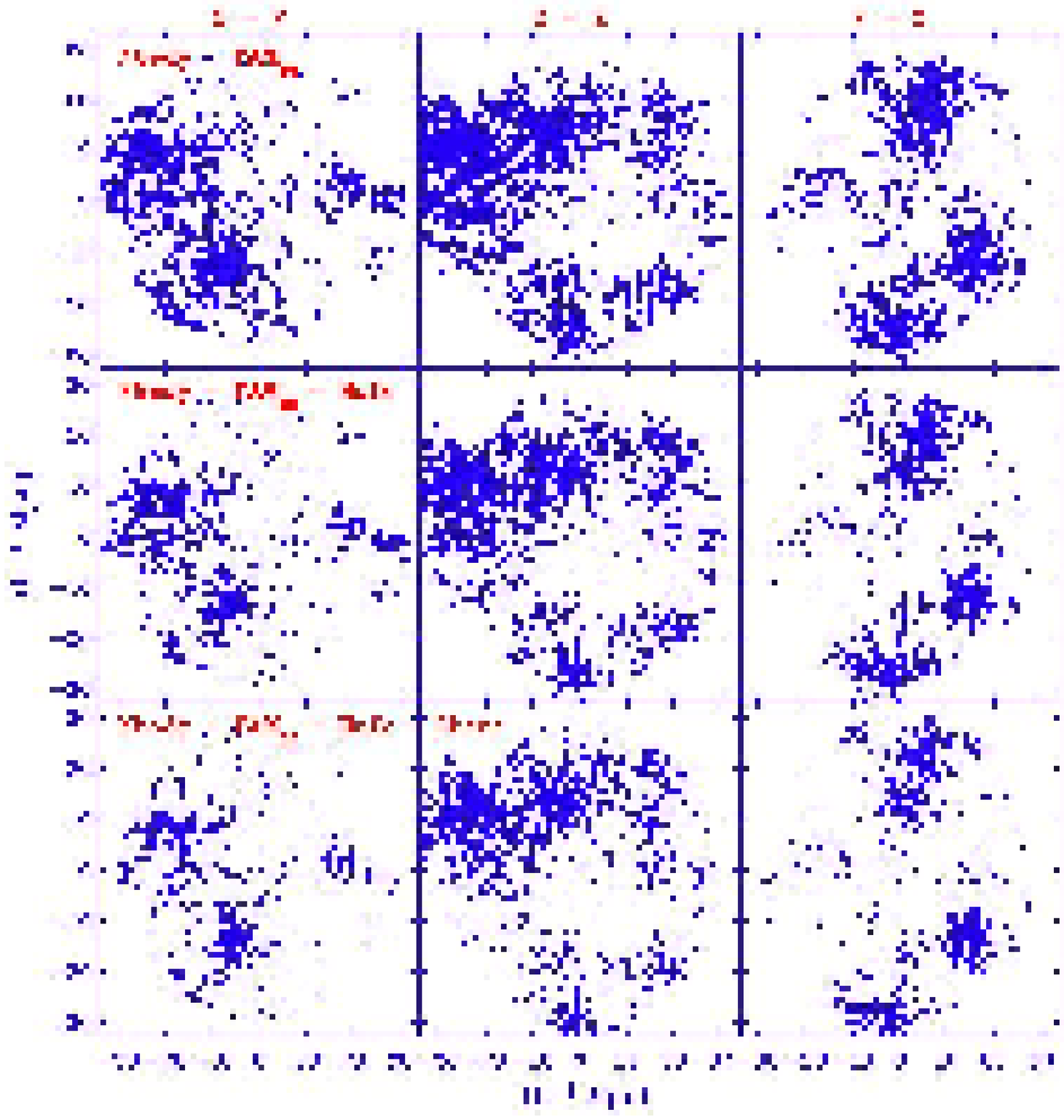}}
  \caption{2D projected peculiar velocities for the same mock catalog 
    as in Fig.~\ref{fig:map30nb} and Fig.~\ref{fig:map100nb} in three
    perpendicular central planes of $10\hmpc$ width. The coordinate
    frame is defined by the three eigenvectors of the shear tensor.
    The $x$-axis is aligned along the direction of the largest
    (stretching) eigenvalue, the $y$-axis is aligned along the middle
    eigenvalue and the $z$-axis along the smallest (compressional)
    eigenvalue. The first two rows are the same as
    Fig.~\ref{fig:nbfam30b}, within the ``shear reference system''.
    The third row depicts the final residual field after subtraction
    of the quadrupolar shear component.}
  \label{fig:nbfam30bs}
\end{figure*}

\subsubsection{Dipolar component: bulk flow}

The externally generated velocity flow is dominated by its bulk flow
component. This is in general true for both cosmologies.  The large
impact of the bulk flow over the local 30$\hmpc$ volume can be
immediately inferred from the values in the first section of
Table~\ref{table:multplbs}, revealing contributions in excess of
$200\kms$.

To facilitate visual appreciation of this observation we have have
reoriented the reference system in Figure~\ref{fig:nbfam30b} such that
the $x$-axis is oriented along the bulk flow. While the original
Cartesian system is an arbitrary one and thus lacks a physical
context, the {\it ``bulk flow reference system''} confines the
inferred bulk flow ${\tilde u}$ exclusively to the $x$-direction.  As
a result there are no bulk flow components in the corresponding $y$-
and $z$-direction (note that within the $y-z$ plane their direction is
arbitrarily defined). The pre-eminence of the bulk flow component can
be immediately seen in the $x-y$ and $x-z$ frames in the top row of
Figure~\ref{fig:nbfam30b}. Note that the same velocity maps, mostly so
the $y-z$ frame, reveal a clear shear pattern.

\subsubsection{Quadrupolar component: velocity shear}
\label{sec:shear_frame}

Seeking to assess the quadrupolar term in the external velocity field
we first remove the remaining expansion term from ${\tilde s}_{ij}$.
Diagonalization of the resulting traceless shear tensor ${\tilde
  s'}_{ij}$ yields the shear eigenvalues and eigenvectors. The
eigenvalues $s_1$, $s_2$ and $s_3$ are indicative for the strength of
the tidal force field induced by the surrounding matter distribution,
while the principal directions of this quadrupolar velocity
perturbation field are indicated by the corresponding eigenvectors
${\hat e}_{s,i}$.

\begin{figure*}[t]
  \vskip 0.5truecm \centering
  \includegraphics[width=6.0in]{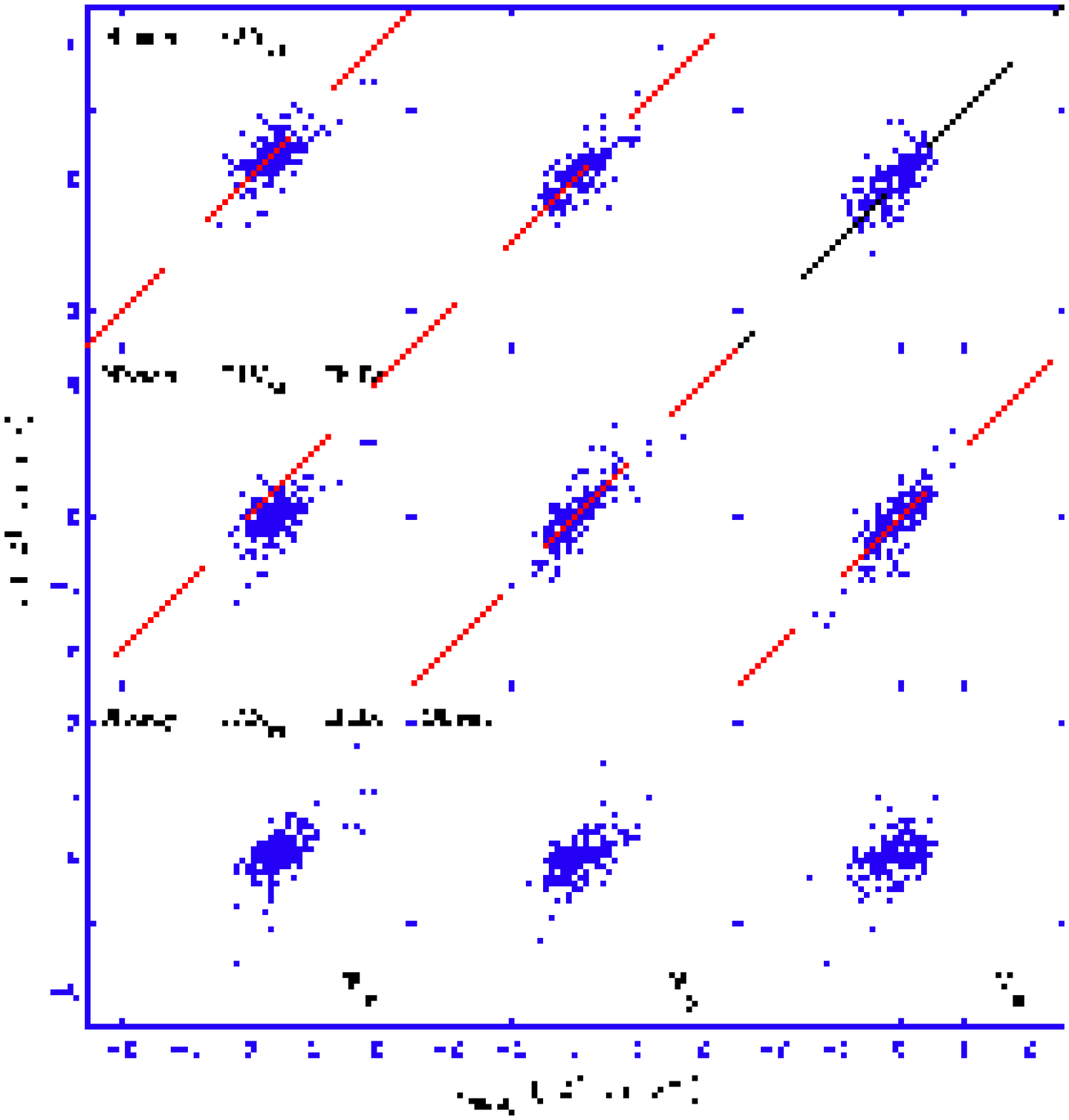}
   \caption{point-to-point comparison between the three successive 
    residual velocities and the corresponding mock catalog $N$-body 
    velocity. The three panels in each row correspond to the $x$, $y$ and 
    $z$ velocity components. The coordinate system is the ``shear reference 
    system''. Top row: the tidal velocity field $N$-body-$FAM$. Middle row: 
    residual velocity field after subtraction bulk flow. Bottom row: 
    residual velocity field after subtraction of both tidal bulk and shear
    components.}
  \label{fig:vvcompc}
\end{figure*} 

The {\it ``shear ellipsoid''}, the quadratic surface defined by the
shear tensor ${\tilde s'}_{ij}$ with principal axes aligned along the
eigenvectors and with axis size set by the corresponding eigenvalue
$s_i$, defines a natural reference system to assess the tidal shear
flow field. The coordinate axes of this {\it ``shear reference
frame''} are identified with the orthonormal basis defined by the
(normalized) eigenvectors. The $x$-axis is chosen to be aligned along
the major axis of the {\it ``shear ellipsoid''}, the direction defined
by the largest (positive) eigenvalue $s_1$ and directed along the
strongest dilational (stretching) motion incited by the external tidal
field.  Likewise the $z$-axis is chosen to coincide with the lowest
(negative) eigenvalue $s_3$, aligned along the strongest
``compressional'' component of the tidal velocity flow. This leaves
the $y$-axis as the one coinciding with the intermediate eigenvalue
$s_2$.

The imprint of the shearing motions can be discerned within the $y-z$
plane and, most prominently, along the ``x-z'' projection of the {\it
``bulk flow reference system''}. After subtraction of the bulk flow
component, i.e. $N$-body-FAM$_{30}$-$v_{bulk}$, the quadrupolar
component of the externally induced velocity flow represents its
principal constituent (Fig.~\ref{fig:nbfam30b}, lower row). This is
confirmed by the values quoted in Table~\ref{table:multplbs} for the
shear contribution.  In particular when stated in the velocity
equivalent unit of $\kms$ these shear values suggest that the
quadrupolar shear contributions are of a comparable magnitude to those
of the bulk flow. The maps in the lower row of
Figure~\ref{fig:nbfam30b} suggest that there are strong dilational and
compressional motions within the $y-z$ plane.  By contrast, the shear
motions in the $x$-direction appear to be uncommonly weak.  Given the
{\it ``bulk flow reference system''}, it implies that for this
particular realization we see a bulk flow directed almost
perpendicular to the shear flow motions.

\begin{table*} 
  \centering 
  \begin{tabular}{lccccc}\hline\hline
    \\
    Cosmology & $R_{Spear}$ &  $R_{lin}$  &  $a_0$ &$a_{lrg}$ & $\sigma_{lrg}$\\
    &            &             &(km s$^{-1}$)&  &(km s$^{-1}$)\\
    \\
    \hline
    \\
    $\Lambda$CDM& 0.62\,\,$\pm$\,\,0.17&0.58\,\,$\pm$\,\,0.22&
    -4.03\,\,$\pm$\,\,75.45&0.83\,\,$\pm$\,\,0.20&246\,\,$\pm$\,\,134\\
    $\tau$CDM & 0.75\,\,$\pm$\,\,0.11&0.71\,\,$\pm$\,\,0.14&
    -9.31\,\,$\pm$\,\,30.19&0.93\,\,$\pm$\,\,0.13&219\,\,$\pm$\,\,\,86\\
    \\
    \hline\hline
  \end{tabular}
  \caption{Average final results for the $N-$body velocities $vs.$ the 
    corrected FAM$_{30}$ after adding to it the tidal Bulk and Shear 
    contributions. 
    The errors represent the $1\sigma$ scatter around the average value.
    Col. 1: Cosmological model. 
    Col. 2: non parametric (Spearman) correlation coefficient.
    Col. 3: linear correlation index.
    Col. 4: zero point of the best linear fit.
    Col. 5: slope of the best fitting line.
    Col. 6: dispersion around the fit.}
  \label{table:nbfam30bs}
\end{table*}

Figure~\ref{fig:nbfam30bs} depicts the same $\Lambda$CDM mock sample
as presented in Figs.~\ref{fig:map30nb}, \ref{fig:map100nb} \&
\ref{fig:nbfam30b}, here in the {\it ``shear reference frame''}. The
top row shows the full externally induced flow field, $N$-body -
FAM$_{30}$, in this reference system. The tidal shear flows are almost
exclusively confined to the $x-z$ plane. This is most evidently
illustrated in the central row of frames showing the velocity field
without its bulk flow component: hardly any systematic flow is
noticeable in the $y$-direction of the intermediate shear eigenvalue.


\subsection{Multipole Scale Dependence}

When turning to the external influences over a large $100\hmpc$
region, we may conclude from the second section of the table that most
of the external contributions are accounted for, both bulk flow and
shear are at least a factor of 3-4 smaller than for the inner
$30\hmpc$ region.
The third and fourth section show that the explicit contributions 
from the regions between $100\hmpc$ and $30\hmpc$ and those beyond
$100\hmpc$ are indeed significantly different, those beyond $100\hmpc$
tending towards zero contributions and as far as the shear is
concerned almost an order of magnitude smaller than the equivalent
contributions by the $30-100\hmpc$ region.

A similar graphical assessment involving the FAM$_{100}$
reconstructions emphasizes the minor significance of tidal
contributions stemming from density fluctuations beyond a radius of
$100\hmpc$. No coherent velocity pattern can be recognized in the
residual velocity field between full $N$-body and FAM$_{100}$
reconstruction. The comparison between this residual velocity field
with the velocity maps including the contributions of the inferred
bulk flow and shear flow do hardly show any difference.  In all cases
the velocity fields are dominated by the same thermal motions.


\subsection{Multipole Velocity Flow Model}
\label{sec:mulvelmodel}

Following our argument that the externally induced velocity flow
within the inner $30\hmpc$ mainly consists of a bulk flow and shear
contribution, we may expect that the effect of the external gravity
field can be sufficiently accounted for by adding these components to
a local velocity field model based on the mass distribution in and
around the Local Superclusters.

By separating the ``internal'' FAM velocity field from the
``external'' multipole contributions of the (monopole,) dipole and
quadrupole components of the {\it ``tidal''} velocity field and adding
the two, we obtain a total {\it ``FAM-multipole''} model velocity
${v_{fam-mpl,i}}$,
\begin{equation}
  v_{fam-mpl,i}\,\approx\,v_{FAM,i}\,+\,v_{exp,i}+v_{bulk,i}\,+\,v_{shear,i}
  \label{eq:vfammpl}
\end{equation}
A visual impression of the extent of the successive multipole
contributions may be obtained from Fig.~\ref{fig:vmapmultp}. The
vector plots of the four velocity contributions to $v_{fam-mpl}$
(Eq.~\ref{eq:vfammpl}) are depicted in four successive rows, each
within the mutually perpendicular three central slices (wrt. the shear
reference system). The top row concerns the FAM$_{30}$ velocity field
reconstruction, followed successively by the expansion/contraction
term (monopole), the bulk flow (dipole) and velocity shear
(quadrupolar).

From fig.~\ref{fig:nbfam30b} and fig.~\ref{fig:nbfam30bs} we conclude
that the differences between the ``full'' $N$-body velocities and
$v_{fam-mpl}$, the total sum of the internal FAM$_{30}$ and external
dipole and quadrupole contributions, do not appear to show systematic
trends as it can be noticed from the residual bulk and shear components in section 4 of Table~\ref{table:multplbs}.  Wherever there are large deviations, these are mainly
confined to the high density virialized regions.


\subsection{Point-to-Point Comparison}

A quantitative quality assessment of the {\it ``FAM-multipole''} model
is offered by the point-to-point comparison between the full $N$-body
velocity and its difference with respect to the successive modes of
the {\it ``FAM-multipole''} velocity in Fig.~\ref{fig:vvcompc}. The
$v_x$, $v_y$ and $v_z$ of the various velocity components refer to the
{\it ``bulk flow reference system''}. The top row, plotting
$v_{Nbody}$ vs. the residual $N$-body-FAM$_{30}$, reveals the expected
systematic differences due to missing externally induced
contributions. Given the fact that the bulk flow in this reference
system is confined to the $x$-component, we may note the uniform
systematic shift of the $x$ residuals with respect to the zeropoint
$(v_{Nbody},v_{res})=(0,0)$ (top lefthand frame). The subsequent
addition of the dipolar bulk flow contribution to FAM$_{30}$ leads to
a systematic downward uniform vertical shift of
$N$-body-FAM$_{30}$-Bulk (middle row Fig.~\ref{fig:vvcompc}): also the
residuals in the $x$-direction now center on $v_{res}=0$ (note that by
virtue of the bulk flow the $N$-body velocities in the $x$-direction
are also skewed to values larger than $v_{Nbody}=0$).

The three point-to-point diagrams in the middle row of
Fig.~\ref{fig:vvcompc} show that even while the bulk flow is taken
into account systematic motions remain in all three directions. The
point-to-point comparisons still follow a strong correlation with
respect to the the $N$-body ~velocities. It mainly involves the
presence of the quadrupolar velocity shear component (in addition to a
minor ingredient contributed by the monopole expansion/contraction
term). This can be immediately inferred from the comparison between
the diagrams in the central and lower row of Fig.~\ref{fig:vvcompc}:
once the quadrupole component ``Shear'' has been added to the
``FAM$_{30}$+Bulk'' velocities the systematic effects seem to have
largely vanished. What remains in the residuals is mainly random
scatter, centered on the $v_{res}=0 \kms$ line, with some exceptional
outliers originating in the virialized regions.

We have quantified the point-to-point comparisons by performing linear
regressions similar to those presented in Section~\ref{sec:pntpntnbg}.
Table~\ref{table:nbfam30bs} summarizes the results of this comparison
for all catalog samples for both cosmologies.  In both cosmological
models the slope of the best fitting line is consistent with unity at
$\sim 1\sigma$ confidence level. As expected, the scatter around the
fit is similar to that of all previous analyses (see
Table~\ref{table:vscatlrg}). Offsets around the zero-point are
consistent with zero, although with a large dispersion. The strength
of the point-to-point correlations has increased considerably with
respect to their FAM$_{30}$ counterpart (Table~\ref{table:vscatlrg})
and it is very similar to the FAM$_{100}$ case.

\begin{figure*}[t]
  \centering
  \vskip -1.0truecm
  \includegraphics[width=6.5in]{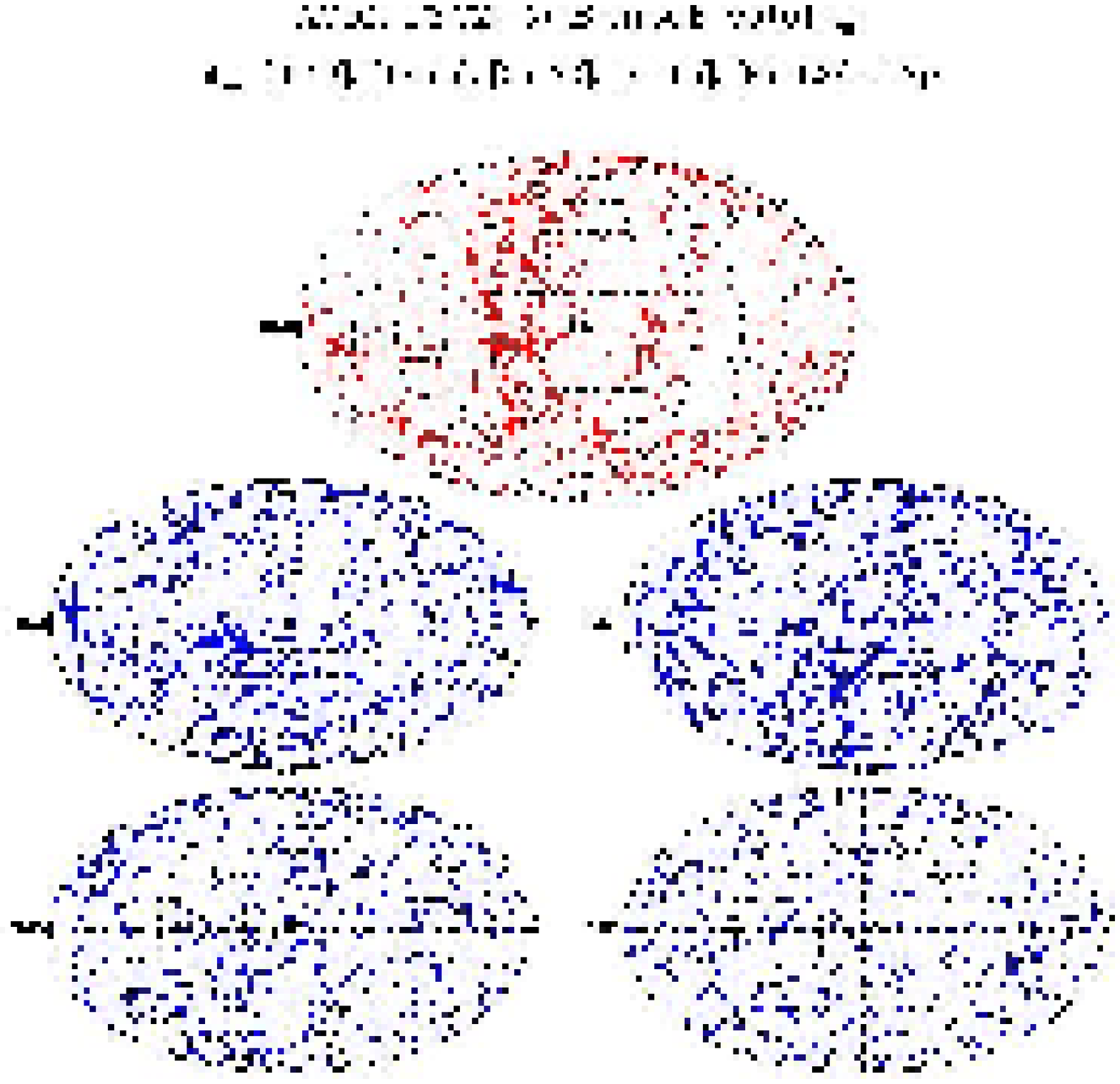}
  \vskip -1.0truecm
  \caption{Aitoff projections of the galaxy distribution of a 
    NBG + PSC$z$ ~$\Lambda$CDM catalog. The top-most panel shows the
    corresponding NBG distribution, $d_{sur}=[0,30]\hmpc$. The
    subsequent panels depict the external galaxy distribution enclosed
    by the shells defined by survey depth $d_{sur}=[30-55]\hmpc$,
    $d_{sur}=[55-70]\hmpc$ $d_{sur}=[70-85]\hmpc$ and
    $d_{sur}=[85-100]\hmpc$.}
  \label{fig:aitofflcdm}
\end{figure*} 

\begin{figure*}[t]
  \centering
  \vskip -1.0truecm
  \includegraphics[width=6.5in]{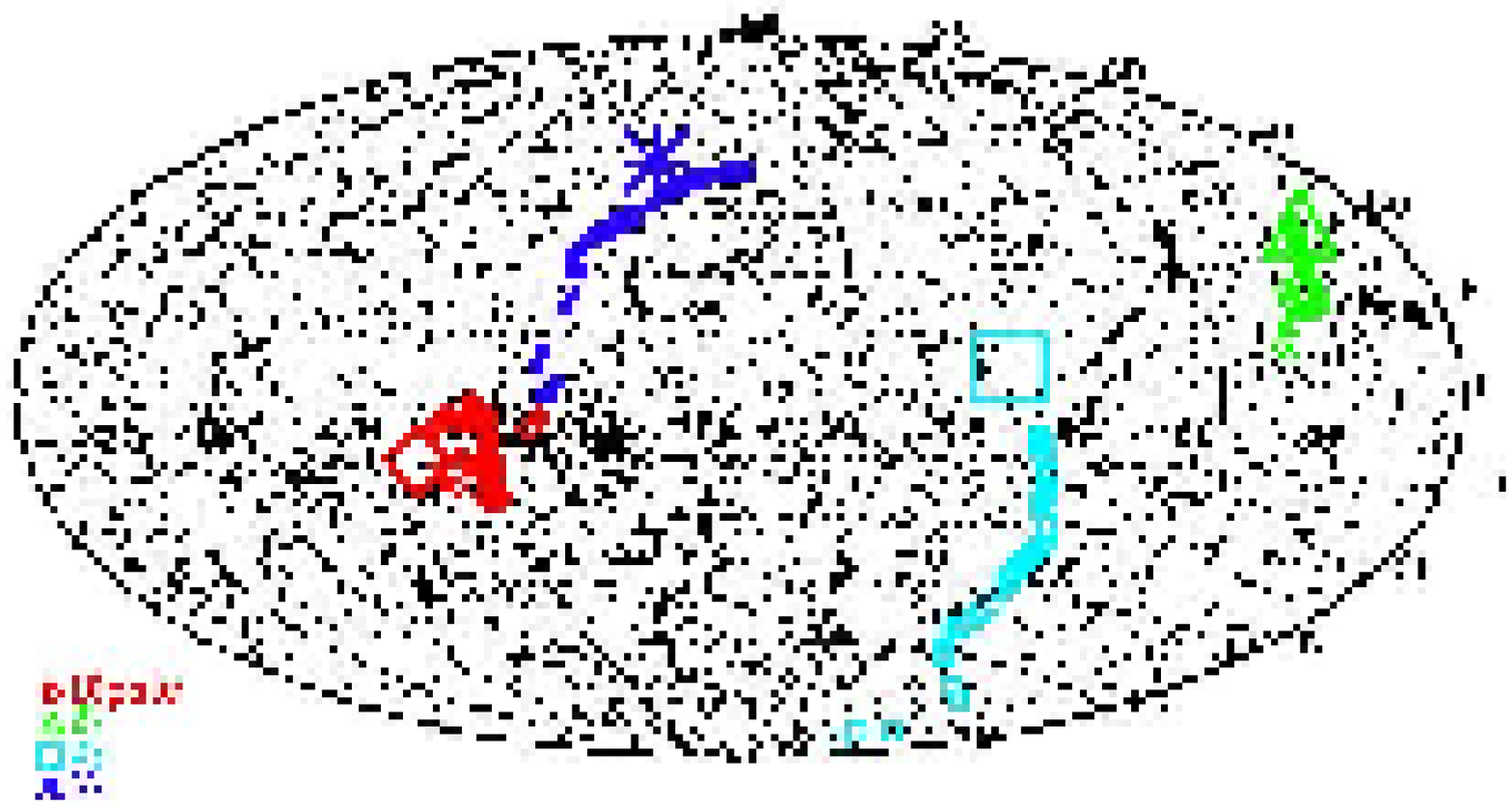}
  \vskip -1.0truecm
  \caption{Sky distribution of galaxies in external shell of survey depth 
    $d_{sur}=30-100\hmpc$. This galaxy distribution should reflect the
    mass distribution inducing the local tidal velocity flow. By means
    of symbols we have indicated the track of the gravity dipole and
    quadrupole eigenvector directions (cf. eqn.~\ref{eq:gbulkdiscr}
    and eqn.~\ref{eq:tidshrdiscr}) on the sky, by radially expanding
    outward the survey depth $d_{sur}$ in steps of $1\hmpc$, from
    $d_{sur}=30\hmpc$ to $d_{sur}=100\hmpc$. For comparison the same
    symbol, but then enlarged, indicates the directions for the
    corresponding bulk velocity and velocity shear eigenvector
    directions.  Diamond: dipole. Triangle: stretching component
    shear. Square: middle component shear. Star: compressional
    component shear.}
  \label{fig:gravbslcdm}
\end{figure*} 


\subsection{Surrounding Matter Distribution: Tidal Source}

The surrounding external matter distribution is the source for the
tidal velocity field which we inferred in the previous sections. For
various purposes we wish to relate the computed dipolar bulk flow and
quadrupolar shear flow components to the surrounding matter
distribution which induced them.

The induced tidal velocities involve spatial scales ranging from
$30\hmpc$ to $100\hmpc$. Over this range the linear theory of
gravitational instability holds to good approximation. This translates
into a direct linear relationship between induced velocity ${\bf
  v}_{ext}$ and the cumulative external gravitational force ${\bf
  g}_{ext}$,
\begin{equation}
  {\bf v}_{ext}({\bf x},t)\,=\,{\displaystyle 2 f(\Omega,\Lambda) \over 
    \displaystyle 3 H \Omega}\, {\bf g}_{ext}({\bf x},t)\,,
  \label{eqn:linvg}
\end{equation}
with $f(\Omega,\Lambda)$ the linear velocity growth factor.  This
linear relationship also holds for every component of the velocity and
gravity fields, and thus also for the individual dipolar and
quadrupolar components of the externally induced velocity field.  They
are directly proportional to equivalent dipole and quadrupolar
components of the gravity field:
\begin{eqnarray}
  {\bf v}_{bulk}({\bf x},t)&\,=\,&\,\,\,\,{\displaystyle 2 f(\Omega,\Lambda) 
    \over \displaystyle 3 H  \Omega}\,\,{\bf g}_{bulk}({\bf x},t)\nonumber\\
  \nonumber\\
  s_{ij}({\bf x},t)&\,=\,&-{\displaystyle 2 f(\Omega,\Lambda) \over 
    \displaystyle 3 H \Omega}\,\,{\cal T}_{ij}({\bf x},t)
  \label{eqn:lindyn}
\end{eqnarray}
with the (external) gravitational tidal shear tensor ${\cal T}_{ij}$
is defined as (see van de Weygaert \& Bertschinger~\cite{rienbert}),
\begin{equation}
  {\cal T}_{ij}\,=\,{\displaystyle \partial^2 \Phi_{tidal} \over \displaystyle 
    \partial x_i \partial x_j}\,.
  \label{eqn:gravshr}
\end{equation}
Notice that because of its external nature, the term ${\displaystyle 1
  \over \displaystyle 3} \nabla^2 \Phi_{tidal} \,\delta_{ij}$ is
always equal to zero.

\begin{figure*}[!t]
  \centering
  \vskip -1.0truecm
  \hskip -0.5truecm
  \includegraphics[width=6.5in]{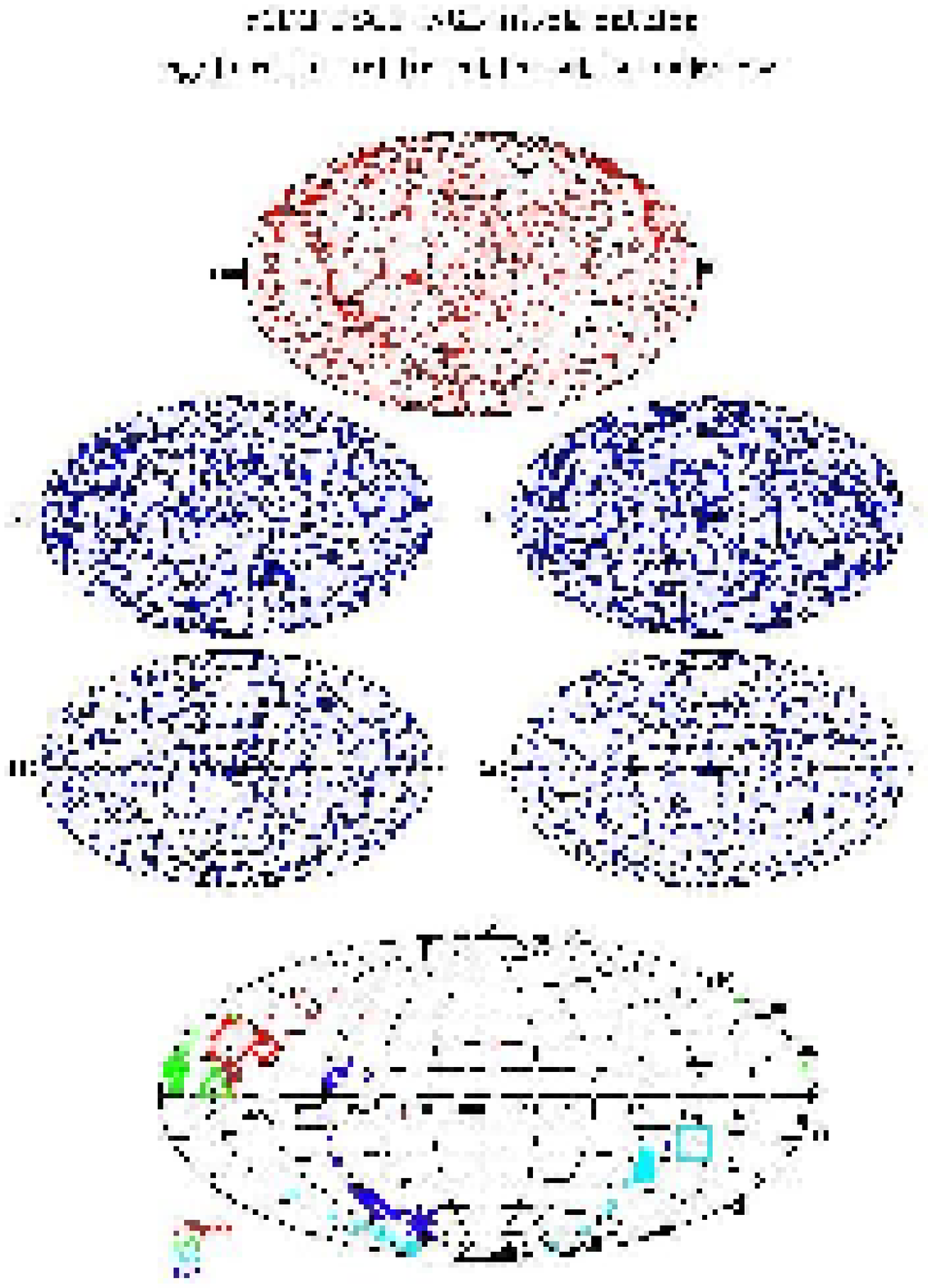}
  \vskip -2.0truecm
  \caption{Aitoff projections of the galaxy distribution of a 
    NBG + PSC$z$ ~$\tau$CDM catalog. The top-most panel shows the
    corresponding NBG distribution, $d_{sur}=[0,30]\hmpc$. The
    subsequent 4 panels depict the external galaxy distribution
    enclosed by the shells defined by survey depth
    $d_{sur}=[30-55]\hmpc$, $d_{sur}=[55-70]\hmpc$
    $d_{sur}=[70-85]\hmpc$ and $d_{sur}=[85-100]\hmpc$.  Bottom panel:
    convergence gravity dipole and quadrupole. By means of symbols we
    have indicated the track of the gravity dipole and quadrupole
    eigenvector directions (cf. eqn.~\ref{eq:gbulkdiscr} and
    eqn.~\ref{eq:tidshrdiscr}) on the sky, by radially expanding
    outward the survey depth $d_{sur}$ in steps of $1\hmpc$, from
    $d_{sur}=30\hmpc$ to $d_{sur}=100\hmpc$. For comparison the same
    symbol, but then enlarged, indicates the directions for the
    corresponding bulk velocity and velocity shear eigenvector
    directions.  Diamond: dipole. Triangle: stretching component
    shear. Square: middle component shear. Star: compressional
    component shear.}
  \label{fig:aitofftcdm}
\end{figure*}

Ideally, we would like to infer the external tidal potential
$\Phi_{tidal}$ directly from the galaxy distribution in a sufficiently
large surrounding region. This is specifically true for its dipolar
and quadrupolar moments, with the intention to insert these terms
directly into the expression for the FAM potential
(eqn.~\ref{phiextsmth} and eqn.~\ref{eqn:sfamtot}). The required
externally induced bulk flow velocity and velocity shear should be the
result. The comparison of the FAM computed velocities for the local
volume, in combination with the computed tidal velocities
(Sec.~\ref{sec:velmultres}), and the observed and measured velocities
would then enable us to determine the amount of mass and average
density in the local volume.

To determine the gravitational influence of the surrounding matter
distribution, we set out to assess the sky distribution of galaxies in
the local NBG volume, out to $r_{NBG}=30\hmpc$, along with the
external mass distribution in radial shells out to a distance
$r_{PSCz} \le 100\hmpc$.  A prominent dipolar matter configuration in
the sky distribution will translate into a strong bulk gravity force.
Similarly, quadrupolar anisotropies will translate into an effective
tidal shear force.  In figure~\ref{fig:aitofflcdm} we have plotted the
galaxies in one of our $\Lambda$CDM mock catalogs in five successive
distance shells. Aitoff projections of the angular positions of the
galaxies, as seen from the centre of the local NBG volume, provide an
impression of the level of anisotropy in the mass distribution at
successive radii.

\begin{figure*}[t]
  \centering
  \mbox{\hskip -0.75truecm\includegraphics[width=7.3in]
    {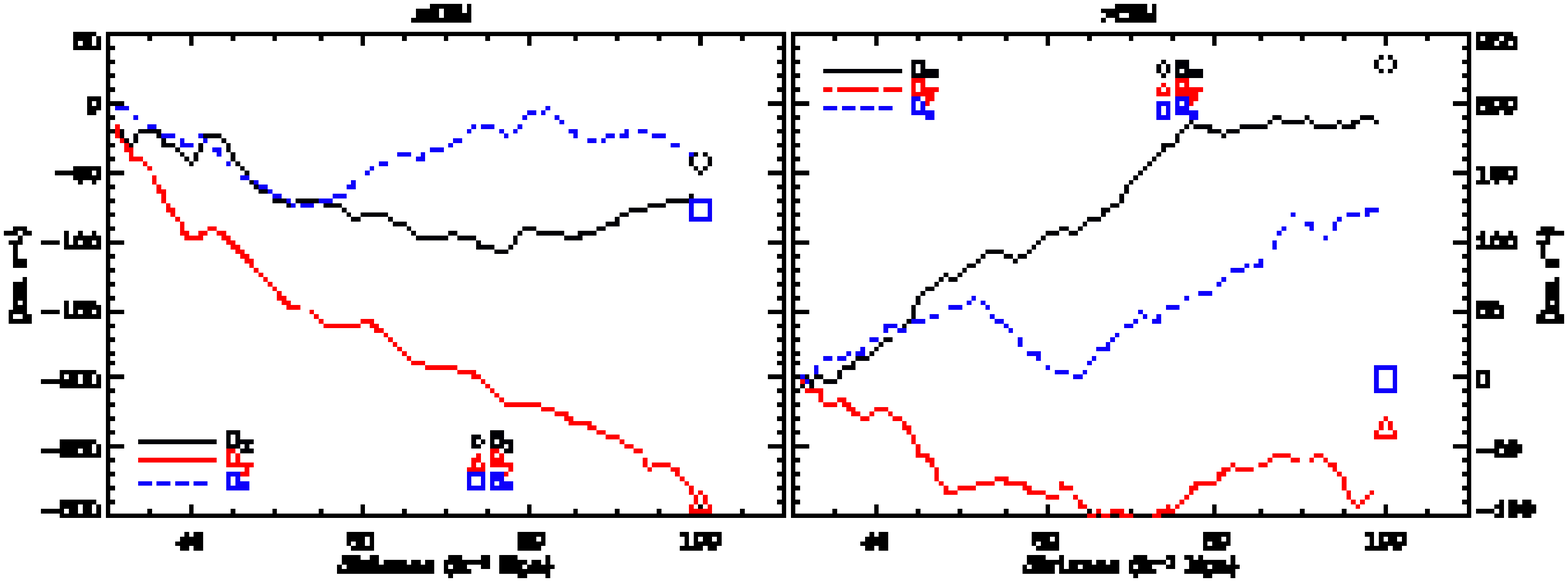}}
  \caption{Cumulative gravity dipole and tidal gravity shear. 
    As a function of survey depth $d_{sur}$ the Cartesian components
    of the gravity dipole, $g_x$, $g_y$ and $g_z$ is followed (in
    equivalent velocity unit $\kms$). The symbols at $d_{sur}$
    indicate the corresponding velocity dipole.  Lefthand panel:
    $\Lambda$CDM. Righthand panel: $\tau$CDM. }
  \label{fig:cumgrav}
\end{figure*} 

The first sky plot (top sphere) depicts the sky position of the
galaxies in the local NBG-mimicking mock sample. It involves a highly
flattened distribution, perhaps reminiscent of the Supergalactic
Plane. The four subsequent shells correspond to successive cuts
through the PSCz mimicking samples, at sampling depths
$d_{sur}=[0-30],\,[30-55],\,[55-70]$ and $[85-100]\,\hmpc$.  The first
and direct observation is the diminishing sample density as a function
of survey depth, in accordance with the selection function
(eqn.~\ref{eq:sel}).  Structure is most prominent in the first shell,
at $d_{sur}=[30-55]\,\hmpc$ (central left sphere). The structure
contained in this shell also shows a clear affiliation with the matter
distribution in the local NBG volume. The compact massive
concentration at $l\approx 220^{\circ}$ is clearly connected to a
dense region in the local ``plane''. A superficial inspection of the
angular galaxy distribution reveals the presence of strong dipolar and
quadrupolar components, effecting considerable tidal forces. Note that
both external shells display a rather strong concentration of galaxies
in their southern hemisphere, in the vicinity of $l \approx
180-200^{\circ}$. Similar but weaker contributions can also be
recognized from the galaxy distribution in the shell between
$d_{sur}\approx 70-85\hmpc$. Beyond $d_{sur}>85\hmpc$, however, the
angular pattern appear to be considerably less pronounced. This is in
line with the earlier findings that there were hardly noticeable tidal
contributions from large distances.

To see to what extent the depicted galaxy distribution can indeed be
held responsible for most of the inferred tidal bulk flow and tidal
shear, we have determined the corresponding bulk force ${\bf
  g}_{bulk}$ (eqn.~\ref{eq:gbulk}) and tidal shear ${\cal T}_{ij}$
(eqn.~\ref{eq:tideij}) evoked by the external galaxy distribution
($r>30\hmpc$). Since we do not have a continuous density field but the
positions of a finite number of objects in our galaxy flux-limited and
full mass distribution catalogs, the bulk acceleration on the LG is
computed from the discrete equivalent. For a sample of galaxies at
locations ${\bf x}_i$, with an average number density $n$ of selected
objects, this leads to
\begin{equation}
  {\bf g}_{bulk}\,=\,{\displaystyle H\, f(\Omega,\Lambda) \over \displaystyle 
    4 \pi n}\,\sum_i\,{1 \over \displaystyle \psi(x_k)}\,{\displaystyle 
    {\bf x}_k \over \displaystyle |{\bf x}_k|^3}\,.
  \label{eq:gbulkdiscr}
\end{equation}
where $\psi(x_k)$ is the sample selection function at distance $x_k$,
whose inverse functions as weighting factor. For practical reasons,
comparison with the inferred bulk flow ${\bf v}_{bulk}$, we have
translated the bulk acceleration into equivalent velocity units by
means of the transformation $H f(\Omega,\Lambda)/{3 \over 2}\Omega
H^2$. The equivalent ``discrete'' expression for the external tidal
shear is
\begin{equation}
  {\cal T}_{ij}\,=\,{\displaystyle H\, f(\Omega,\Lambda) \over \displaystyle 
    4 \pi n}\,\sum_i\,{\displaystyle 1 \over \displaystyle \psi(x_k)} 
  {\displaystyle 3\,x_{ki}\,x_{kj} \over \displaystyle |{\bf x}_k|^5}\,.
  \label{eq:tidshrdiscr}
\end{equation}

For the $\Lambda$CDM mock galaxy sample depicted in
Fig~\ref{fig:aitofflcdm} we determine the gravity dipole by computing
for a set of spherical external shells the resulting bulk flow
acceleration (eq.~\ref{eq:gbulkdiscr}) and the gravity quadrupole by
computing external tidal shear (eq.~\ref{eq:tidshrdiscr}). Recently, a
similar approach was followed by Teodoro et al.~\cite{teod2004}. The
spherical shell volumes are defined by an inner radius
$r_{inn}=30\hmpc$ and an outer radius $r_{out}$. The width of the
shell is gradually enlarged by increasing $r_{out}$ from
$r_{out}=30\hmpc$ to $r_{out}=100\hmpc$. The convergence of the
resulting gravity dipole direction on the sky can be observed in
Fig.~\ref{fig:gravbslcdm}. The small red diamonds are consistently
located near $l\approx 230^{\circ}$, and converge at a sky location
close to the direction of the velocity dipole (large red diamond). To
get an idea of the amplitudes involved, Fig.~\ref{fig:cumgrav} (top
panel) shows the development of the cumulative gravity dipole as a
function of external distance $d_{sur}$ ($30\hmpc<d_{sur}<100\hmpc$).
By means of symbols the corresponding velocity dipole, for each of the
three directions $x$, $y$ and $z$, are inserted at the outer radius of
$d_{sur}\approx 100 \hmpc$. Note that we have restricted ourselves to
a case study example. A more extensive and proper assessment, including
a proper error estimate of both gravity dipole and quadrupole as well
as the bulk and shear flow, is beyond the scope of the present
argument. This issue, involving the shot noise influence on gravity
dipole and quadrupole and the role of FAM uncertainties on the
velocity flow components, will be treated in more detail in a
forthcoming study.

We see that in the $x$-, $y$ and $z$-directions of the gravity and
velocity dipoles are in reasonable agreement, within a margin of
$\approx 30 \kms$. This observation justifies our expectation that the
dipole can be estimated to sufficient accuracy from the surrounding
external galaxy distribution. The dipole may then be estimated from
the surrounding external galaxy distribution, so that the latter can
be invoked to correct for the influence of the external tidal field in
the dynamics of the local volume.

The situation is comparable for the cumulative tidal shear, in terms
of its three eigenvalues and eigenvectors. Also the gravity quadrupole
appears to converge relatively smoothly towards the velocity shear.
This may be inferred from the plotted directions of the eigenvectors
${\hat {\bf e}}_{{\cal T}1}$, ${\hat {\bf e}}_{{\cal T}2}$ and ${\hat
  {\bf e}}_{{\cal T}3}$ of the tidal shear ${\cal T}_{ij}$. They are
indicated by means of three symbols, the triangle corresponding to the
stretching component ${\cal T}_1$, the star the middle component
${\cal T}_2$ and the square the compressional component ${\cal T}_3$.
The tidal shear tensor wanders extensively across the ``sky'' as we
push the outer radius of the external shell outward, as is shown by
the paths of the corresponding eigenvectors. Interestingly, once the
shell radius starts to approach $100\hmpc$, each of the eigenvectors
appear to converge near the location of the corresponding stretching,
central and compressing velocity shear tensor eigenvectors.  However,
also here we notice significant deviations in individual cases.

For comparison, we can appreciate the role of the external tidal field
on local dynamics for the case of the $\tau$CDM cosmology.
Figure~\ref{fig:aitofftcdm} combines the galaxy sky distribution for a
$\tau$CDM mock galaxy sample, in the same radial shells as in
Fig.~\ref{fig:aitofflcdm}. The final frame shows the Aitoff projection
of the gravity dipole and gravity quadrupole eigenvectors for a set of
gradually increasing radial shells. From the galaxy sky distribution
in the four slices at sampling depths
$d_{sur}=[0-30],\,[30-55],\,[55-70]$ and $[85-100]\,\hmpc$ we notice
that these involve considerably more isotropic distributions. Hardly
any prominent patterns can be discerned in the sky distribution. This
is expressed in a more erratic wandering of gravity dipole and
quadrupole directions (lower frame, fig.~\ref{fig:aitofflcdm}). This
also implies a more substantial contribution of shot noise effects.
The latter represent a major source for deviations between the
velocity dipole and shear flow quadrupole and the gravity dipole and
quadrupole. The smaller coherence length of the $\tau$CDM fluctuations
and the more randomly oriented contributions by the individual
external matter concentrations may therefore be directly related to
the lower level of coincidence between velocity and gravity directions
than in the case of the more prominent anisotropies in the
$\Lambda$CDM cosmology. In this, we have to realize that the amplitude
of dipole and quadrupolar contributions between the two scenarios are
not too different (cf. table~\ref{table:nbfam30bs}). The less
prominent anisotropies in the $\tau$CDM catalogues are therefore
compensated by a higher average matter density.


\subsection{Multipole Components: Summary}

The above results reassure the fact that the external tidal field can
be well characterized by its main multipole components, the bulk flow
and velocity shear. In terms of multipole amplitude convergence, these
results show a better agreement for the $\Lambda$CDM model than for
the $\tau$CDM one.  This is due to the intrinsic characteristics of
both cosmic models. As has been discussed in Section~\ref{sec:setupc},
and may be directly appreciated from Fig.~\ref{fig:pksig}, the
relatively lower amplitude of the $\tau$CDM perturbations is
compensated by a higher mass content. It leads to an equally strong
external gravitational influence. On the other hand, the smaller
spatial coherence of density features in the $\tau$CDM scenario causes
the orientation of the gravity dipole and quadrupoles to be rather
jittery. The direction of the cumulative gravitational force in the
$\tau$CDM scenario wanders erratically over the sky as we move further
out from the local volume. This differs from the situation in the
$\Lambda$CDM samples, where we observe a consistent, systematic and
coherent convergence towards the final dipole direction.

The above results confirm the fact that the external tidal velocity
field can be well characterized by its main multipole components, the
bulk flow and velocity shear. This depends to some extent on the
cosmology. In terms of multipole amplitude convergence, these results
show a better agreement for the $\Lambda$CDM than for the $\tau$CDM
model.


\section{Conclusions}

In this work we have applied the FAM technique to construct model
velocity fields using mock catalogs resembling the NBG and IRAS-PSC$z$
galaxy catalogs.  The mock catalogs were extracted from $N$-body
simulations in which the central observer mimics some of the
properties of the Local Group environment.  Comparing FAM velocities
obtained from the NBG mock catalogs with those obtained from the
larger PSC$z$ mock catalogs and, finally, to the $N$-body velocities,
allowed us to quantify the importance of the gravity field generated
by the mass distribution within and beyond the LS.

Neglecting the mass distribution outside the LS leads to a systematic
underestimate of the gravity field. The amplitude of this bias depends
on the amount of power on scales larger than the LS, and thus on the
cosmological models. In a $\tau$CDM universe model peculiar velocities
are $\sim 20 \%$ smaller than the true ones.  In the case of a
$\Lambda$CDM model, which has more power on large scales, model
velocities are underestimated by $\sim 35 \%$.

The results of the described FAM analyses are encouraging in the sense
that the presently available all-sky, flux limited catalogs such as
PSC$z$ appear to be capable of accounting for the major share of the
velocity field on the scale of the Local Supercluster. While the
$30\hmpc$ restricted NBG sample showed a substantial deficiency in its
capacity to generate the local cosmic motions, in particular in the
case of the $\Lambda$CDM Universe models, in both cases the $100\hmpc$
mock samples appear to embody nearly all matter concentrations
responsible for the generated velocities in our local (NBG catalog)
neighbourhood.

Also we notice a telling difference between the performance of both
FAM$_{30}$ and FAM$_{100}$ reconstructions for the case of the
$\Lambda$CDM cosmology catalogs on the one hand and the $\tau$CDM
model catalogs on the other hand. The fact that the $\Lambda$CDM model
involves substantially more power on large scales, $r > 30\hmpc$, than
the $\tau$CDM model is reflected in the better quality of the
FAM$_{30}$ reconstructions for the $\tau$CDM catalogs. The presence of
substantial mass inhomogeneities with a scale in excess of that of the
local Universe regions implies a larger external contribution to the
local velocity field. This is also borne out by the fact that for the
$\Lambda$CDM catalogs we see a considerable improvement in velocity
field reconstruction quality going from the FAM$_{30}$ to the
FAM$_{100}$ reconstructions (see Table~\ref{table:multplbs}), while
this is far less so for the $\tau$CDM catalogs.

Of course, whether the resulting models do indeed form an unbiased
representation of the actual velocity field will to some extent also
depend on whether the galaxy distribution in the flux limited galaxy
catalogs does represent an unbiased reflection of the actual
(external) mass distribution surrounding the Local Supercluster
resembling region. The results of recent studies (Verde et al.
\cite{verde2002}, Lahav et al. \cite{lahav2002}, Tegmark, Zaldarriaga
\& Hamilton \cite{tegzalham2001} and Branchini, Dekel \& Sigad
\cite{brandeksig2002}) are quite encouraging in this respect. They
seem to indicate, certainly on scales larger than $\approx 5 \hmpc$,
that both IRAS and 2dF galaxies trace the underlying mass distribution
in an unbiased fashion.

Nonetheless, observations along the lines of the presented mock
catalogs seem to suggest that a proper analysis of Local Universe
dynamics based on a combination of information of local small-scale
(peculiar) galaxy velocities and a rough yet well-founded idea of the
matter distribution on scales of a few hundred $100\hmpc$ may help us
towards acquiring far more insight into the dynamical history of the
emergence and assembly of the striking nonlinear patterns we have
discovered in the large scale matter distribution.  Moreover, we have
uncovered evidence that a meticulous point-to-point analysis of such
velocity samples may help towards modelling the total local force
field, including a proper model for the external forces.

When modeling the peculiar velocity of a LS look-alike region by only
considering the matter distribution within $30 \hmpc$, the end product
is a biased velocity field lacking of any large scale signature.
This bias can be eliminated by accounting for the mass distribution
beyond the LS. Our experiments demonstrate that sampling the mass
distribution out to scales of $100\hmpc$, in a flux limited fashion,
is sufficient to account for the large scale contribution to the
peculiar velocities in our cosmological neighborhood.  More precisely,
we have found that the cosmic velocity field within the LS, modeled by
FAM using the mass distribution traced by PSC$z$ galaxies out to
$100\hmpc$ is unbiased.  The differences between true and FAM velocity
field are random and mainly occur in high density environments which
are dominated by virial motions that are not modeled correctly by FAM.

The gravity and velocity fields generated by the mass distribution
beyond the scale of the Local Supercluster are well characterized by
their bulk flow and shear components.  Therefore, one can obtain an
unbiased model velocity field by superimposing a local model velocity
field within the Local Supercluster to the bulk flow and shear
components of the velocity field generated by the mass distribution
between 30 and $100\hmpc$.

These considerations suggest that velocity models which only consider
the dynamics within the Local Superclusters might have been affected
by systematic errors. In particular, our work suggests that, when
compared with observed velocities, they might have underestimated the
value of the density parameter, $\Omega_m$, by 15-25 \%. 
However, the analysis of Shaya, Tully and Peebles (\cite{shaya}), based
on the galaxy distribution in the Local Supercluster, shows that a
lower, not a larger, value of $\Omega_m$ is found when complementing
the local mass distribution with the large scale one traced by rich
Abell clusters.  A more precise evaluation of this bias will be
performed in a future work in which we will perform the same analysis
presented here using a new set of mock catalogs that are constrain to
reproduce the distribution of the mass in our local Universe (see
e.g. van de Weygaert \& Hoffman
\cite{vdwhof}, Mathis {\it et al.}  \cite{mathis2002}, Klypin {\it et
  al.} \cite{klypin}).

Furthermore, our analysis shows that all model velocity fields of the
Local Supercluster which are based on the PSC$z$ catalog (e.g.
Branchini {\it et al.} \cite{branchini99}, Schmoldt {\it et al.}
\cite{schmoldt99}, Valentine {\it et al.} \cite{valentine}, Sharpe
{\it et al.} \cite{sharpe}) are free from systematic biases arising
from having neglected the large scale contribution from scales beyond
its realm. Moreover, since the IRAS PSC$z$ survey is considerably
deeper than $100\hmpc$, it is reasonable to assume that the PSC$z$
catalog can be used to predict unbiased velocities well beyond our
Local Supercluster
that, if compared with observed galaxy peculiar velocities,
can discriminate among different cosmologies 
characterized by different values of $\Omega_m$, like the $\Lambda$CDM and 
$\tau$CDM models.

The plausibility of this hypothesis has been
recently confirmed by the analysis of Hoffman {\it et al.}
(\cite{hoffman01}) that shows that the bulk and shear components of
the external velocity field in the local universe inferred from the
peculiar velocities in the Mark III catalog (Willick {\it et al.}
\cite{willicka}, \cite{willickb}) are qualitatively consistent with
those expected from the mass distribution traced by IRAS PSC$z$
galaxies. On the other hand, the claim on the basis of the SMAC
cluster peculiar velocity sample (Hudson et al.~\cite{hudsv}) of an
extra $225 \kms$ bulk flow component generated by matter
concentrations on a scale exceeding $100\hmpc$ should issue some
caution with respect to claims of having accounted for all external
influences on the local cosmic flow.

Coupling the local velocity model provided by FAM to the large scale
contribution provided by linear theory allows to obtain a model
velocity field which is unbiased, nonlinear and fast to compute.  This
means that, for the first time, we are in the position of performing a
large number of experiments aimed at studying the nonlinear evolution
of cosmic structures, such as filaments and clusters, and explore the
role of tidal fields during their gravitational collapse. This relates
to the observation that filaments are forming as a consequence of
anisotropic collapse, induced a compressional tidal force acting
perpendicular to the ``axis'' of the filament. By tracing out the
coherent paths of the compressional modes of the primordial tidal
field one can identify the sites of the later nonlinear filaments
(Bond et al.~\cite{cosmweb}, see Van de Weygaert~\cite{weyfoam}). In
turn this is directly related to cluster locations: the strong
primordial tidal shear is the result of a local quadrupolar mass
distribution. The corresponding overdensities tend to evolve into rich
clusters, explaining the intimate link of clusters and filaments in
the cosmic web.

Finally, it is worth stressing that in this work we have neglected the
fact that we measure galaxy redshifts rather then positions. By means
of an elegant formalism, Phelps~\cite{phelps2000} demonstrated the
feasibility of working out the action principle in redshift space.
With respect to FAM, Nusser \& Branchini (\cite{nusbran}) have shown
that it can be easily implemented in redshift space and Branchini,
Eldar \& Nusser (\cite{branchini02}) demonstrated that it performs
equally well in real and redshift space. Therefore, our unbiased,
nonlinear model velocity field also allows to perform an accurate
correction for redshift space distortions and thus lead to a precise
reconstruction of the mass distribution in real space.  Mapping the
mass in the local universe down to nonlinear scales and comparing it
with the distribution of baryonic mass (in form of stars or diffuse,
ionized gas) is of considerable astrophysical interests as it will
constrain and help understanding the process of galaxy formation and
evolution within the Universe


\begin{acknowledgements}
  
  The authors thank Shaun Cole for allowing the use of his $N$-body
  simulations. E.R.D. thanks W.E.~Schaap for stimulating discussions.
  E.R.D. thanks the Universita' di Roma tre for its hospitality
  while part of this work was done.  E.R.D.~has been supported by The
  National Council for Research and Technology (CONACyT, M\'exico)
  through a scholarship. San Crispino provided unique and inspiring
  guidance.

\end{acknowledgements}


\end{document}